\documentclass[a4paper,12pt]{article}
\usepackage{epsfig}
\usepackage{amssymb,subfigure}
\usepackage{amsfonts}
\usepackage{amsmath}
\usepackage{euscript}
\usepackage{verbatim}
\usepackage{latexsym}
\usepackage[compat=1.0.0]{tikz-feynman}
\usepackage{graphicx,color}
\usepackage{caption}
\usepackage{float}
\usepackage{slashed}
\usepackage{graphicx}
\graphicspath{ {images/} }
\usepackage{ytableau}
\usepackage{mathtools}
\usepackage{mdframed}

\usepackage{mathrsfs} 

\usepackage[colorinlistoftodos]{todonotes}
\usepackage[colorlinks=true, allcolors=blue]{hyperref}

\usepackage{wrapfig}
	\usepackage[T1]{fontenc}
	\usepackage{tikz}
	\usetikzlibrary{decorations.pathmorphing}
\usepackage{tikz}
\usetikzlibrary{shapes.geometric, arrows,patterns,snakes}
\tikzstyle{ellip} = [ellipse, minimum width=3cm, minimum height=1cm,text centered, draw=black]

\newskip\humongous \humongous=0pt plus 1000pt minus 1000pt

\newif\ifdtup

\allowdisplaybreaks[1]

\catcode`\@=11

\@addtoreset{equation}{section}

\def\@normalsize{\@setsize\normalsize{15pt}\xiipt\@xiipt
\abovedisplayskip 14pt plus3pt minus3pt%
\belowdisplayskip \abovedisplayskip
\abovedisplayshortskip \z@ plus3pt%
\belowdisplayshortskip 7pt plus3.5pt minus0pt}

\def\small{\@setsize\small{13.6pt}\xipt\@xipt
\abovedisplayskip 13pt plus3pt minus3pt%
\belowdisplayskip \abovedisplayskip
\abovedisplayshortskip \z@ plus3pt%
\belowdisplayshortskip 7pt plus3.5pt minus0pt
\def\@listi{\parsep 4.5pt plus 2pt minus 1pt
     \itemsep \parsep
     \topsep 9pt plus 3pt minus 3pt}}

\relax

\catcode`@=12

\catcode`\@=11

\def\section{\@startsection{section}{1}{\z@}{3.5ex plus 1ex minus
   .2ex}{2.3ex plus .2ex}{\large\bf}}

\def\SymBoxes#1#2#3#4{\newdimen\un@t \un@t#3%
\raisebox{#1}{\rule{#2\un@t}{#4}\hskip-#2\un@t
\@tempdimb\un@t \advance\@tempdimb by-#4\@tempcntb#2\relax%
\@whilenum{\@tempcntb>0}\do{
\rule{#4}{\un@t}\hskip\@tempdimb \advance\@tempcntb by\m@ne}%
\hskip-#2\un@t \rule[\un@t]{#2\un@t}{#4}%
\rule[\un@t]{#4}{#4}\hskip-#4
\rule{#4}{\un@t}}\hskip-#4}                

\DeclareMathAlphabet{\pazocal}{OMS}{zplm}{m}{n}

\begin{document}

\newcommand{\beq}{\begin{equation}}
\newcommand{\eeq}{\end{equation}}
\newcommand{\bea}{\begin{eqnarray}}
\newcommand{\eea}{\end{eqnarray}}
\newcommand{\beas}{\begin{eqnarray*}}
\newcommand{\eeas}{\end{eqnarray*}}
\newcommand{\defi}{\stackrel{\rm def}{=}}
\newcommand{\non}{\nonumber}
\newcommand{\bquo}{\begin{quote}}
\newcommand{\enqu}{\end{quote}}
\renewcommand{\(}{\begin{equation}}
\renewcommand{\)}{\end{equation}}
\def \eqn#1#2{\begin{equation}#2\label{#1}\end{equation}}
\def\IZ{{\mathbb Z}}
\def\IR{{\mathbb R}}
\def\IC{{\mathbb C}}
\def\IQ{{\mathbb Q}}
\def\de{\partial}
\def\Tr{ \hbox{\rm Tr}}
\def\H{ \hbox{\rm H}}
\def\HE{ \hbox{$\rm H^{even}$}}
\def\HO{ \hbox{$\rm H^{odd}$}}
\def\K{ \hbox{\rm K}}
\def\Im{ \hbox{\rm Im}}
\def\Ker{ \hbox{\rm Ker}}
\def\const{\hbox {\rm const.}}
\def\o{\over}
\def\im{\hbox{\rm Im}}
\def\re{\hbox{\rm Re}}
\def\bra{\langle}\def\ket{\rangle}
\def\Arg{\hbox {\rm Arg}}
\def\Re{\hbox {\rm Re}}
\def\Im{\hbox {\rm Im}}
\def\exo{\hbox {\rm exp}}
\def\diag{\hbox{\rm diag}}
\def\longvert{{\rule[-2mm]{0.1mm}{7mm}}\,}
\def\a{\alpha}
\def\dag{{}^{\dagger}}
\def\tq{{\widetilde q}}
\def\p{{}^{\prime}}
\def\W{W}
\def\N{{\cal N}}
\def\hsp{,\hspace{.7cm}}

\def\br{\nonumber\\}
\def\IZ{{\mathbb Z}}
\def\IR{{\mathbb R}}
\def\IC{{\mathbb C}}
\def\IQ{{\mathbb Q}}
\def\IP{{\mathbb P}}
\def \eqn#1#2{\begin{equation}#2\label{#1}\end{equation}}

\newcommand{\sgm}[1]{\sigma_{#1}}
\newcommand{\idd}{\mathbf{1}}

\newcommand{\C}{\ensuremath{\mathbb C}}
\newcommand{\Z}{\ensuremath{\mathbb Z}}
\newcommand{\R}{\ensuremath{\mathbb R}}
\newcommand{\rp}{\ensuremath{\mathbb {RP}}}
\newcommand{\cp}{\ensuremath{\mathbb {CP}}}
\newcommand{\vac}{\ensuremath{|0\rangle}}
\newcommand{\vact}{\ensuremath{|00\rangle}}
\newcommand{\oc}{\ensuremath{\overline{c}}}

\newcommand{\scrip}{\mathscr{I}^{+}}
\newcommand{\scrim}{\mathscr{I}^{-}}

\begin{titlepage}
\begin{flushright}
CHEP XXXXX
\end{flushright}
\bigskip
\def\thefootnote{\fnsymbol{footnote}}

\begin{center}
{\large
{\bf HKLL for the Non-Normalizable Mode
}
}
\end{center}

\bigskip
\begin{center}
{Budhaditya BHATTACHARJEE$^a$\footnote{\texttt{budhadityab@iisc.ac.in}}, \  Chethan KRISHNAN$^a$\footnote{\texttt{chethan.krishnan.physics@gmail.com}},  \& \ Debajyoti SARKAR$^b$\footnote{\texttt{dsarkar@iiti.ac.in}} }
\vspace{0.1in}


\end{center}

\renewcommand{\thefootnote}{\arabic{footnote}}

\begin{center}
$^a$ {Center for High Energy Physics,\\
Indian Institute of Science, Bangalore 560012, India}
\vspace{0.2in}

$^b$ {Department of Physics, \\ 
Indian Institute of Technology Indore, \\
Khandwa Road, 453552 Indore, India}

\end{center}

\noindent
\begin{center} {\bf Abstract} \end{center}
We discuss various aspects of HKLL bulk reconstruction for the free scalar field in AdS$_{d+1}$. First, we consider the spacelike reconstruction kernel for the non-normalizable mode in global coordinates. We construct it as a mode sum. In even bulk dimensions, this can be reproduced using a chordal Green's function approach that we propose. This puts the global AdS results for the non-normalizable mode on an equal footing with results in the literature for the normalizable mode. In Poincar\'e AdS, we present explicit mode sum results in general even and odd dimensions for both normalizable and non-normalizable kernels. For generic scaling dimension $\Delta$, these can be re-written in a form that matches with the global AdS results via an antipodal mapping, plus a remainder. We are not aware of a general argument in the literature for dropping these remainder terms, but we note that a slight complexification of a boundary spatial coordinate (which we call an $i \epsilon$ prescription) allows us to do so in cases where $\Delta$ is (half-) integer. Since the non-normalizable mode turns on a source in the CFT, our primary motivation for considering it is as a step towards understanding linear wave equations in general spacetimes from a holographic perspective. But when the scaling dimension $\Delta$ is in the Breitenlohner-Freedman window, we note that the construction has some interesting features within AdS/CFT.

\vspace{1.6 cm}
\vfill

\end{titlepage}

\setcounter{page}{2}
\tableofcontents

\setcounter{footnote}{0}

\section{Introduction}

The holographic correspondence \cite{Witten:1998qj} between AdS and CFT \cite{Maldacena:1997re, Gubser:1998bc} is remarkable because it provides an apparently complete definition of quantum gravity in asymptotically AdS spacetimes. Since the correspondence is highly non-local, AdS/CFT shifts the mystery of quantum gravity to the question of how the bulk seems to have a local description in terms of the dual holographic variables. Ultimately, we would like to have an intrinsically CFT answer to this question, but a good first step is to write fields that solve the semi-classical bulk equations of motion in terms of boundary operators. At the semi-classical level, this can be accomplished by inverting the usual extrapolate AdS/CFT dictionary for bulk fields and was done in a series of papers \cite{Banks:1998dd, Bena:1999jv} culminating in the celebrated work of Hamilton, Kabat, Lifshitz, and Lowe (HKLL) \cite{Hamilton:2005ju, Hamilton:2006az, Hamilton:2006fh, Hamilton:2007wj}. See \cite{Kabat:2012hp,Sarkar:2014jia, Sarkar:2014dma} for extensions, and \cite{Kajuri:2020vxf, DeJonckheere:2017qkk} for reviews. These papers write local bulk operators containing only the normalizable mode as an integral of local CFT operators on (sub-)regions of the AdS boundary. In other words, local bulk operators can be described using certain non-local operators in the boundary theory. 

HKLL construction is typically done for the normalizable mode \cite{Balasubramanian:1998sn}. This is natural because the non-normalizable bulk solution is best thought of as a deformation of the CFT rather than as an operator in the spectrum of the CFT. Despite this, at the level of free probe fields, nothing prevents us from doing an analogue of HKLL construction for the non-normalizable mode as well -- it can be viewed as an exercise in solving bulk wave equations with non-standard boundary conditions. A further fact that motivates such a calculation is that within the Breitenlohner-Freedman (BF) window of masses \cite{Breitenlohner:1982bm, Klebanov:1999tb} both solutions of the wave equation are acceptable as genuine operators in the CFT. So it is useful to develop the formalism for the ``other'' mode as well. This is the context of the present paper. While this is of intrinsic technical interest in AdS/CFT, as we have just outlined, we also have other (more conceptual) motivations for doing this. These motivations have their origins in questions of flat space holography that have come up in \cite{Krishnan:2016mcj, Basu:2016srp, Bhattacharjee:2019xhb, Krishnan:2019ygy, Krishnan:2020oun}. The way the two modes of a wave equation are organized in flat space is seemingly distinct from that in AdS. Depending on whether we choose the holographic screen to be $\mathscr{I}$ \cite{Krishnan:2019ygy}, or a timelike cut-off \cite{Bhattacharjee:2019xhb}, the data can be stored in terms of ingoing/outgoing modes \cite{Krishnan:2020oun}, or in terms of a {\em bulk} source and a homogeneous mode \cite{Bhattacharjee:2019xhb}. This is to be contrasted with the normalizable and non-normalizable solutions\footnote{Note that the latter corresponds to a {\em boundary} source in AdS.} that arise in AdS. The re-organization of holographic data in flat space makes it interesting to understand the bulk-reconstruction aspects of even the non-normalizable mode in AdS. In any event, the HKLL kernel for the non-normalizable mode will be a primary object of interest in this paper, and some related questions in flat space holography will be discussed elsewhere.

Our goal, then, is simply to write down the bulk field in terms of the two independent boundary modes in the schematic form 
\bea
\Phi(b) = \int K_n(b;x)\ \phi_n(x) +\int K_{nn}(b;x)\ \phi_{nn}(x) , \label{genstruc}
\eea
where $b$ stands for a bulk location, and the integrals are over the boundary (schematically captured by the coordinate $x$). The existence of the two independent modes is a property of second-order PDEs, and in the context of AdS, the subscript $n$ denotes the normalizable mode and  $nn$, the non-normalizable one. Our task is to identify the corresponding kernels -- the various subtleties will be elucidated as we proceed. Note that it is crucial for our discussion here that we are working with linear wave equations so that we can simply sum the two modes together. We will conduct our discussion at the level of these two boundary modes, and they will implicitly determine the holographic data, namely the expectation value and the CFT source. When the non-normalizable mode is set to zero, as is well-known, the expectation value is simply the normalizable mode \cite{Balasubramanian:1998sn, Kleb}. But when there is a non-normalizable mode, extracting the expectation value is less trivial, see, for e.g. the discussion in section 2.2 of \cite{Bala2}. 

It was noted in \cite{Bhattacharjee:2019xhb} that in flat space, the solutions of linear wave equations lead to a similar structure to \eqref{genstruc}, where the analogue of the non-normalizable mode is a {\em bulk source localized on a holographic screen}, and the normalizable mode is replaced by the homogeneous solution. It was pointed out that the structure is, in fact, identical in AdS as well, with the nice extra property that when the screen is moved to the AdS boundary, this {\em bulk} source turns into the {\em boundary} source after the usual radial scaling of the non-normalizable mode. In other words, the structure of the two modes has a nice understanding in spacetimes more general than AdS, with the structure reducing to the usual story in AdS when we take the source to the AdS boundary. This is one of our motivations for believing that it is worthwhile understanding the general structure \eqref{genstruc} better.

\subsection{Summary of the Paper}
 
In constructing the kernel for the non-normalizable mode, we find natural variations of results for the normalizable case. We try to give a unified presentation where (hopefully) the context and general ideas are also clear because the subject is riddled with various technicalities and special cases. We first consider global AdS and define the reconstruction kernel for the non-normalizable mode in two ways -- using a mode sum approach as well as a spacelike Green's function approach. The mode sum approach proceeds analogously to the normalizable case \cite{Hamilton:2006az}, and we obtain explicit kernels in even and odd dimensions. The spacelike Green's function approach relies on first constructing a Green's function in terms of the chordal distance and applies only in even-dimensional AdS.\footnote{We are not aware of a compelling discussion in the literature of why the chordal distance method only applies in even dimensions. It seems plausible to us that it is related to the fact that the Huygens principle for wave propagation applies only in even dimensions. In an appendix, we show that the normalizable HKLL kernel obtained via the chordal distance approach vanishes in any real non-even AdS dimension.} For the normalizable mode, it matched with the (even-dimensional) mode sum construction  \cite{Hamilton:2006az}. In this paper, we develop a similar spacelike Green's function approach using the chordal distance for the non-normalizable mode. We do this in even dimensions, where we expect it to be reliable. We show that the result indeed matches the explicit mode sum result. A notable feature of the spacelike chordal Green's function approach is that, unlike in Euclidean signature \cite{Witten:1998qj}, the normalization is to be fixed by integrating only over the spacelike separated region of the boundary. This is natural, and this is necessary for the matching to work.


Going ahead to the Poincar\'e patch, in both even and odd-dimensional cases, we write down formulas for the kernels by explicitly doing the mode sum integrals. Explicit mode sum integrals have previously been written down in some specific dimensions for the normalizable mode \cite{Bena:1999jv, Hamilton:2006fh}, we generalize them to arbitrary even and odd dimensions and also present the expressions for the non-normalizable mode.\footnote{These results hold for generic values of the scalar field mass $m$. In an appendix, we also write down explicit evaluations of the non-normalizable mode sum integrals for special values of the mass when $\nu \equiv \sqrt{\frac{d^2}{4}+m^2}$  is an integer. In this special case, the general solution of the scalar field in AdS contains Bessel functions of the second kind.} Using some hypergeometric identities, we can re-write these results in a form that makes the connection with the global AdS results more plausible. In particular, for any value of the
scaling dimension of the scalar field $\Delta$, the mode summed Poincar\'e expressions can be written in a form that matches precisely with the global AdS kernel via an antipodal matching, plus some remainder terms. As far as we are aware, these remainder terms have not been investigated in the literature, except in the case of AdS$_3$ for integer scaling dimensions $\Delta \ge 2$. In that case, an argument was provided in appendix C of \cite{Hamilton:2006az} for why these terms can be safely omitted from the kernel.\footnote{The restriction that the lowest value $\Delta$ can take is 2 for the argument to go through in AdS$_3$, was not emphasized there.} We will not settle this issue here for all values of $\Delta$ and $d$, but we find that slightly complexifying a suitable radial spatial direction of the boundary (we will call it an $i \epsilon$-prescription) leads to an immediate generalization of the argument in appendix C of \cite{Hamilton:2006az} that applies to all half-integer $\Delta \ge \frac{d}{2}$ in even-dimensional AdS$_{d+1}$ and to all integer $\Delta \ge d$ in odd-dimensional AdS$_{d+1}$, for the normalizable mode. Our observation can be viewed as a natural generalization of appendix C of \cite{Hamilton:2006az}. 

The Poincar\'e kernels that we write down via our hypergeometric identities are initially supported over the entire Poincar\'e boundary. But in even-dimensional AdS, as we mentioned above, they can be restricted to the spacelike separated region of the Poincar\'e boundary. This is via an argument that is closely related to an antipodal identification argument for the normalizable mode that was presented in \cite{Hamilton:2006az} for relating the global and Poincar\'e kernels. We show that this argument can be extended to the non-normalizable mode as well, where the phases involved are different but are precisely suited for the matching to work. We also demonstrate the matching between the Poincar\'e and global non-normalizable kernels in odd AdS via a straightforward adaptation of the normalizable results of \cite{Hamilton:2006az}. 



In the next few sections, first, we develop the mode sum, and chordal Green's function approaches for global AdS. Then we turn to the mode sum kernels for the Poincar\'e patch and then discuss aspects of the antipodal mapping, which helps to connect to the global results. We will also show that when the scaling dimensions are within the BF window, our results have some particularly nice features. Various appendices are dedicated to exploring various ideas and technicalities not emphasized in the main body of the paper. In appendix A, we demonstrate that the spacelike Green's function approach of \cite{Hamilton:2006az, Heemskerk:2012mn} for the normalizable mode leads to a trivial kernel if the AdS is not even-dimensional. Appendix D discusses the $i \epsilon$-prescription in a spacelike boundary coordinate that is natural in some of these discussions. The argument is of some elegance, and we feel that it may be of broader interest. Appendix E writes down explicit formulas for bulk reconstruction kernels in arbitrary dimensions for the normalizable and non-normalizable modes by making boundary coordinates imaginary -- this generalizes the AdS$_3$ results for the normalizable mode in \cite{Hamilton:2006fh}. Appendix F writes down the kernel for the special case when the mass of the scalar is non-generic, $\nu \equiv \sqrt{\frac{d^2}{4}+m^2} \in \IZ$.  Other appendices contain technical results, including evaluations of some integrals, which are useful in the main body of the paper. Throughout the paper, we have tried to present explicit formulas and also to emphasize ambiguities and open problems. See \cite{group} for some recent papers that are on the topic of bulk reconstruction.


\section{Mode-Sum Kernel in Global AdS}

In this section, we derive the expressions for the spacelike bulk reconstruction kernel corresponding to the non-normalizable mode in global AdS as a mode sum. This is a close adaptation of the procedure outlined in \cite{Hamilton:2006az} for the normalizable mode. We present it in some detail to establish notation and because some of these expressions will be useful later. 

The bulk wave equation in global AdS is 
\begin{align}
    -\partial^{2}_{\tau}\Phi + \partial^{2}_{\rho}\Phi + (d-1)\sec\rho\csc\rho\partial_{\rho}\Phi - \csc^{2}\rho \nabla^{2}_{\Omega}\Phi - m^{2}R^{2}\sec^{2}\rho \Phi = 0
\end{align}
The solution to this wave equation is
\begin{align}
    \Phi(\tau,\rho, \Omega) = \Phi_{1}(\tau, \rho, \Omega) + \Phi_{2}(\tau, \rho,\Omega)
\end{align}
where $\Phi_{1}$ is the normalizable mode and $\Phi_{2}$ is the non-normalizable mode. Their explicit expressions are
\begin{align}
    \Phi_{1}(\tau, \rho, \Omega) &= \sum_{n = 0}^{\infty}\sum_{l, m}a_{n l m}e^{- i \omega_{n, 1} \tau}(\cos\rho)^{\Delta}(\sin \rho)^{l} P_{n}^{\Delta - \frac{d}{2}, l + \frac{d}{2} - 1}(-\cos 2\rho)Y_{l, m}(\Omega) + \text{c.c} \\
    \Phi_{2}(\tau, \rho, \Omega) &= \sum_{n = 0}^{\infty}\sum_{l, m}b_{n l m} e^{- i \omega_{n, 2} \tau}(\cos\rho)^{d - \Delta}(\sin \rho)^{l} P_{n}^{\frac{d}{2} - \Delta, l + \frac{d}{2} - 1}(-\cos 2\rho)Y_{l, m}(\Omega) + \text{c.c}
\end{align}
in terms of Jacobi polynomials. The quantization conditions for the normalizable and non-normalizable modes\footnote{Note that the quantization condition on the non-normalizable mode is a restriction we are choosing to impose for aesthetic reasons. This is unlike in the case of the normalizable mode, where such a condition is $necessary$ -- the normalizable solutions have a basis of normal modes. In general, especially when outside the BF window, it is not necessary that such a condition be imposed on the non-normalizable solution. But we will find that imposing such a restriction results in a final kernel which matches nicely with appropriate expressions obtained via the chordal Green function approach, Poincare patch expresions, etc. It may be interesting to investigate this point further, but we will not undertake it here.} give the following expressions for $\omega$ \cite{Balasubramanian:1998sn}:
\begin{align}
    \omega_{n, 1} &= \Delta + l + 2 n \\
    \omega_{n, 2} &= d - \Delta + l + 2 n
\end{align}
We will focus on the non-normalizable mode in what follows.

\subsection{Even AdS}

In order to obtain the expression for the kernel in even-dimensional\footnote{We work with AdS$_{d+1}$, so even AdS corresponds to odd $d$.} AdS, we focus first on the center of AdS ($\rho=0$) where only the s-wave ($l= 0$) contributes. The result can be extended to arbitrary bulk points using AdS isometries. 
\begin{align}
    \Phi_{2}(\tau, \rho = 0, \Omega) = \sum_{n = 0}^{\infty}b_{n} e^{-i(2n + d - \Delta)\tau}P_{n}^{\frac{d}{2} - \Delta, \frac{d}{2} - 1}(-1) + \text{c.c}
\end{align}
The other ingredient required is the s-wave part of the boundary field, obtained by extracting the non-normalizable scaling $(\cos \rho)^{d-\Delta}$. This is given by 
\begin{align}
    \Phi_{0}(\tau) \equiv \Phi_{0 +}(\tau) + \Phi_{0 -}(\tau)
\end{align}
where the boundary field is split into its positive and negative frequency modes $\Phi_{0 \pm}(\tau)$ which are given by
\begin{align}
    \Phi_{0 +} &= \sum_{n = 0}^{\infty}b_{n} e^{-i(2n + d - \Delta)\tau}P_{n}^{\frac{d}{2} - \Delta, \frac{d}{2} - 1}(1)  \label{positiveglobal}\\
    \Phi_{0 -} &= \sum_{n = 0}^{\infty}b^{*}_{n} e^{i(2n + d - \Delta)\tau}P_{n}^{\frac{d}{2} - \Delta, \frac{d}{2} - 1}(1)
\end{align}
In terms of $\Phi_{0 +}(\tau)$, $b_{n}$ can be written as (where $V_{d-1}$ is the volume of the sphere $S^{d-1}$)
\begin{align}
    b_{n} = \frac{1}{\pi V_{d-1} P_{n}^{\frac{d}{2}-\Delta, \frac{d}{2} - 1}(1)}\int_{-\pi/2}^{\pi/2}\mathrm{d}\tau \int \mathrm{d}\Omega \sqrt{g_{\Omega}}\,e^{i(2n + d - \Delta)\tau}\Phi_{0 +}(\tau)
\end{align}
The bulk field s-wave at the origin ($\tau' = 0, \rho'= 0$) is then written as
\begin{align}
    \Phi_{2}|_{\text{origin}} = \int_{-\pi/2}^{\pi/2} \mathrm{d}\tau \int \mathrm{d}^{d-1}\Omega \sqrt{g_{\Omega}}\, K_{+}(\rho', \tau', \Omega' | \tau, \Omega)\Phi_{0 +}(\tau, \Omega) + \text{c.c} \label{kernelcenter}
\end{align}
where 
\begin{align}
    K_{+} = \frac{1}{\pi V_{d-1}}\sum_{n = 0}^{\infty}e^{i(2n + d -\Delta)\tau}\frac{P_{n}^{\frac{d}{2} - \Delta, \frac{d}{2} - 1}(-1)}{P_{n}^{\frac{d}{2} - \Delta, \frac{d}{2} - 1}(1)}
\end{align}
This summation can be performed by using the explicit form of the Legendre $P$ functions in terms of $\Gamma$ functions, and using the series representation of Hypergeometric $\;_{2}F_{1}$
\begin{align}\label{Kplus}
    K_{+} = \frac{1}{\pi V_{d-1}} z^{\frac{d-\Delta}{2}}\;_{2}F_{1}(1, \frac{d}{2}; \frac{d}{2}  -\Delta + 1; - z)
\end{align}
using the notation $z = e^{2 i \tau}$. Using the hypergeometric identity \eqref{id2} which is valid in even AdS, 
we can re-write \eqref{Kplus} as  
\begin{align}\label{result}
    K_{+} = \frac{1}{\pi V_{d-1}} &z^{\frac{d - \Delta}{2}}\Bigg( \frac{\Gamma(\frac{d}{2} - \Delta + 1)\Gamma(\frac{d}{2}-1)}{\Gamma(\frac{d}{2})\Gamma(\frac{d}{2}-\Delta)}z^{-1}\;_{2}F_{1}(1, 1 + \Delta - \frac{d}{2}; 2 - \frac{d}{2}; -\frac{1}{z}) \notag\\ &+ \frac{\Gamma(\frac{d}{2} - \Delta + 1)\Gamma(1 - \frac{d}{2})}{\Gamma(1-\Delta)}z^{-\frac{d}{2}}\;_{2}F_{1}(-\Delta, \frac{d}{2}; \frac{d}{2};-\frac{1}{z}) \Bigg)
\end{align}

The first term in \eqref{result} can be expanded in a series in $z$. The series has the form $z^{\frac{d-\Delta}{2}}\sum_{n = 0}c_{n} z^{-n}$. It can be shown that each term in this series will vanish when integrated against $\Phi_{0 +}(\tau)$, so it can be dropped from the full kernel expression. The surviving term in \eqref{result} simplifies to
\begin{align}
    K_{+} = \frac{1}{\pi V_{d-1}} z^{\frac{d-\Delta}{2}}\frac{\Gamma(\frac{d}{2} - \Delta + 1)\Gamma(1 - \frac{d}{2})}{\Gamma(1 - \Delta)}z^{-\frac{d}{2}}(1 + \frac{1}{z})^{-\Delta}
\end{align}
Note that 
\bea
\sqrt{z} + \frac{1}{\sqrt{z}} = 2 \cos\tau = \lim_{\rho \rightarrow \frac{\pi}{2}}2 \sigma(\tau, \Omega| \tau' = 0, \rho'=0, \Omega') \cos\rho, \label{ind}
\eea 
where $\sigma$ is the AdS covariant length, see \eqref{chordform} for the explicit formula. In terms of $\sigma$, the final kernel becomes (using the expression for $V_{d-1}$)
\begin{align}
    K_{2}^G(\tau, \Omega| \tau' = 0, \rho'=0, \Omega') &= - \frac{2^{-\Delta}\Gamma(\Delta)\tan \pi \Delta}{2 \pi^{\frac{d}{2}}\Gamma(\Delta - \frac{d}{2})}\lim_{\rho \rightarrow \frac{\pi}{2}}(\sigma \cos\rho)^{-\Delta}\theta(\text{spacelike}) \notag \\
    &\equiv a'_{d \Delta}\lim_{\rho \rightarrow \frac{\pi}{2}}(\sigma \cos\rho)^{-\Delta}\theta(\text{spacelike}) \label{evencheck}
\end{align}
Since this expression is constructed out of the AdS covariant length, it also holds for arbitrary bulk points. We have introduced the notation $a'_{d \Delta}$ to avoid wasting electrons, later.

\subsection{Odd AdS}

For odd-dimensional AdS, again we can re-write \eqref{Kplus}, but it is important here that $\frac{d}{2} \in \mathbb{Z}$. The following identity is useful
\begin{align}\label{transf2}
    \;_{2}F_{1}(a, a + m; c; z) &= \frac{\Gamma(c)(-z)^{-a -m}}{\Gamma(a+m)\Gamma(c-a)}\sum_{n = 0}^{\infty}\frac{(a)_{n + m}(1- c + a)_{n + m} z^{-n}}{n! (n+m)!}\left(\ln(-z) + h_{n}\right)\notag \\ &+ \frac{\Gamma(c)(-z)^{-a}}{\Gamma(a+m)}\sum_{n = 0}^{m - 1}\frac{\Gamma(m-n)(a)_{n}}{\Gamma(c - a - n)n!}z^{-n}
\end{align}
where $h_{n} = \psi(1+m+n) + \psi(1 + n) - \psi(a+m+n) - \psi(c-a-m-n)$.
This identity does not hold for $c - a \in \mathbb{Z}$. Note that this identity is distinct from \eqref{id4}.

By a similar argument as in the previous subsection, it can be shown that the only term that contributes to the kernel is the one proportional to $\ln(-z)$, the other terms vanish when integrated against $\Phi_{0 +}(\tau)$. Using the values of $a, b, c$ from \eqref{Kplus} and \eqref{form0} we get
\begin{align}
    K_{+} &= \frac{1}{\pi V_{d-1}}z^{\frac{d-\Delta}{2}}\frac{\ln(z)\Gamma(\frac{d}{2} - \Delta + 1)}{\Gamma(\frac{d}{2})\Gamma(\frac{d}{2}-\Delta)\Gamma(1 - \frac{d}{2} + \Delta)}\frac{\pi z^{-d/2}}{\sin\pi\Delta \Gamma(1 - \Delta)}\left(1 + \frac{1}{z}\right)^{-\Delta} \notag \\
    &= \frac{\Gamma(\frac{d}{2} - \Delta + 1)\sin\pi(\frac{d}{2} - \Delta)}{\Gamma(\frac{d}{2})\sin\pi\Delta\Gamma(1-\Delta)}(2\sigma \cos\rho)^{-\Delta}\ln z \label{KplusOdd}
\end{align}
In the second line we have partially re-written some of the terms using the invariant chordal distance \eqref{ind}. Our goal is now to write the entire expression in this way, so that we can invoke AdS isometries to move away from the center of AdS.
 
We first observe that the series expansion of $(\sigma \cos\rho)^{-\Delta}$ in powers of $z$ can be re-written in the two forms 
\begin{align}\label{ser1&2}
    \lim_{\rho \rightarrow \pi/2}(2 \sigma \cos\rho)^{-\Delta} = z^{-\Delta/2}\sum_{n = 0}^{\infty}c_{n}z^{-n} = z^{\Delta/2}\sum_{n = 0}^{\infty}d_{n}z^{n}
\end{align}
The coefficients can be determined, but are not important. The point is that the first form vanishes when integrated against positive frequency boundary modes and the second form vanishes when integrated against negative frequency modes. Therefore, we obtain
\begin{align}\label{vanish}
    \int_{-\pi/2}^{\pi/2}\mathrm{d}\tau \int \mathrm{d}\Omega\, \sqrt{g_{\Omega}} \,(\sigma \cos\rho)^{-\Delta}(\phi_{0 +}(\tau) + \phi_{0-}(\tau)) = 0
\end{align}
Since $z^{*} = \frac{1}{z}$, we can also re-write \eqref{kernelcenter} as 
\begin{align}
    \Phi|_{\text{origin}} = A\int_{-\pi/2}^{\pi/2}\mathrm{d}\tau \int \mathrm{d}\Omega\, \sqrt{g_{\Omega}}\, (2 \sigma \cos\rho)^{-\Delta}\ln z (\phi_{0 +}(\tau) - \phi_{0-}(\tau))
\end{align}
where $A = \frac{\Gamma(\frac{d}{2} - \Delta + 1)\sin\pi(\frac{d}{2} - \Delta)}{\Gamma(\frac{d}{2})\sin\pi\Delta\Gamma(1-\Delta)}$. Following \cite{Hamilton:2006az} and differentiating \eqref{vanish} with respect to $\Delta$, and using \eqref{ser1&2}, we find
\begin{align}
    \int_{-\pi/2}^{\pi/2}\mathrm{d}\tau \int \mathrm{d}\Omega \,\sqrt{g_{\Omega}}\,& (\sigma \cos\rho)^{-\Delta}\ln z (\phi_{0 +}(\tau) - \phi_{0-}(\tau)) \notag \\ &= 2\int_{-\pi/2}^{\pi/2}\mathrm{d}\tau \int \mathrm{d}\Omega \,\sqrt{g_{\Omega}}\lim_{\rho \rightarrow \pi/2}(\sigma \cos\rho)^{-\Delta}\ln(\sigma \cos\rho) \phi_{0}(\tau)
\end{align}
This lets us express the value of the field at the origin of AdS in terms of an integral over points on the boundary that are spacelike separated from the origin in an AdS covariant form
\begin{align}
    \Phi|_{\text{origin}} = 2 A\int_{-\pi/2}^{\pi/2}\mathrm{d}\tau \int \mathrm{d}\Omega \sqrt{g_{\Omega}}\lim_{\rho \rightarrow \pi/2}(2 \sigma \cos\rho)^{-\Delta}\ln(\sigma \cos\rho) \phi_{0}(\tau)
\end{align}
This form allows us to extend the result to arbitrary bulk points via AdS isometry. The final form of the kernel is
\begin{align}
    K_{2}^G &= \frac{(-1)^{\frac{d}{2} + 1}2^{-\Delta}\Gamma(\frac{d}{2} - \Delta + 1)}{\pi^{\frac{d}{2} + 1}\Gamma(1- \Delta)}\lim_{\rho \rightarrow \pi/2}(\sigma \cos\rho)^{-\Delta}\ln(\sigma \cos\rho)\theta(\text{spacelike}) \notag \\
    &\equiv c'_{d \Delta}\lim_{\rho \rightarrow \pi/2}(\sigma \cos\rho)^{-\Delta}\ln(\sigma \cos\rho)\theta(\text{spacelike}) \label{KplusOddFinal}
\end{align}
We introduce the notation $c'_{d \Delta}$ 
to reduce future clutter. Note that as a result of the manipulations we have done above to write the kernel in an AdS covariant form, in the odd AdS case, the kernel is to be integrated against the full boundary mode and not just its positive frequency part. This will be important when we try to relate the global result here with the Poincar\'e result later.

\section{Chordal Green's Function in Global AdS}\label{chordal1}

We will consider the following bulk-to-bulk Green's function, where the $\theta(\mathrm{spacelike})$ indicates the region spacelike separated from the unprimed bulk point in global coordinates:\footnote{The superscript $G$ denotes that the object is defined in global coordinates. Note that we are working in the large $\sigma$-limit when near the boundary, so delta functions arising from radial derivatives acting on the step function can be ignored.}
\begin{align}
	\mathcal{G}^G_{\Delta}(\sigma) = \frac{2^{-\Delta}C^G_{\Delta}}{2\Delta - d}\sigma^{-\Delta}\;_{2}F_{1}(\frac{\Delta}{2}, \frac{\Delta + 1}{2};\Delta - \frac{d}{2} + 1; \frac{1}{\sigma^{2}})\,\theta(\mathrm{spacelike})\label{greenNNormGAdS2}
\end{align}
The claim is that this is a natural Green's function to be used for the non-normalizable mode -- this will be explained further when we re-visit this discussion in Poincar\'e coordinates in section \ref{chordal2}, see also the discussion of the complementary (normalizable) spacelike Green's function in appendix A of \cite{Hamilton:2006az}.
Our goal in this section is to use the above Green's function to reproduce the global HKLL kernel that we arrived at in the last section via mode sum. As we have briefly alluded to before (and will discuss in more detail in section \ref{chordal2}), the chordal distance Green's function we are using above is expected to yield the right answer only in even-dimensional AdS. 

One can use Green's theorem to relate the above Green's function to the kernel and we will do so momentarily. But in order to get a precise match with the mode sum result, we need to fix the normalization factor $C^G_\Delta$. This is what we turn to first. 

The strategy for fixing the normalization is an adaptation of the Euclidean argument due to Witten \cite{Witten:1998qj, Erbin}. 
We first define the bulk-to-boundary propagator. In Poincar\'e coordinates this is defined via
\begin{align}
	\mathcal{K}^P_{\Delta}(z, x; x') = \lim_{z' \rightarrow 0}\sqrt{\gamma_{z'}}(z')^{d-\Delta}n^{z'}\partial_{z'}\mathcal{G}_{\Delta}(z, x; z', x') = 
	\lim_{z' \rightarrow 0}(z')^{-\Delta}z' \partial_{z'}\mathcal{G}^P_{\Delta}(z, x; z', x') \label{Kzzp}
\end{align}
where we have $n^{z'} = \frac{1}{\sqrt{g_{z' z'}}}$ and $\sqrt{\gamma_{z'}} = \frac{1}{z'^{d}}$. We will use the above expression in section \ref{chordal2}.
The analogous definition in global coordinates is (using $n^{\rho'} = \frac{1}{\sqrt{g_{\rho' \rho'}}}$)
\begin{align}
	\mathcal{K}^G_{\Delta}(\rho,\tau,\Omega;\tau',\Omega') &= \lim_{\rho' \rightarrow \pi/2} \sqrt{\gamma_{\rho'}}(\cos\rho')^{d - \Delta}n^{\rho'}\partial_{\rho'}\mathcal{G}^G_{\Delta}(\rho, \tau,\Omega; \tau', \Omega') \notag \\ 
	&= \lim_{\rho' \rightarrow \pi/2}(\cos\rho')^{-\Delta}\cos\rho' \partial_{\rho'}\mathcal{G}^G_{\Delta}(\rho, \tau,\Omega; \tau', \Omega')\label{Krrp}
\end{align}
We will elevate the relations in \eqref{Kzzp} and \eqref{Krrp} to a covariant statement. Such a relation is best understood in terms of the product $\mathcal{K}_{\Delta}\times \ j_{0}$, where $j_{0}$ is the non-normalizable mode on the boundary. In terms of a bulk-to-bulk Green's function $\mathcal{G}_{\Delta}$, this product can be written as 
\begin{align}
	\mathcal{K}_{\Delta}(r, x; x')j_{0}(x') = \lim_{r' \rightarrow r'_{\partial}}\sqrt{\gamma'_{\partial}}j(r', x')n^{r'}\partial_{r'}\mathcal{G}_{\Delta}(r, x; r', x')\label{defK}
\end{align}
where we use the notation $r'$ (and $r$) to denote the ``radial coordinate'' in the chosen coordinate system (for example, $z$ in Poincar\'e and $\rho$ in global coordinates). We use the bulk field $j(r',x')$ instead of $\phi(r',x')$ to instruct the reader to pick  the non-normalizable mode when taking the limit in \eqref{defK}. The subscript $\partial$ implies the value of the respective function on the AdS boundary (for example, the boundary metric is denoted by $\gamma_\partial$ and the boundary value of the radial coordinate is $r_\partial$). All the other coordinates are represented collectively by the $x'$ coordinates. The vector $n^{r'}$ denotes the normal vector to the $r' = \mathrm{constant}$ surface.
Applying this to the global coordinates by using \eqref{Krrp} and the using relation $\cos\rho' \partial_{\rho'}\mathcal{G}_{\Delta}|_{\rho' \rightarrow \pi/2} = - \Delta \mathcal{G}_{\Delta}|_{\rho' \rightarrow \pi/2}$ for the Green's function \eqref{greenNNormGAdS2}, we obtain the following expression 
\begin{align}
	\mathcal{K}^G_{\Delta}(\rho, \tau, \Omega; \tau', \Omega') = \lim_{\rho' \rightarrow \pi/2} - \frac{(2\Delta - d)}{(\cos\rho') ^{\Delta}}\,\mathcal{G}^G_{\Delta}(\rho, \tau,\Omega; \tau', \Omega')
\end{align}
Using this allows us to write the expression for the kernel as
\begin{align}
	\mathcal{K}^G_{\Delta}(\rho, \tau, \Omega; \tau', \Omega') = \lim_{\rho' \rightarrow \pi/2} - C^G_\Delta \,(2\sigma \cos\rho')^{-\Delta}\theta(\mathrm{spacelike})\label{Kfinrho}
\end{align}

In order to properly normalize $\mathcal{G}^G_{\Delta}$, we demand a $\delta$ function normalization for the bulk-boundary propagator in \eqref{Krrp}, in the limit that both points go to the boundary. In Euclidean signature, this was implemented in a somewhat magical way by Witten in \cite{Witten:1998qj}. We will remove the magic by writing the normalization condition in the explicit form
\bea
\lim_{z\rightarrow 0} \int \mathrm{d}^{d}x' \mathcal{K}^P_{\Delta}(z, x; x')j_{0}(x') = \lim_{z\rightarrow 0} j(z, x) 
\eea
in Poincar\'e patch or as
\bea
\lim_{\rho \rightarrow \pi/2} \int \mathrm{d}\tau \mathrm{d}^{d-1}\Omega'\, \mathcal{K}^G_{\Delta}(\rho, \tau, \Omega; \tau',\Omega')\, j_{0}(\tau',\Omega') = \lim_{\rho\rightarrow \pi/2} j(\rho, \tau, \Omega) 
\eea
in global coordinates. These demands fix the corresponding normalization constants. Note that we have not introduced superscripts $P$ or $G$ for the fields $j$ or sources $j_0$, they will be distinguishable by their arguments. We have explicitly done this integral in the Poincar\'e case in section \ref{chordal2} to determine $C_\Delta^P$. To compute the normalization in global coordinates, we need to evaluate 
\bea
C^{G}_{\Delta}\int_{\text{global spacelike}}\mathrm{d}\tau'\mathrm{d}^{d-1}\Omega'(2 \sigma \cos\rho')^{-\Delta}j_{0}(\tau',\Omega') 
\eea
The integration domain is the region of the global boundary that is spacelike separated from the bulk point. 
It turns out that this integral is easiest to evaluate by doing a coordinate change to Poincar\'e coordinates. This turns the above expression into (here $x' = \{\vec{x}',t'\}$)
\bea
C^{G}_{\Delta}\int_{\text{global spacelike}}\mathrm{d}t'\mathrm{d}^{d-1}\vec{x}'\,(2 \sigma z')^{-\Delta}\, j_{0}(t',x') 
= 2C^{G}_{\Delta}\int_{\text{Poincar\'e spacelike}}\mathrm{d}t'\mathrm{d}^{d-1}\vec{x}\, (2 \sigma z')^{-\Delta}\, j_{0}(t',x') \nonumber \\ \label{changevar}
\eea  
In the first expression we have used the Jacobian connecting the measures in the two coordinates as well as the relation between the boundary modes in the two coordinates: 
\begin{align}
\frac{\mathrm{d}\tau' \mathrm{d}^{d-1}\Omega'}{(\cos\rho')^{d}} &= \frac{\mathrm{d}t' \mathrm{d}^{d-1}\vec{x}'}{(z')^{d}} \\
(\cos\rho')^{d-\Delta}j_{0}(\tau',\Omega') &=  (z')^{d-\Delta}j_{0}(t',\vec{x}')
\end{align}
Analogous relations were also used in section 3.1 of \cite{Hamilton:2006az} to connect the normalizable boundary modes in global and Poincar\'e coordinates.
The second expression in \eqref{changevar} restricts the integration range to the spacelike part of the Poincar\'e boundary, and follows from an antipodal identification - this is discussed in great detail in Section \ref{spacelikekernel}.
 The final integral is precisely one that is done in section \ref{chordal2}, to determine $C_\Delta^P$, see equation \eqref{PoinC}. Together with this, we have therefore fixed both $C_{\Delta}^P$ and $C_{\Delta}^G$.  The final expression  for $C^G_{\Delta}$ is
\begin{align}
	C^G_{\Delta} = \frac{\Gamma(\Delta)\tan\pi\Delta}{2\pi^{d/2}\Gamma(\Delta - \frac{d}{2})}
\end{align}

With the normalization at hand, we now proceed to determine the global kernel using Green's theorem starting from \eqref{greenNNormGAdS2}. 
Using the asymptotic behavior of the bulk fields
\begin{align}
	\Phi_{1}(\tau,\rho,\Omega)|_{\rho \rightarrow \frac{\pi}{2}} &\rightarrow (\cos\rho)^{\Delta}\phi_{0}(\tau,\Omega)\\
	\Phi_{2}(\tau, \rho,\Omega)|_{\rho \rightarrow \frac{\pi}{2}} &\rightarrow (\cos\rho)^{d - \Delta}j_{0}(\tau,\Omega)
\end{align}
and the expression for the chordal distance $\sigma$
\begin{align}
	\sigma(\tau, \rho, \Omega| \tau', \rho', \Omega') = \frac{\cos(\tau - \tau') - \sin\rho \sin\rho' \cos(\Omega - \Omega')}{\cos\rho \cos\rho'} \label{chordform}
\end{align}
we observe the following limits (using $k_{\Delta} = \frac{2^{-\Delta-2 }\Gamma(\Delta)\tan\pi\Delta}{\Gamma(\Delta - \frac{d}{2}+1)\pi^{\frac{d}{2}}}$)
\begin{align}
	\mathcal{G}^{G}_{\Delta}(\sigma)|_{\rho' \rightarrow \pi/2} &=k_{\Delta}\sigma^{-\Delta} \\
	\partial_{\rho'}\mathcal{G}^{G}_{\Delta}(\sigma)|_{\rho' \rightarrow \pi/2} &= -\Delta k_{\Delta}\frac{\sigma^{-\Delta}}{\cos\rho'}
\end{align}
Similarly, the bulk solution ($\Phi(\tau,\rho,\Omega) = \Phi_{1}(\tau,\rho,\Omega) + \Phi_{2}(\tau,\rho,\Omega) $) and its derivative behave in the following way at the boundary limit
\begin{align}
	\Phi(\tau,\rho,\Omega)|_{\rho \rightarrow \pi/2} &= (\cos\rho)^{\Delta}\phi_{0}(\tau,\Omega) + (\cos\rho)^{d-\Delta}j_{0}(\tau,\Omega)\\
	\partial_{\rho}\Phi(\tau,\rho,\Omega)|_{\rho \rightarrow \pi/2} &= -\Delta(\cos\rho)^{\Delta-1}\phi_{0}(\tau,\Omega) - (d-\Delta)(\cos\rho)^{d-\Delta - 1}j_{0}(\tau,\Omega)
\end{align}
Using Green's theorem
\begin{align}
	\Phi(\tau,\rho,\Omega) = \int \mathrm{d}\tau' \mathrm{d}^{d-1}\Omega' \sqrt{g'}(\Phi(\tau',\rho',\Omega')\partial_{\rho'}\mathcal{G}^{G}_{\Delta}(\sigma) - \mathcal{G}^{G}_{\Delta}(\sigma)\partial_{\rho'}\Phi(\tau',\rho',\Omega'))|_{\rho' \rightarrow \pi/2}\label{green00GAdSv2}
\end{align}
we get
\begin{align}
	\Phi(\tau,\rho,\Omega) &= \int \mathrm{d}\tau' \mathrm{d}^{d-1}\Omega' \sqrt{g'}(\Phi(\tau',\rho',\Omega')\partial_{\rho'}\mathcal{G}^{G}_{\Delta}(\sigma) - \mathcal{G}^{G}_{\Delta}(\sigma)\partial_{\rho'}\Phi(\tau',\rho',\Omega'))|_{\rho' \rightarrow \pi/2} \nonumber \\
	&= -\int \mathrm{d}\tau' \mathrm{d}^{d-1}\Omega' \sqrt{g'}\Big(((\cos\rho')^{\Delta}\phi_{0}(\tau',\Omega') + (\cos\rho')^{d-\Delta}j_{0}(\tau',\Omega'))\Delta k_{\Delta} \frac{\sigma^{-\Delta}}{\cos\rho'} \nonumber \\ &+ k_{\Delta}\frac{\sigma^{-\Delta}}{\cos\rho'}(\Delta (\cos\rho')^{\Delta}\phi_{0}(\tau',\Omega') + (d-\Delta)(\cos\rho')^{d - \Delta}j_{0}(\tau',\Omega'))\Big) |_{\rho' \rightarrow \pi/2} \nonumber \\
	&= -\int \mathrm{d}\tau' \mathrm{d}^{d-1}\Omega' (\cos\rho')^{-d + 1}\Big((2\Delta - d)k_{\Delta}\frac{\sigma^{-\Delta}}{\cos\rho'}(\cos\rho')^{d-\Delta}j_{0}(\tau',\Omega')\Big) |_{\rho' \rightarrow \pi/2} 
\end{align}
This gives us the following reconstruction kernel (noting that $(2\Delta - d)k_{\Delta} = -a'_{d \Delta}$) for the non-normalizable mode
\begin{equation}
	K^G_{2}(\tau,\rho,\Omega| \tau',\Omega') = a'_{d \Delta}\lim_{\rho' \rightarrow \pi/2}(\sigma \cos\rho')^{-\Delta}  \theta(\text{spacelike})\label{ker2greenGAdSv2}
\end{equation}
This reproduces \eqref{evencheck} precisely.  

Note that in doing the above Green's theorem calculation we could have set the normalizable mode to zero. We have retained it anyway, because the Green's function we are working with is the non-normalizable one, and it precisely picks out the right answer. A further observation that is worth noting is that the Poincar\'e coordinates result comes with an additional factor of $2$, which is expected due to the antipodal mapping between the spacelike regions of the two coordinate systems. We will see this again elsewhere.

\section{Poincar\'e Mode-Sum Kernels}

In this section, we switch gears and consider Poincar\'e AdS. We will write down the mode expansions for the scalar field (for generic and special masses), and write down the kernels as formal inversions of these expressions.

\subsection{Mode Expansions}

We begin by considering a probe scalar field of generic mass in Lorentzian $AdS_{d+1}$. The metric in Poincar\'e coordinates is given by
\begin{equation}
ds^{2} = \frac{-dt^{2} + dz^{2} + d\vec{x}^{2}_{d-1}}{z^{2}}
\end{equation}
The solution to the wave equation in this background is

\begin{equation}\label{a1}
\Phi(x,z) = \int \frac{\mathrm{d}^{d}q}{(2\pi)^{d}}e^{i q.x}z^{\frac{d}{2}}\Big( a(q)J_{\nu}(|q|z) + b(q)J_{-\nu}(|q|z)\Big)
\end{equation}
where $x\equiv (t,\vec{x})$, $\nu = \sqrt{\frac{d^{2}}{4} + m^{2}}$ and $q = (\omega,\vec{k})$, with $|q| = \sqrt{\omega^{2} - |k|^{2}}$. 
The near-boundary behavior of this solution can be see from the asymptotic expansion of the Bessel functions near $z = 0$. 
The coefficients of each $z$ term can be written as a function of $x$. Using the notation $\nu=\Delta - \frac{d}{2} $, we have the following series
\begin{align}\label{exp1}
    \Phi(x,t,z) &=  z^{d- \Delta}j_{0}(x) + z^{d -\Delta + 2}j_{2}(x) + \cdots + z^{d - \Delta + 2n}j_{2n}(x) + \cdots \nonumber  \\
    &+ z^{\Delta}\phi_{0}(x)+z^{\Delta + 2}\phi_{2}(x) + \cdots + z^{\Delta + 2n}\phi_{2n} + \cdots \nonumber \\
    &= \sum_{n=0}^{\infty} z^{d-\Delta + 2n}j_{2n}(x) +  z^{\Delta + 2n}\phi_{2n}(x)
\end{align}
The coefficients at each order are
\begin{align}
	j_{2n}(x) &= \frac{1}{2^{-\nu}}(-1)^{n}\frac{1}{4^{n}\Gamma(n+1)\Gamma(n-\nu + 1)}\int \frac{\mathrm{d}^{d}q}{(2\pi)^{d}}b(q)e^{i q.x}|q|^{2 n - \nu} \label{bx1}\\
    \phi_{2 n}(x) &= \frac{1}{2^{\nu}}(-1)^{n}\frac{1}{4^{n}\Gamma(n+1)\Gamma(n+\nu + 1)}\int \frac{\mathrm{d}^{d}q}{(2\pi)^{d}}a(q)e^{i q.x}|q|^{2 n + \nu} \label{ax1.1}    
\end{align}

The above discussion applies when the mass of the scalar is generic. As customary, this is the case that we will mostly be concerned with in this paper. But in the case where $\nu \in \text{Integer} \equiv p$, the solution of the bulk wave equation involves Bessel functions of the second kind as well. We will present some of the details of the HKLL kernels for the $\nu = p$ case in an appendix. The mode expansion in this case takes the form
\begin{equation}\label{a2}
    \Phi(x,z) = \int \frac{\mathrm{d}^{d}q}{(2\pi)^{d}}e^{i q.x}z^{\frac{d}{2}}\Big( a(q)J_{p}(|q|z) + b(q)Y_{p}(|q|z)\Big)
\end{equation}
The asymptotic expansions of the Bessel $J$ and $Y$ allows us to write the solution again as an expansion in $z$.
\begin{align}\label{phiseries1full}
    \Phi(x,z) &= z^{d - \Delta}j_{0}(x) + z^{d - \Delta + 2}j_{2}(x) + \cdots + z^{d-\Delta + 2n}j_{2n}(x) + \cdots + z^{d - \Delta + 2p - 2}j_{2p-2}(x) \nonumber  \\ 
    &+ \ln(z)\Big(z^{\Delta}\tilde{\phi}_{0}(x) + z^{\Delta + 2}\tilde{\phi}_{2}(x) + \cdots + z^{\Delta + 2n}\tilde{\phi}_{2n}(x) + \cdots)\nonumber \\ 
    &+  z^{\Delta}\phi_{0}(x) + z^{\Delta + 2}\phi_{2}(x) + \cdots + z^{\Delta + 2n}\phi_{2n}(x) + \cdots \nonumber\\
    &= \sum_{n = 0}^{p -1}z^{d - \Delta + 2n}j_{2n}(x)+ \sum_{n=0}^{\infty}\ln(z)z^{\Delta + 2n}\tilde{\phi}_{2n}(x) + \sum_{n=0}^{\infty}z^{\Delta + 2 n}\phi_{2n}(x) 
\end{align}
To get to the above form, we have defined $B_{k} = \frac{(-1)^{k}}{4^{k}\Gamma(k+p+1)\Gamma(k+1)}\Big(\psi(k+1) + \psi(k + p + 1)\Big)$,  $\; D_{k} = \frac{\Gamma(p-k)}{4^{k}\Gamma(k+1)}$ and $A_{k} = \frac{(-1)^{k}}{4^{k}\Gamma(k+1)\Gamma(k+ p + 1)} $,
and we have the following expressions for the $z$-independent coefficients:
\begin{align}
    \phi_{2 n}(x) &= \frac{1}{2^{p}}A_{n}\int \frac{\mathrm{d}^{d}q}{(2\pi)^{d}}a(q)e^{i q.x}|q|^{2 n + p} - \frac{1}{2^{p}\pi}B_{n}\int \frac{\mathrm{d}^{d}q}{(2\pi)^{d}}b(q)e^{i q.x}|q|^{2 n + p}\nonumber \\  &+ \frac{2}{2^{p}\pi}A_{n}\int\frac{\mathrm{d}^{d}q}{(2\pi)^{d}}b(q)e^{i q.x}|q|^{2 n + p}\ln(\frac{|q|}{2})\label{ab2}\\
    \tilde{\phi}_{2 n}(x) &=  \frac{2}{2^{p}\pi}A_{n}\int \frac{\mathrm{d}^{d}q}{(2\pi)^{d}}b(q)e^{i q.x}|q|^{2 n + p} \label{b2}\\
    j_{2n}(x) &= -\frac{1}{2^{-p}\pi}D_{n}\int \frac{\mathrm{d}^{d}q}{(2\pi)^{d}}b(q)e^{i q.x}|q|^{2 n - p} \; \forall \; n \in \{0,p-1\} \label{b3}
\end{align}
Note that none of the $j_{2n}(x)$ combine with $\phi_{2n}(x)$. This is because the highest power of $z$ arising in that term is $d- \Delta + 2(p-1)$. Now, we know that $p = \Delta - \frac{d}{2}$. Therefore, the highest power is $d - \Delta + 2(\Delta - \frac{d}{2} - 1) = \Delta - 2$, which is less than the lowest power of $z$ on the $\phi_{2n}(x)$ terms (which is $\Delta$).


\subsection{Kernels as Formal Mode-Sum Integrals}

The integrals in  \eqref{ax1.1}-\eqref{bx1} and \eqref{ab2}-\eqref{b3} can be inverted to find the expressions for $a(q)$ and $b(q)$. 
To do this, it is necessary to pick two independent pieces of data, one from the $\phi$ side (any $\phi_{2n}$) and one from the $j$ side (any $j_{2n}$). Once such pieces are chosen  (say using the expressions for $n=0$), then the rest of the terms ($\phi_{2n}, j_{2n}$) can be evaluated in terms of them. The resultant expressions for $a(q)$ and $b(q)$ are obtained as Fourier transforms of the boundary fields.

Let's begin by looking at \eqref{ax1.1}. 
Fixing $n = 0$ gives 
\begin{equation}
    \phi_{0}(x) = \frac{1}{2^{\nu}}\frac{1}{\Gamma(\nu + 1)}\int \frac{\mathrm{d}^{d}q}{(2\pi)^{d}}a(q)e^{i q.x}|q|^{\nu}
\end{equation}
An inverse Fourier transform extracts $a(q)$ in terms of $q$ and $\phi_0$.
Therefore, we can write
\begin{equation}
    a(k) = 2^{\nu}\frac{\Gamma(1+\nu)}{|k|^{\nu}}\int \phi_{0}(x)e^{-i k.x}\mathrm{d}^{d}x
\end{equation}
The calculation for $b(q)$ follows in a similar fashion to give us
\begin{equation}
    b(k) = |k|^{\nu}\frac{\Gamma(1-\nu)}{2^{\nu}}\int j_{0}(x)e^{-i k.x}\mathrm{d}^{d}x
\end{equation}
Using this, we can write all the other $\phi_{2n}$ and $j_{2n}$ as follows
\begin{align}
    \phi_{2 n}(x) &= (-1)^{n}\frac{\Gamma(1+n)}{4^{n}\Gamma(n+1)\Gamma(n+\nu + 1)}\int \mathrm{d}^{d}x' \int \frac{\mathrm{d}^{d}q}{(2\pi)^{d}}|q|^{2 n}e^{i q.(x-x')}\phi_{0}(x') \label{phi2n1}\\
    j_{2n}(x) &= (-1)^{n}\frac{\Gamma(1-\nu)}{4^{n}\Gamma(n+1)\Gamma(n-\nu + 1)}\int \mathrm{d}^{d}x' \int \frac{\mathrm{d}^{d}q}{(2\pi)^{d}}|q|^{2 n}e^{i q.(x-x')}j_{0}(x') \label{j2n1}
\end{align}
In terms of $\phi_{0}(x)$ and $j_{0}(x)$ the bulk solution is
\begin{align}
    \Phi(x,z) = &\int \mathrm{d}^{d}x' K_1 (z,x; x')\phi_{0}(x')\nonumber + \int \mathrm{d}^{d}x'K_2 (z,x;x')j_{0}(x)
\end{align}
where we have the following integral representations of the bulk reconstruction kernels\footnote{There are many different kernels we work with in this paper, distinguished by the fact that they are for the normalizable mode or non-normalizable mode, Poincar\'e or global patches, odd or even AdS, etc. We will distinguish normalizable and non-normalizable kernels by the subscripts 1 and 2 respectively. Global kernels will carry the superscript $G$ and Poincar\'e will carry none. Odd and even AdS kernels should be clear from the context, and so we do not distinguish them via notation.}
\begin{align}
    K_{1}(z,x;x') &= \int \frac{\mathrm{d}^{d}q}{(2\pi)^{d}}\frac{2^{\nu}\Gamma(1+\nu)}{|q|^{\nu}}e^{i q.(x-x')}z^{\frac{d}{2}}J_{\nu}(|q|z) \label{ker1} \\
    K_{2}(z,x;x') &= \int \frac{\mathrm{d}^{d}q}{(2\pi)^{d}}\frac{|q|^{\nu}\Gamma(1-\nu)}{2^{\nu}}e^{i q.(x-x')}z^{\frac{d}{2}}J_{-\nu}(|q|z) \label{ker2}
\end{align}

We can write down similar expressions for the non-generic mass as well. We begin by looking at \eqref{ab2}-\eqref{b3}. These are the expressions for $\nu = p \in \text{Integers}$. It is clear that $\tilde{\phi}_{2n}(x)$ and $j_{2n}(x)$ are related to each other, since both terms arise from the Bessel $Y$ function. Note however, that this is true only up to $n = p-1$, which is the number of $j_{2n}$ that exist. Similar to the previous case, two independent pieces of data on the boundary are required. The rest of the fields at the boundary can then be written in terms of those two. Inverting the expression \eqref{b3} for $n=0$ gives an expression for $b(k)$ in terms of $j_{0}(x)$. This expression can be used in \eqref{ab2} with $n=0$ to determine  $a(k)$ in terms of the boundary fields $\phi_{0}$ and $j_{0}$. 
\begin{align}
b(k) &= -\frac{|k|^{p}\pi}{2^{p}\Gamma(p)}\int j_{0}(x)e^{- i k.x}\mathrm{d}^{d}x \label{bk} \\
a(k) &= \frac{2^{p}\Gamma(p+1)}{|k|^{p}}\int \phi_{0}(x)e^{-i  k.x}\mathrm{d}^{d}x + \frac{(\gamma - \psi(p+1) + 2\ln\left(\frac{|k|}{2} \right))|k|^{p} }{2^{p}\Gamma(p)}\int j_{0}(x)e^{- i k.x}\mathrm{d^{d}}x \label{ak} 
\end{align}
Using the equations \eqref{ak} and \eqref{bk}, $\Phi(x,z)$ \eqref{a2} can be written in terms of $\phi_{0}$ and $j_{0}$. 
\begin{align}
\Phi(x,z) = \int \mathrm{d}^{d}x' \tilde{K}_1 (z,x; x') \phi_0 (x') + \int \mathrm{d}^{d}x' \tilde{K}_2 (z, x; x') j_0 (x')
\end{align}
where the integral representations of the bulk reconstruction kernels corresponding to $\phi_0$ (denoted by $\tilde{K}_1$) and $j_0$ (by $\tilde{K}_2$) are
\begin{align}
\tilde{K}_{1}(z,x;x') &= \int \frac{2^{p}\Gamma(p+1)}{|q|^{p}}e^{i q.(x-x')}z^{\frac{d}{2}}J_{p}(|q|z)\frac{\mathrm{d}^{d}q}{(2\pi)^{d}} \label{K1int} \\
\tilde{K}_{2}(z,x;x') &= -\int \frac{|q|^{p}\pi}{2^{p}\Gamma(p)}e^{i q.(x-x')}z^{\frac{d}{2}}\left\{\frac{-\gamma + \psi(p+1)- 2 \ln(|q|/2)}{\pi}J_{p}(|q|z) + Y_{p}(|q|z)\right\}\frac{\mathrm{d}^{d}q}{(2\pi)^{d}} \label{K2int}
\end{align}

\section{Poincar\'e Kernel Integral: Explicit Evaluation}

In this section, we evaluate the integral for the reconstruction kernels in \eqref{ker1}-\eqref{ker2} by generalizing the argument of \cite{Hamilton:2006az, Bena:1999jv} to arbitrary dimensions. We start with
\begin{equation}
     K_{1}(z,x;x') = \int \frac{\mathrm{d}^{d}q}{(2\pi)^{d}}\frac{2^{\nu}\Gamma(1+\nu)}{|q|^{\nu}}e^{i q.(x-x')}z^{\frac{d}{2}}J_{\nu}(|q|z)
\end{equation}
Following the result in appendix \ref{BApp} \eqref{loreqfinal}, this integral can be simplified (by noting that the $\zeta_{\nu} = J_{\nu}$ in \eqref{loreqfinal}) to the following 
\begin{align}\label{ker1.5}
K_1 (z,x;x') = \frac{2^{\nu}\Gamma(1+\nu)}{\pi (2\pi)^{\frac{d}{2}}}\frac{z^{\frac{d}{2}}}{X^{\frac{d}{2}-1}}\int_{a = 0}^{\infty}  a^{\mu + \frac{d}{2}}J_{\nu}(a z)K_{\frac{d}{2}-1}(a X) \mathrm{d}a
\end{align}
where the notation $X = \sqrt{\Delta x^{2} - \Delta t^{2}}$ is used, see appendix \ref{BApp} for more detail on the notation. Note that, $\Delta t$ comes with a $+i\epsilon$ to handle singularities. 
For this integration, we use the identity \cite{grad}
\begin{align}\label{iden}
    \int_{0}^{\infty}x^{-\lambda}K_{\mu}(a x)J_{\nu}(b x) \mathrm{d}x &= \frac{b^{\nu}}{2^{\lambda + 1}a^{\nu - \lambda + 1}\Gamma(1+\nu)}\Gamma(\frac{\nu - \lambda + \mu + 1}{2})\Gamma(\frac{\nu - \lambda - \mu + 1}{2})\nonumber \\ &\times\;_{2}F_{1}(\frac{\nu - \lambda + \mu + 1}{2}, \frac{\nu - \lambda - \mu + 1}{2}; \nu + 1; -\frac{b^{2}}{a^{2}}) \nonumber \\
    &\;\;\;\;\;\; \forall \;\; \text{Re}(a \pm i b) > 0 \;\;\&\;\; \text{Re}(\nu - \lambda + 1) > |\text{Re}(\mu)|
\end{align}
Using this identity, and identifying that $\mu = \frac{d}{2}-1$ and $\lambda = \nu - \frac{d}{2}$, we get the following result (with $C_{3} = \frac{\Gamma(\frac{d}{2})}{2^{\nu - \frac{d}{2}+1}\Gamma(1+\nu)}$, and also noting that $\nu - \lambda + 1 = \nu - \nu + \frac{d}{2} + 1 = \frac{d}{2} + 1 > \frac{d}{2} - 1 = |\mu|$ and $\text{Re}(a \pm i b) = \text{Re}{X} > 0$)
\begin{align}\label{ker1.7.1}
    K_{1}(z,x;x') = \frac{\Gamma(d/2)}{2 \pi ^{d/2}}\frac{z^{\nu + \frac{d}{2}}}{(\sqrt{\Delta x^{2} - \Delta t^{2}})^{d}}\;_{2}F_{1}(\frac{d}{2},1;\nu+1; - \frac{z^{2}}{\Delta x^{2} - \Delta t^{2}})
\end{align}
This is the result obtained in \cite{Bena:1999jv} for the specific case of $d=4$. \par 
The kernel for the non-normalizable mode for the non-integer $\nu = \Delta - \frac{d}{2}$ case, given in \eqref{ker2}, is obtained similarly: 
\begin{align}
    K_{1}(z,x;x') &= \frac{\Gamma(d/2)}{2 \pi ^{d/2}}\frac{z^{\Delta}}{(\sqrt{\Delta x^{2} - \Delta t^{2}})^{d}}\;_{2}F_{1}(\frac{d}{2},1;1- \frac{d}{2} +  \Delta; - \frac{z^{2}}{\Delta x^{2} - \Delta t^{2}}) \label{ker1.7}\\
    K_{2}(z,x;x') &= \frac{\Gamma(d/2)}{2 \pi ^{d/2}}\frac{z^{d -\Delta }}{(\sqrt{\Delta x^{2} - \Delta t^{2}})^{d}}\;_{2}F_{1}(\frac{d}{2},1;1 + \frac{d}{2}-\Delta; - \frac{z^{2}}{\Delta x^{2} - \Delta t^{2}}) \label{ker1.8}
\end{align}
These expressions hold for both even and odd AdS, and for the normalizable mode in both integer and non-integer case. The only case that is excluded from this is the non-normalizable mode for integer value of $\nu$. This special case is studied in appendix \ref{integernu}. \par 
We would like to cast these kernels in an AdS covariant form. To achieve this, we use transformation formulas for the hypergeometric functions in \eqref{ker1.7}-\eqref{ker1.8}. 
The two cases of even and odd AdS have to be treated separately, since the transformation identities of the hypergeometric functions are different when the parameters for the function are integers as opposed to when they are not. In the following, we study the two cases separately.

\subsection{Even AdS case}

First, we study even AdS. Since $d$ is odd, we can directly employ the relation \eqref{id2} in \eqref{ker1.7}. For convenience, we shall denote $\sqrt{\Delta x^{2} - \Delta t^{2}}$ by $X$ and $c_d = \frac{\Gamma(d/2)}{2\pi^{d/2}}$.
\begin{align}
    K_{1}(z,x;x') &= c_{d}\frac{z^{\Delta}}{X^{d}}\Big(\frac{X^{2}}{z^{2}}\Big) \frac{\Gamma(\nu+1)\Gamma(\frac{d}{2}-1)}{\Gamma(\frac{d}{2})\Gamma(\nu)}\;_{2}F_{1}(1, 1 -\nu ; 2 - \frac{d}{2}; - \frac{X^{2}}{z^{2}}) \nonumber \\ 
    &+c_{d}\frac{z^{\Delta}}{X^{d}}\Big(\frac{X^{2}}{z^{2}}\Big)^{\frac{d}{2}}\frac{\Gamma(1+\nu)\Gamma(1 - \frac{d}{2})}{\Gamma(1 + \nu - \frac{d}{2})}\;_{2}F_{1}(\frac{d}{2}, \frac{d}{2} - \nu; \frac{d}{2}; - \frac{X^{2}}{z^{2}}) \label{evenhyp}
\end{align}
Using the relation that $\;_{2}F_{1}(\beta, \alpha; \alpha; z) = (1-z)^{-\beta}$, we get
\begin{align}
    K_{1}(z,x;x') &= c_{d}\frac{z^{\Delta - 2}}{X^{d-2}} \frac{\Gamma(\Delta -\frac{d}{2}+1 )\Gamma(\frac{d}{2}-1)}{\Gamma(\frac{d}{2})\Gamma(\Delta -\frac{d}{2})}\;_{2}F_{1}(1, 1 -\Delta +\frac{d}{2}; 2 - \frac{d}{2}; - \frac{X^{2}}{z^{2}}) \nonumber \\ 
    &+c_{d}\frac{\Gamma(1+\Delta -\frac{d}{2})\Gamma(1 - \frac{d}{2})}{\Gamma(1 + \Delta -d)}\Big(\frac{z^{2} + X^{2}}{z}\Big)^{\Delta - d} \label{K1_3.9}
\end{align}\par
We make the following two observations regarding this kernel. First, we note that (using the series representation of the hypergeometric $_{2}F_{1}$) the powers of $z$ arising from the first line of the above equation go as $z^{\Delta - 2 - 2n}$. These powers do not occur in the series expansion (in $z$) of the normalizable mode \eqref{exp1}. This is a suggestion that these terms should vanish when integrated against the positive energy mode. Let us also note that an identical hypergeometric function showed up in the global AdS kernel expression in equation (14) of \cite{Hamilton:2006az}.\footnote{See also related discussion near our \eqref{result}.} Even though structurally different,\footnote{The argument of the hypergeometric function in our case contains {\em bulk} coordinates.} this term was found to vanish when integrated against the positive frequency mode.

We will take these observations as circumstantial evidence that the first line in the expression above ought to vanish when integrated against the positive frequency boundary mode (and hence can be dropped from the expression). This turns out to also be natural for matching with the global kernel, as we will see in the next section. Similar suggestions have appeared previously \cite{Hamilton:2006az}, but we are not aware of a universal statement of this type that is demonstrably valid for all $d$ and $\Delta$. It may be necessary that some form of analytic continuation of coordinates (see eg.  \cite{Hamilton:2006fh}) is necessary before this expectation can be made fully precise and established.

In this paper, we will not state or prove these statements for general values of $\Delta$. But when $\Delta$ is a half-integer $\ge \frac{d}{2}$, the first line of \eqref{evenhyp} reduces to a polynomial. In this case, we can give a precise meaning to the statement in terms of an $i \epsilon$-prescription in the spatial direction and prove that it drops out. We present the details in appendix \ref{iepsilon}. 
This argument can be viewed as a generalization of the argument used in appendix C of \cite{Hamilton:2006az} to argue that the extra terms are indeed vanishing. The discussion in appendix C of \cite{Hamilton:2006az} was for an odd AdS case (specifically, AdS$_3$), but the generalization works in both even and odd dimensions, as we will see in the next subsection. 

In any event, the remaining terms can be written in terms of the chordal distance $\sigma$ as (see appendix \ref{chordalwaveeqn} for some definitions):
\begin{align}
    K_{1}(z,x;x') &= \lim_{z' \rightarrow 0}\frac{(-1)^{\frac{d-1}{2}}2^{\Delta - d}\Gamma(1 + \Delta - \frac{d}{2})}{2 \pi^{\frac{d}{2}} \Gamma(1 + \Delta - d)}\lim_{z' \rightarrow 0}(\sigma z')^{\Delta - d} \notag \\
    &\equiv  a_{d \Delta}\lim_{z' \rightarrow 0}(\sigma z')^{\Delta - d} \label{3p10}
\end{align}
where we have used the expression for $c_{d}$. We will later write this result as a kernel with support only on the spacelike region of Poincar\'e. The result will lead to a precise match including a factor of 2 with the global spacelike kernel \cite{Hamilton:2006az}. We also introduce the notation $a_{d\Delta}$. 

Similar to the normalizable mode, for generic $\Delta$ and for the non-normalizable case, we can transform \eqref{ker1.8} using hypergeometric identities to
\begin{align}
    K_{2}(z,x;x') &= c_{d}\frac{z^{d -\Delta - 2}}{X^{d-2} } \frac{\Gamma(1 - \Delta + \frac{d}{2})\Gamma(\frac{d}{2}-1)}{\Gamma(\frac{d}{2})\Gamma(\frac{d}{2} - \Delta)}\;_{2}F_{1}(1, 1 +\Delta - \frac{d}{2} ; 2 - \frac{d}{2}; - \frac{X^{2}}{z^{2}}) \nonumber \\ 
    &+c_{d}\frac{z^{ d-\Delta}}{X^{d}}\Big(\frac{X^{2}}{z^{2}}\Big)^{\frac{d}{2}}\frac{\Gamma(1-\Delta + \frac{d}{2})\Gamma(1 - \frac{d}{2})}{\Gamma(1 - \Delta )}\Big(1 + \frac{X^{2}}{z^{2}}\Big)^{-\Delta}
\end{align}
Similar to the kernel for the normalizable mode, we observe that the first line in the above expression does not contain the correct powers of $z$. The series expansion of the first term appears with the powers of $z$ as $z^{d - \Delta  - 2 - 2n}$. These powers do not appear in the $j_{2n}$ part in \eqref{exp1}. 
The remaining term is then given by
\begin{align}
    K_{2}(z,x;x') = \lim_{z' \rightarrow 0}c_{d}\frac{\Gamma(1-\Delta + \frac{d}{2})\Gamma(1 - \frac{d}{2})}{\Gamma(1 - \Delta )}(2 \sigma z')^{-\Delta}
\end{align}
Plugging in the value of $c_{d}$, we get
\begin{align}
    K_{2}(z,x;x') &= \frac{(-1)^{\frac{d-1}{2}}2^{-\Delta}\Gamma(1 - \Delta + \frac{d}{2})}{2\pi^{\frac{d}{2}}\Gamma(1-\Delta)}\lim_{z' \rightarrow 0}(\sigma z')^{-\Delta} \notag \\ &= -\frac{2^{-\Delta}\Gamma(\Delta)\tan\pi\Delta}{2\pi^{\frac{d}{2}}\Gamma(\Delta - \frac{d}{2})}\lim_{z' \rightarrow 0}(\sigma z')^{-\Delta} \notag\\
   & =  a'_{d \Delta}\lim_{z' \rightarrow 0}(\sigma z')^{-\Delta}\label{3p14}
\end{align}
In the next section, we will write this result in a form that has support only on the spacelike separated region of the Poincar\'e boundary. It will match with the spacelike Green's function in global coordinates except for a factor of 2 in the coefficient. This factor of 2 is expected \cite{Hamilton:2006az}, because the regions of integration in the global and Poincar\'e kernels are different. Let us also note that the result that we have obtained above, after using the explicit form of the Poincare chordal distance \eqref{chordPformG}, turns into expression (14) of \cite{Bala2}. This can be viewed as another argument for dropping the extra terms we mentioned earlier. 

For the non-normalizable case, we will not discuss the analogues of the half-integer $\Delta$ cases we discussed above for the normalizable mode. This is because it turns out that the $i \epsilon$-prescription is of use only when $\nu$ is an integer, but the  expressions \eqref{ker1.7} and \eqref{ker1.8} do not apply for integer $\nu$.

\subsection{Odd AdS case}

This case needs to be treated separately since $\frac{d}{2} = m+1 \in \IZ$ and we need the transformation \eqref{id4}. Employing it in \eqref{ker1.7}, we get the result
\begin{align}
    K_{1}(z,x;x') &= c_{d}\frac{z^{\Delta}}{X^{d}}\;_{2}F_{1}(1, m + 1; \Delta - m; -\frac{z^{2}}{X^{2}}) \nonumber \\
    &= c_{d}\frac{z^{\Delta}}{X^{d}}\frac{(-1)^{m}\Gamma(\Delta - m)\ln(1 + \frac{z^{2}}{X^{2}})}{\Gamma(\Delta - 2m-1)\Gamma(\Delta - m - 1)\Gamma(m+1)}\sum_{k=0}^{\infty}\frac{\Gamma(k - m - 1 + \Delta)}{\Gamma(k+1)}\Big(\frac{X^{2}}{X^{2} + z^{2}}\Big)^{k + \frac{d}{2}} \nonumber \\
    &+ c_{d}\frac{z^{\Delta}}{X^{d}}\frac{(-1)^{m}\Gamma(\Delta - m)}{\Gamma(\Delta - 2m-1)\Gamma(\Delta - m - 1)\Gamma(m+1)}\sum_{k=0}^{\infty}\frac{\Gamma(k - m - 1 + \Delta)h_{k}}{\Gamma(k+1)}\Big(\frac{X^{2}}{X^{2} + z^{2}}\Big)^{k+\frac{d}{2}} \nonumber \\
    &+ c_{d}\frac{z^{\Delta}}{X^{d}}\frac{X^{2}}{z^{2}+X^{2}}\frac{\Gamma(\Delta - m)}{\Gamma(m+1)\Gamma(\Delta - m - 1)\Gamma(\Delta - 2m - 1)}\nonumber \\ &\times \sum_{k=0}^{m-1}\Gamma(\Delta - 2m - 1 + k)\Gamma(m-k)(-1)^{k}\Big(\frac{X^{2}}{X^{2}+z^{2}}\Big)^{k} \label{oddc1}
\end{align}
where we have used the notation $X^{2} = \Delta x^{2} - \Delta t^{2}$ and $h_k$ is given in \eqref{id4}.
Using \eqref{form0}, we can see that \eqref{oddc1} reduces to the following simpler expression
\begin{align}
    K_{1}(z,x;x') &= \lim_{z'\rightarrow 0}c_{d}\frac{(-1)^{\frac{d}{2}-1}2^{\Delta - d}\Gamma(\Delta - \frac{d}{2}+1)}{\Gamma(\Delta - d + 1)\Gamma(\frac{d}{2})}(\sigma z')^{\Delta - d}(\ln(\sigma z') + \ln(\frac{2 z}{X^{2}})) \nonumber \\
    &+ c_{d}\frac{z^{\Delta}}{X^{d}}\frac{(-1)^{\frac{d}{2}-1}\Gamma(\Delta - \frac{d}{2} + 1)}{\Gamma(\Delta - d + 1)\Gamma(\Delta - \frac{d}{2})\Gamma(\frac{d}{2})}\sum_{k=0}^{\infty}\frac{\Gamma(k - \frac{d}{2} + \Delta)h_{k}}{\Gamma(k+1)}\Big(\frac{X^{2}}{X^{2} + z^{2}}\Big)^{k + \frac{d}{2}} \nonumber \\
    &+ c_{d}\frac{z^{\Delta}}{X^{d}}\frac{\Gamma(\Delta - m)}{\Gamma(m+1)\Gamma(\Delta - m - 1)\Gamma(\Delta - 2m - 1)}\nonumber \\ &\times \sum_{k=0}^{m-1}\Gamma(\Delta - 2m - 1 + k)\Gamma(m-k)(-1)^{k}\Big(\frac{X^{2}}{X^{2}+z^{2}}\Big)^{k+1}\label{oddc2}
\end{align} 
For reasons similar to the discussions in the even AdS case, we expect that all the terms except the first term $\propto (\sigma z')^{\Delta - d}\ln(\sigma z')$ can be set to zero after suitable re-interpretation. Note in particular that the construction of Poincar\'e kernel starting from global modes, does not lead to such terms \cite{Hamilton:2006az}. 
As in the even AdS case, we can explicitly show the absence of the extra terms for an infinite sub-class of cases using the spatial $i \epsilon$-prescription: in odd AdS, this happens when $\Delta \ge d$ is an integer. In this case, we again argue that the extra terms reduce to polynomials whose pole structure is easily handled via our $i \epsilon$ prescription.


In any event, assuming that the kernel $K_{1}$ can be written as
\begin{align}
    K_{1}(z,x;x') = \lim_{z'\rightarrow 0}\frac{(-1)^{\frac{d}{2}-1}2^{\Delta - d - 1}\Gamma(\Delta - \frac{d}{2}+1)}{\Gamma(\Delta - d + 1)\pi^{\frac{d}{2}+1}}(\sigma z')^{\Delta - d}\ln(\sigma z') \label{oddc3}
\end{align}
we can note the following. It is possible to always add any extra piece to the kernel as long as it vanishes when integrated against the boundary field. Therefore, in this regard, the kernel $K_{1}$ can be modified by adding its complex conjugate. The complex conjugate will vanish when integrated against the positive energy modes, since that integration is done by adding a positive $i \epsilon$ piece to $t - t'$ in \eqref{ker1}-\eqref{ker2}. But since the kernel above is real, adding the complex conjugate simply doubles it. This is similar to another argument presented in appendix C of \cite{Hamilton:2006az}. The basic point here is that even though the kernel is real, it is not analytic, and therefore it picks up different pieces when multiplied against the positive and negative frequency modes.\footnote{We thank Dan Kabat for a helpful discussion on this argument.}  
 
As a result of this we obtain $2$ times the result of \eqref{oddc3}, where it is understood that now we are multiplying the kernel against the full boundary mode (which is the sum of the positive and negative energy modes). This factor of 2 should not be confused with a factor of 2 that will arise in some discussions due to antipodal mapping relating global and Poincar\'e kernels. 
In any event, an equally viable kernel for the normalizable mode would be the following
\begin{align}
    K_{1}(z,x;x') &= \lim_{z'\rightarrow 0}\frac{(-1)^{\frac{d}{2}-1}2^{\Delta - d}\Gamma(\Delta - \frac{d}{2}+1)}{\Gamma(\Delta - d + 1)\pi^{\frac{d}{2}+1}}(\sigma z')^{\Delta - d}\ln(\sigma z') \notag \\
    &\equiv c_{d \Delta} \lim_{z'\rightarrow 0}(\sigma z')^{\Delta - d}\ln(\sigma z')\label{oddc3.1}
\end{align}
where we have introduced the notation $c_{d\Delta}$.

For the non-normalizable mode \eqref{ker1.8}, one can retrace some of the same steps as above. This gives us the following expression
\begin{align}
    K_{2}(z,x;x')&= c_{d}\frac{z^{d-\Delta}}{X^{d}}\frac{\Gamma(m + 2 - \Delta)(-1)^{m}\ln(1 + \frac{z^{2}}{X^{2}})}{\Gamma(m+1)\Gamma(1 - \Delta)\Gamma(1 + m - \Delta)}\sum_{k=0}^{\infty}\frac{\Gamma(1 + m - \Delta + k)}{\Gamma(1+k)}\Big( \frac{X^{2}}{X^{2} + z^{2}}\Big)^{k + \frac{d}{2}} \nonumber  \\
    &+ c_{d}\frac{z^{d-\Delta}}{X^{d}}\frac{\Gamma(m + 2 - \Delta)(-1)^{m}}{\Gamma(m+1)\Gamma(1 - \Delta)\Gamma(1 + m - \Delta)}\sum_{k=0}^{\infty}\frac{\Gamma(1 + m - \Delta + k)h_{k}}{\Gamma(1+k)}\Big( \frac{X^{2}}{X^{2} + z^{2}}\Big)^{k + \frac{d}{2}} \nonumber \\
    &+ c_{d}\frac{z^{d-\Delta}}{X^{d}}\frac{\Gamma(2 + m - \Delta)}{\Gamma(m+1)\Gamma(m - \Delta + 1)}\sum_{k=0}^{\frac{d}{2}-2}(-1)^{k}\Gamma(m-k)\Gamma(1-\Delta)\Big( \frac{X^{2}}{X^{2} + z^{2}}\Big)^{k+1}\label{oddc4}
\end{align}
Using \eqref{form0}, we note that \eqref{oddc4} reduces to the following
expression
\begin{align}
    K_{2}(z,x;x') &= \lim_{z' \rightarrow 0}c_{d}\frac{\Gamma(1 + \frac{d}{2} - \Delta)(-1)^{\frac{d}{2}-1}2^{-\Delta}}{\Gamma(\frac{d}{2})\Gamma(1-\Delta)}(\sigma z')^{-\Delta}(\ln(\sigma z') + \ln (\frac{ 2 z}{X^{2}})) \nonumber \\
    &+ c_{d}\frac{z^{d-\Delta}}{X^{d}}\frac{\Gamma(1 + \frac{d}{2} - \Delta)(-1)^{\frac{d}{2} - 1}}{\Gamma(\frac{d}{2})\Gamma(1 - \Delta)\Gamma(\frac{d}{2} - \Delta)}\sum_{k=0}^{\infty}\frac{\Gamma(\frac{d}{2} - \Delta + k)h_{k}}{\Gamma(1+k)}\Big( \frac{X^{2}}{X^{2} + z^{2}}\Big)^{k + \frac{d}{2}} \nonumber \\
    &+ c_{d}\frac{z^{d-\Delta}}{X^{d}}\frac{\Gamma(1 + \frac{d}{2} - \Delta)}{\Gamma(\frac{d}{2})\Gamma(\frac{d}{2} - \Delta )}\sum_{k=0}^{\frac{d}{2}-2}(-1)^{k}\Gamma(\frac{d}{2}-k - 1)\Gamma(1-\Delta)\Big( \frac{X^{2}}{X^{2} + z^{2}}\Big)^{k+1}
\end{align}
Using similar arguments, we are again lead to 
\begin{align}
    K_{2}(z,x;x') &= \lim_{z' \rightarrow 0}\frac{\Gamma(1 + \frac{d}{2} - \Delta)(-1)^{\frac{d}{2}-1}2^{-\Delta }}{\pi^{\frac{d}{2}+1}\Gamma(1-\Delta)}(\sigma z')^{-\Delta}\ln(\sigma z') \notag \\
    &= c'_{d \Delta}\lim_{z' \rightarrow 0}(\sigma z')^{-\Delta}\ln(\sigma z')\label{oddPoincareK}
\end{align}
As in the even AdS case, it turns out that for the non-normalizable mode, the cases for which the $i \epsilon$-prescription applies occur when $\nu$ is an integer and therefore $\eqref{ker1.8}$ does not apply.

\section{Spacelike Kernel: Antipodal Mapping}\label{spacelikekernel}

The mode-sum kernels we have written down in the previous section have support everywhere on the boundary of the Poincar\'e patch, and not just on the points that are spacelike separated from the bulk point that we are trying to reconstruct. In this section, we will show that in the even-dimensional case (at least), we can restrict ourselves to the spacelike region. This will help us also connect with the global kernels from earlier sections. 


For the normalizable and non-normalizable modes, we reproduce the expressions derived in the previous section \eqref{3p10} and \eqref{3p14} here for convenience:
\begin{align}
	K_{1} = a_{d \Delta}\lim_{z' \rightarrow 0}(\sigma z')^{\Delta - d}, \ \ \
	K_{2}  = a'_{d \Delta}\lim_{z' \rightarrow 0}(\sigma z')^{-\Delta}.
\end{align}
We start by reviewing the normalizable mode discussion of \cite{Hamilton:2006az} from a purely Poincar\'e perspective. The bulk field contribution at a point $P = (z, x, t)$ (in the Poincar\'e  patch) from the normalizable mode  is written as an integral over the boundary field $\phi^{\text{Poincar\'e}}_{0}$ as follows (all integrals considered in this section are over the Poincar\'e patch, and let us emphasize that we will often suppress $\lim_{z' \rightarrow 0}$ to avoid clutter):
\begin{align}
	\Phi_{\text{normalizable}}(z,x,t) = \int \mathrm{d}t' \mathrm{d}^{d-1}x' a_{d \Delta} (\sigma z')^{\Delta - d}\phi^{\text{Poincar\'e}}_{0}(x',t') \label{4p31}
\end{align}
\begin{figure}[t]
		\centering
		\includegraphics[width=0.5\textwidth]{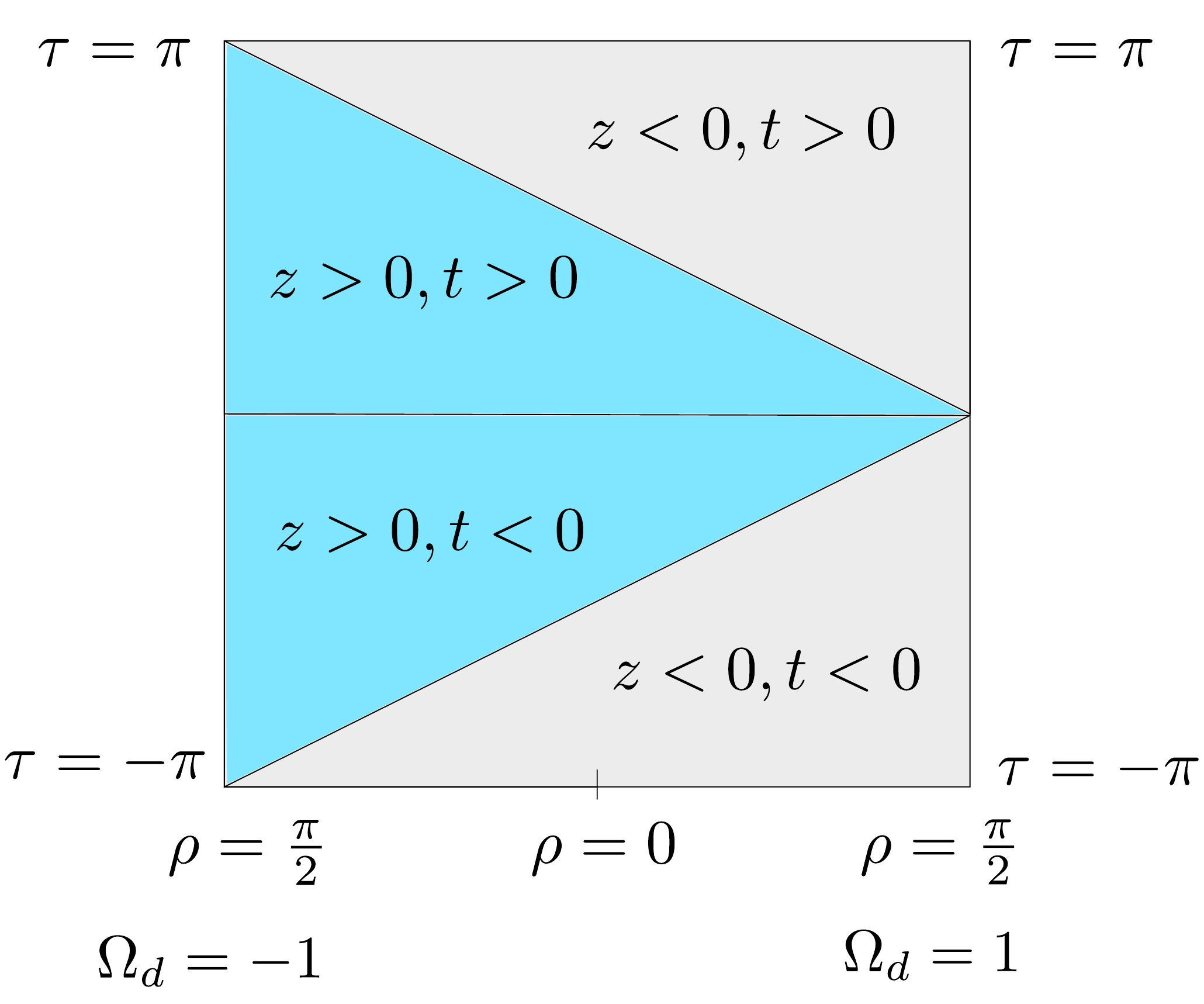}
		\caption{The Penrose diagram for Lorentzian AdS space. The region in blue is spanned by the Lorentzian Poincar\'e coordinates (known as the Poincar\'e patch). The region in grey is the exempted region. The full box is spanned by the Lorentzian global coordinates. As is evident, half of the AdS boundary (denoted by $\rho = \pi/2$ and $\Omega_{d} = 1$) is not covered by the Poincar\'e coordinates.}
		\label{figomegac}
\end{figure}
The kernel $\lim_{z' \rightarrow 0}(\sigma z')^{\Delta - d}$ has a lightcone singularity, and needs a prescription to make it fully well-defined. Motivated by its connection to the global coordinates discussion in \cite{Hamilton:2006az}, we will take it to be defined via 
\begin{align}
		\Phi_{\text{normalizable}}(z,x,t) = \int \mathrm{d}t' \mathrm{d}^{d-1}x' a_{d \Delta} |\sigma z'|^{\Delta - d}\begin{rcases}
			\begin{dcases}
				e^{i \pi \Delta}\tilde{\phi}^{}_{0}(x',t')  & (x',t') \in \mathbf{I} \! \! \\ 
				\tilde{\phi}^{}_{0}(x',t') & (x',t') \in \mathbf{II}	\!\! \\ 
				e^{-i \pi \Delta}\tilde{\phi}^{}_{0}(x',t') & (x',t') \in \mathbf{III} \!\!   
			\end{dcases}
		\end{rcases} \label{4p31redef}
\end{align}
We can motivate this as a phase arising from an antipodal mapping of the field from the region complementary to the Poincar\'e patch in global coordinates (see \cite{Hamilton:2006az} for more details) 
\begin{align}
\label{redef}
	\phi^{\text{Poincar\'e}}_{0}(x',t') = \begin{cases*}
		e^{i \pi \Delta} \tilde{\phi}^{}_{0}(x',t') & \text{future timelike region of $P$ ($\mathbf{I}$)} \\
		\tilde{\phi}^{}_{0}(x',t') & \text{spacelike region of $P$ ($\mathbf{II}$)}\\
		e^{-i \pi \Delta} \tilde{\phi}^{}_{0}(x',t') & \text{past timelike region of $P$ ($\mathbf{III}$)}
	\end{cases*}
\end{align}
and then absorbing the phase into a redefinition of the kernel. This essentially shifts the boundary field (which is integrated over in \eqref{4p31}) from $\phi^{\text{Poincar\'e}}_{0}$ to $\tilde{\phi}^{}_{0}$. Viewing the phase as being part of the kernel from this point on, we will call it $\tilde K_1$ instead of $K_1$. 


Now, consider the following function
\begin{align}
    F(z, t , x ; t' , x') = \Big( \frac{1}{2 z}(z^{2} + |x - x'|^{2} - (t' -t - i \epsilon)^{2}) \Big)^{\Delta - d} \label{f1}
\end{align}
The pole/branch point of this function is at $t ' = t \pm \sqrt{z^{2} + |x-x'|^{2}} + i \epsilon$. Therefore, $F$ is analytic in the lower half $t'$ plane. When $F$ is integrated against $\tilde{\phi}^{}_{0}(x',t')$, it is easy to see that the condition $\omega > |k|$ implies that the integral vanishes:
\begin{align}
	\int \mathrm{d}t' \mathrm{d}^{d-1}x' F(z,t,x; x', t')\tilde{\phi}^{}_{0}(x',t') = 0
\end{align}
Therefore we can add multiples of this function to the kernel without affecting its bulk reconstruction properties.  Now, the $i\epsilon$ prescription ensures that the function $F$ takes the following forms in the three regions of the point $P$:
\begin{align}
    F = \begin{cases}
    -e^{i \pi \Delta}|\sigma z'|^{\Delta - d}  &\text{future timelike region of $P$ ($\mathbf{I}$)} \\
    |\sigma z'|^{\Delta - d}  &\text{spacelike region of $P$ ($\mathbf{II}$)} \\
    -e^{- i \pi \Delta }|\sigma z'|^{\Delta - d} &\text{past timelike region of $P$ ($\mathbf{III}$)}
    \end{cases}
\end{align}
where we have used the fact that $d$ is odd. 

Therefore, if we modify the kernel $\tilde K_{1}$ as
\begin{align}
    \tilde K_{1} \rightarrow \tilde K_{1} + a_{d \Delta}F
\end{align}
then the modified $\tilde K_{1}$ is perfectly acceptable as a bulk reconstruction kernel, but has the advantage that it vanishes in the past and future timelike regions (while resulting in a factor of $2$ in the spacelike region). The final kernel is therefore spacelike as we wanted (we again suppress the tilde)
\begin{align}
    K_{1} =  2 a_{d \Delta} \lim_{z' \rightarrow 0}(\sigma z')^{\Delta - d}\theta(\text{spacelike}) \label{spclkK1}
\end{align}

The above argument is directly motivated by the discussion in section 3.1 of \cite{Hamilton:2006az}. But let us emphasize that it is conceptually different. In \cite{Hamilton:2006az}, the goal was to get to a spacelike Poincar\'e kernel by starting with a spacelike global kernel. In our case, we started with a Poincar\'e kernel, but one that was {\em not} spacelike. We introduced ingredients that are naturally motivated from the global picture and the antipodal map to restrict our kernel to the spacelike region of Poincar\'e. Satisfyingly, this enabled us to argue that we can restrict our kernel entirely to the spacelike Poincar\'e region, if we simply double the overall coefficient. In the end, all these perspectives are mutually consistent including the precise normalization factors.

\subsection{Non-normalizable Mode in Even AdS}
We now turn to the restriction of the kernel corresponding to the non-normalizable mode to the spacelike region. The general idea is parallel, even though the details of the phase are different. The contribution of the non-normalizable mode in the bulk field can be written analogous to \eqref{4p31} as 
\begin{align}
	\Phi_{\text{non-normalizable}}(z,x,t) = \int \mathrm{d}t' \mathrm{d}^{d-1}x' a'_{d \Delta} (\sigma z')^{-\Delta}j^{\text{Poincar\'e}}_{0}(x',t') \label{4p32}
\end{align} 
In this case, the following redefinition of the boundary non-normalizable mode $j^{\text{Poincar\'e}}_{0}$ is useful
\begin{align}\label{redef2}
	j^{\text{Poincar\'e}}_{0} = \begin{cases*}
		-e^{-i \pi \Delta} \tilde{j}^{}_{0}(x',t') & \text{future timelike region of $P$ ($\mathbf{I}$)} \\
		\tilde{j}^{}_{0}(x',t') & \text{spacelike region of $P$ ($\mathbf{II}$)}\\
		-e^{i \pi \Delta} \tilde{j}^{}_{0}(x',t') & \text{past timelike region of $P$ ($\mathbf{III}$)}
	\end{cases*}
\end{align}
The non-normalizable contribution to the bulk field at $P$ is then written in terms of the redefined boundary field $\tilde{j}^{}_{0}$ as
\begin{align}
	\Phi_{\text{non-normalizable}}(z,x,t) = \int \mathrm{d}t' \mathrm{d}^{d-1}x' a'_{d \Delta} |\sigma z'|^{-\Delta}\begin{rcases}
		\begin{dcases}
			-e^{-i \pi \Delta}\tilde{j}^{}_{0}(x',t')\!\!\!\!  & (x',t') \in \mathbf{I} \! \! \\ 
			\tilde{j}^{}_{0}(x',t') & (x',t')\! \in \mathbf{II}	\!\! \\ 
			-e^{i \pi \Delta}\tilde{j}^{}_{0}(x',t')\!\!\!\! & (x',t') \in \mathbf{III} \!\!   
		\end{dcases}
	\end{rcases} \label{4p32redef}
\end{align}
We can again define a tilded kernel that absorbs the phases, $\tilde K_2$. Exercising the freedom to add terms to the kernel that vanish when integrated against the boundary field, we now add the following function to the kernel
\begin{align}
	F'(z, t , x ; t' , x') &= \Big( \frac{1}{2 z}(z^{2} + |x - x'|^{2} - (t' -t - i \epsilon)^{2}) \Big)^{-\Delta}\label{f2}
\end{align}
It is straightforward to see that similar to \eqref{f1}, $F'$ will also vanish when integrated against $ \tilde{j}^{}_{0}(x',t')$. The $i\epsilon$ prescription ensures that the function $F'$ takes the following forms in the three regions of the point $P$:
\begin{align}
	F' = \begin{cases}
		e^{-i \pi \Delta}|\sigma z'|^{-\Delta}  &\text{future timelike region of $P$ ($\mathbf{I}$)} \\
		|\sigma z'|^{-\Delta}  &\text{spacelike region of $P$ ($\mathbf{II}$)} \\
		e^{i \pi \Delta }|\sigma z'|^{-\Delta} &\text{past timelike region of $P$ ($\mathbf{III}$)} \label{Fprime}
	\end{cases}
\end{align} 
Now we modify $\tilde K_{2}$ as $\tilde K_{2} \rightarrow \tilde K_{2} + a'_{d \Delta}F'$. This has no effect on the bulk field at point $P$. The modified $K_{2}$ vanishes in the past and future timelike regions  and gives a factor of $2$ in the spacelike region. So we have
\begin{align}
	K_{2} = 2 a'_{d \Delta}\lim_{z' \rightarrow 0}(\sigma z')^{-\Delta}\theta(\text{spacelike}) \label{spclkK2}
\end{align}
In analogy with section 3.1 of \cite{Hamilton:2006az}, the above result can be given a natural interpretation (including the factor of 2) in terms of the global spacelike Green's function, via an antipodal identification argument. In fact this was our motivation fo the phases we chose in \eqref{redef2}, as we illustrate in the next subsection.


 
In the previous subsection, the antipodal map from the global boundary to the Poincar\'e patch \cite{Hamilton:2006az} was viewed as the motivation for the choice of phases in the normalizable mode. Let us exhibit the origin of the analogous choice of phases \eqref{4p32} for the non-normalizable mode. The positive frequency part of the non-normalizable mode was written down in section 3. 
From that it is clear that under the antipodal mapping
\cite{Hamilton:2006az}
\begin{align}
	A : \tau \rightarrow \tau \pm \pi \;\;\;\;\;\;\;\;\; \rho \;\; \text{invariant} \;\;\;\;\;\;\;\;\; \Omega \rightarrow \Omega_A \;\;\;\;
\end{align}
the positive frequency non-normalizable mode transforms as
\begin{align}
    j^{\text{global}}_{0+}(A x) = e^{\pm i \pi (\Delta-d)}j^{\text{global}}_{0 + }(x) \label{jAx}
\end{align}
where we have renamed $\Phi_{0 +}$ from section 3 to  $j^{\text{global}}_{0+}$ here, for clarity in the present setting.  
This serves as inspiration for the phases for the field in \eqref{redef2}, where we have used the fact that $d$ is odd.

\subsection{Connection to Global: Non-normalizable Mode}

As we discussed below \eqref{spclkK1}, the discussion in this section has been about restricting the Poincar\'e kernel to a spacelike region in even-dimensional AdS. The methods we used were inspired by the global-to-Poincar\'e connection in even AdS for the normalizable mode \cite{Hamilton:2006az}. In this subsection we will make the connection between global and Poincar\'e explicit for the non-normalizable mode as well, and also discuss the connection between Poincar\'e and global non-normalizable kernels in odd AdS. These are a direct adaptation of the discussion in section 3 of  \cite{Hamilton:2006az}, and we include it here only for completeness.

In even-dimensional AdS, the non-normalizable global kernel is given by the following spacelike expression \eqref{evencheck} 
\begin{align}
  K_{2}^G  = a'_{d \Delta}\lim_{\rho' \rightarrow \frac{\pi}{2}}(\sigma \cos\rho')^{-\Delta}\theta(\text{spacelike})
\end{align}
From \eqref{jAx}, we find that for $d \in \text{odd}$, the antipodal map acts on the positive frequency non-normalizable mode as
\begin{align}
	 j^{\text{global}}_{0+}(A x) = -e^{\pm i \pi \Delta}j^{\text{global}}_{0 + }(x)
\end{align}
The past timelike region (region $\mathbf{I}$) of the Poincar\'e patch is mapped into the spacelike patch (region $\mathbf{II}$) by the transformation $\tau \rightarrow \tau + \pi$ and the future timelike region (region $\mathbf{III}$) is mapped into the spacelike patch by $\tau \rightarrow \tau - \pi$. Therefore, the global smearing function can be written as follows (where the bulk field is evaluated at the point $P$)
\begin{align}
	\phi(P) &= \int \mathrm{d}\tau' \mathrm{d}^{d-1}\Omega' K_{2}^{G}(\tau',\Omega'| P)(j^{\text{global}}_{0+} + j^{\text{global}}_{0-}) \notag \\
	&= \int_{\text{Poincar\'e patch}} \mathrm{d}\tau' \mathrm{d}^{d-1}\Omega' a'_{d \Delta}|\sigma \cos\rho'|^{-\Delta} \begin{cases*}
		-e^{i \pi \Delta} \tilde{j}^{\text{global}}_{0} & \text{in image of $\mathbf{I}$}\\
		\tilde{j}^{\text{global}}_{0} & \text{in region $\mathbf{II}$ }\\
		-e^{-i \pi \Delta} \tilde{j}^{\text{global}}_{0} & \text{in image of $\mathbf{III}$} 
	\end{cases*}
\end{align}
From here, the steps described in \eqref{4p32redef} to \eqref{spclkK2} can be followed, again with the function \eqref{Fprime}, and the end result is precisely \eqref{spclkK2} as we want. 
Let us emphasize that the $\theta(\text{spacelike})$ in this final result covers the Poincar\'e spacelike region.

We turn our attention now to the odd AdS case (i.e. $d \in \text{even}$). The global AdS kernel is now given by \eqref{KplusOddFinal}
\bea
	K^G_{2}  =  c'_{d \Delta}\lim_{\rho' \rightarrow \pi/2}(\sigma \cos\rho')^{-\Delta}\ln(\sigma \cos\rho') \theta(\text{spacelike})
\eea
Note that $d \in \text{even}$ changes the antipodal mapping for $j_{0+}^{\text{global}}$ \eqref{jAx}. We have
\begin{align}
	j^{\text{global}}_{0+}(A x) = e^{\pm i \pi \Delta}j^{\text{global}}_{0 + }(x)
\end{align}
and for the smearing function we get
\begin{align}
	\phi(P) &= \int \mathrm{d}\tau' \mathrm{d}^{d-1}\Omega' K_{2}^{G}(\tau',\Omega'| P)(j^{\text{global}}_{0+} + j^{\text{global}}_{0-}) \notag \\
&= \int_{\text{Poincar\'e patch}} \mathrm{d}\tau' \mathrm{d}^{d-1}\Omega' c'_{d \Delta}|\sigma \cos\rho'|^{-\Delta}\ln|\sigma \cos\rho'| \begin{cases*}
	e^{i \pi \Delta} {j}^{\text{global}}_{0 +} & \text{in image of $\mathbf{I}$}\\
	{j}^{\text{global}}_{0 +} & \text{in region $\mathbf{II}$ }\\
	e^{-i \pi \Delta} {j}^{\text{global}}_{0 +} & \text{in image of $\mathbf{III}$} 
\end{cases*}
\end{align}
We have suppressed the complex conjugate in the last expression to reduce clutter. 
Regarding the phases as part of the smearing function rather than as part of the boundary non-normalizable mode, we get the following expression in Poincar\'e coordinates
\begin{align}
	\phi(P) = \int \mathrm{d}t' \mathrm{d}^{d-1}x'
	\begin{rcases}
		\begin{dcases}
			\;\; e^{i \pi \Delta} \\ 
		 	\;\;\;\;1\\ 
			\;\; e^{-i \pi \Delta} 
		\end{dcases}
	\end{rcases} c'_{d \Delta}|\sigma z'|^{-\Delta}\ln|\sigma z'|\ j_{0+} + {\rm c.c}
\end{align}
where we have followed appendix B of \cite{Hamilton:2006az} to replace $\ln|\sigma \cos\rho'|$ by $\ln|\sigma z'|$, and the $j_0$ is the Poincar\'e non-normalizable mode. The phases in the smearing function can now be absorbed via the same $i\epsilon$ prescription as suggested in section 3.2 of \cite{Hamilton:2006az} to get 
$	\phi(P) = \int \mathrm{d}t' \mathrm{d}^{d-1}x'K_{2}\ j_{0+}+ {\rm c.c} $
where
\begin{align}
	K_{2} =  c'_{d \Delta}|\sigma z'|_{t' \rightarrow t' - i\epsilon}^{-\Delta}\ln|\sigma z'|
\end{align}
This is the same result that we found via mode sum in \eqref{oddPoincareK}, with a specific understanding for the timelike $i \epsilon$ prescription. 

\section{Chordal Green's Functions in Poincar\'e AdS}\label{chordal2}

In this section, we will present a direct method for constructing the spacelike kernel, where we start with a Green's function that is a function of the chordal distance. We have briefly discussed this method in the context of the global kernel in section \ref{chordal1}, but here we will present more motivation and details, as well as present the details of the Poincar\'e calculation. This will bring the discussion full circle.

The chordal distance approach (for the normalizable mode) was considered in appendix A of \cite{Hamilton:2006az} (see also \cite{Heemskerk:2012mn}). We will make a brief comment about the normalizable mode later in this section (as well as discuss some aspects of it in an appendix) but our primary goal here is to discuss the chordal distance approach for the non-normalizable mode. 

Before we proceed, let us emphasize a comment made in a footnote in our Introduction. The chordal distance approach works only in even dimensions. We will see in the appendix that the normalizable spacelike kernel constructed this way, vanishes identically in any real value of the AdS dimension other than when it is even (which was the case considered in appendix A of \cite{Hamilton:2006az}). The distinction between waves in even and odd dimensions is well-known already in flat space -- Huygens' principle applies only in even dimensions, and it is therefore natural to suspect that Green's functions cannot simply be functions of the chordal distance in odd dimensions.\footnote{As a result of (the absence of) Huygens' principle, the flat space odd-dimensional Green's function gets support not just from the lightcone, but also from points inside it. We are not aware of a satisfying intuitive explanation for the dimension-dependence of Huygens' principle, but see a discussion in \cite{Stack}. It will be nice to study this in detail in AdS, because the background is now curved, and also we allow massive fields.} Because of this, it is natural to expect that our calculation below for the non-normalizable mode, should also be trusted only in even dimensions. In fact we will see that in even dimensions, the chordal function matches with the spacelike kernel of the previous section, but in odd dimensions there is no simple comparison. 

In order to construct the normalizable kernel, appendix A of \cite{Hamilton:2006az} started with a Green's function that had a delta function divergence in the bulk. This is related to the fact that the normalizable mode dies down at infinity. When looking for the kernel for the non-normalizable mode therefore, it is natural to start instead with Witten's original Green's function in \cite{Witten:1998qj} which had a divergence at the boundary. This was done initially in Euclidean AdS space. We will quickly review this following \cite{Witten:1998qj, Erbin, DHoker:2002nbb}, before adapting it to the Lorentzian signature we need.\footnote{See also our discussion in section \ref{chordal1} which is complementary.} We start with the bulk-bulk propagator in the form
\begin{align}
    G_{\Delta}(\sigma) = \frac{2^{-\Delta}C_{\Delta}}{2\Delta - d}\sigma^{-\Delta}\;_{2}F_{1}(\frac{\Delta}{2},\frac{\Delta+1}{2}; \Delta - \frac{d}{2}+1; \frac{1}{\sigma^{2}}). \label{g1}
\end{align}
Note that this corresponds to the Associated Legendre function $Q^{\mu}_\nu$ solution of the chordal wave equation. The other $P^\mu_\nu$ solution was the one used in \cite{Hamilton:2006az}.
In the $\sigma \rightarrow \infty$ limit (i.e. $z' \rightarrow 0$), we get the following relation
\begin{align}
    G_{\Delta}(\sigma)|_{z' \rightarrow 0} = \frac{C_{\Delta}}{2\Delta - d}z'^{\Delta}\Big(\frac{z}{z^{2} + |\vec{x}|^{2}}\Big)^{\Delta} \equiv \frac{z'^{\Delta}}{2\Delta - d}K_{\Delta}(z,x)
\end{align}
The function $K_\Delta$ is the bulk-boundary propagator -- it is related to (but should not be confused with) the kernel. Now, $K_{\Delta}$ is required to have a normalized $\delta$-function behavior \cite{Witten:1998qj, Erbin}, and we integrate $K_{\Delta}$ over $x$ to fix the normalization:
\begin{align}
    \int \mathrm{d}^{d}x K_{\Delta}(x) &= C_{\Delta}z^{\Delta}\int \mathrm{d}^{d}\vec{x} \frac{1}{(z^{2} + |\vec{x}|^{2})^{\Delta}} \nonumber \\
    &=  C_{\Delta}z^{\Delta}\Omega_{d-1}\int \mathrm{d}x \frac{x^{d-1}}{(z^{2} + x^{2})^{\Delta}} \nonumber \\
    &= C_{\Delta}z^{d - \Delta}\Omega_{d-1} \int_{0}^{\infty} \mathrm{d}t \frac{t^{d-1}}{(1 + t^{2})^{\Delta}} \;\;\text{with}\;\; |\vec{x}| = x = t z  \nonumber \\
    &= C_{\Delta}z^{d - \Delta}\Omega_{d-1}\frac{\Gamma (\frac{d}{2}) \Gamma (\Delta -\frac{d}{2})}{2 \Gamma (\Delta )}
\end{align}
This gives us the coefficient $C_{\Delta} = \frac{\Gamma(\Delta)}{\pi^{\frac{d}{2}}\Gamma(\Delta  - \frac{d}{2})}$. And with this, we may write
\begin{align}
    K_{\Delta}(z,x;x') = z^{d - \Delta}\delta^{d}(x-x')
\end{align}

This was in Euclidean signature. We are interested in Lorentzian signature, and we now adopt a similar procedure. Note that since $x$ now contains both timelike and spacelike coordinates, the same integral as given above, will not hold. However, it now gets modified as follows (denoting this Lorentzian function by $\mathcal{K}_{\Delta}$)
\begin{align}
    \int_{\text{spacelike}} \mathrm{d}^{d}x \mathcal{K}^P_{\Delta}(x) &= C^P_{\Delta}z^{\Delta}\int_{\text{spacelike}} \mathrm{d}^{d}x \frac{1}{(z^{2} + x^{2})^{\Delta}} \nonumber \\
    &=  C^P_{\Delta}z^{\Delta}\Omega_{d-2}\int_{\text{spacelike}} \mathrm{d}x\mathrm{d}t \frac{x^{d-2}}{(z^{2} + x^{2} - t^{2})^{\Delta}} \nonumber \\
    &= C^P_{\Delta}z^{d - \Delta}\Omega_{d-2} \int_{u = 0}^{\infty} \int_{v= - \sqrt{1 + u^{2}}}^{\sqrt{1 + u^{2}}} \mathrm{d}u \mathrm{d}v \frac{u^{d-2}}{(1 + u^{2} - v^{2})^{\Delta}} \;\;\text{with}\;\; x = z u \;\;\&\;\; t = z v \nonumber \\
    &= C^P_{\Delta}z^{d - \Delta}\Omega_{d-2} \int_{u = 0}^{\infty}\mathrm{d}u u^{d-2}\frac{2\;_{2}F_{1}(\frac{1}{2},\Delta; \frac{3}{2}; 1)}{(1 + u^{2})^{\Delta  - \frac{1}{2}}} \nonumber \\
    &= C^P_{\Delta}z^{d - \Delta}\Omega_{d-2}\frac{\Delta  \Gamma \left(\frac{d-1}{2}\right) \cos (\pi  \Delta ) \Gamma (-\Delta ) \Gamma \left(\Delta -\frac{d}{2}\right)}{2 \sqrt{\pi }} \nonumber \\
    &=  -C^P_{\Delta}z^{d-\Delta}\pi^{\frac{d}{2}-1} \Gamma(1 - \Delta)\Gamma(\Delta - \frac{d}{2})\cos\pi\Delta \label{ker2norm}
\end{align}
Note that in the third step in this calculation, we have restricted $v \in (-\sqrt{1 + u^{2}}, \sqrt{1 + u^{2}})$. This is because we are doing this integral only over the \textit{spacelike} region, since the Green's function is taken to vanish everywhere else.\par
This gives us the following coefficient
\begin{align}
    C^P_{\Delta} =  -\frac{1}{\pi^{\frac{d}{2}-1} \Gamma(1 - \Delta)\Gamma(\Delta - \frac{d}{2})\cos\pi\Delta} = -\frac{ \Gamma(\Delta)\tan\pi\Delta}{\pi^{\frac{d}{2}}\Gamma(\Delta - \frac{d}{2})}  \label{PoinC}
\end{align}
Hence we get the following expression for the spacelike Poincar\'e Green's function in terms of the chordal distance:
\begin{align}
    \mathcal{G}^P_{\Delta}(\sigma) = -\frac{2^{-\Delta-1 }\Gamma(\Delta)\tan\pi\Delta}{\Gamma(\Delta - \frac{d}{2}+1)\pi^{\frac{d}{2}}}\sigma^{-\Delta}\;_{2}F_{1}(\frac{\Delta}{2}, \frac{\Delta + 1}{2};\Delta - \frac{d}{2} + 1; \frac{1}{\sigma^{2}}) \theta({\rm spacelike})\label{greenNNorm}
\end{align} 
Note that this is the Poincar\'e spacelike bulk-to-bulk propagator, to be distinguished from the corresponding global object we discussed in section \ref{chordal1}. The normalizations of the two were related there, and the resulting kernel precisely matched the global mode sum result.

Note that the argument of the Hypergeometric$\;_{2}F_{1}(\xi)$ function satisfies the constraint $|\xi| < 1$. This is because after analytic continuation we will be restricting to the spacelike region ($\sigma > 1$). It is well known  \cite{Arthur, BECKEN2000449} that in this region, the Hypergeometric$\;_{2}F_{1}$ is analytic for all real/complex values. Therefore, the functional form of the Green's function in \eqref{greenNNorm} carries through to Lorentzian case from the Euclidean case.\par
In the limit $z' \rightarrow 0$ (i.e. $\sigma \rightarrow \infty$), we get the following result (with $k_{\Delta} = -\frac{2^{-\Delta - 1}\Gamma(\Delta)\tan\pi\Delta}{\Gamma(\Delta - \frac{d}{2}+1)\pi^{\frac{d}{2}}}$)
\begin{align}
    \mathcal{G}^P_{\Delta}(\sigma)|_{z' \rightarrow 0} &= k_{\Delta}\sigma^{-\Delta} \label{gm1}\\
    \partial_{z'}\mathcal{G}^P_{\Delta}(\sigma)|_{z' \rightarrow 0} &= \Delta k_{\Delta} \frac{\sigma^{-\Delta}}{z'} \label{gm2}
\end{align}
For the field $\Phi(x,t,z)$, from Green's identity
\begin{align}
\Phi(x,t,z) = \int \mathrm{d}t' \mathrm{d}^{d-1}x' \sqrt{g'}(\Phi(x',t',z')\partial_{z'}\mathcal{G}^P_{\Delta}(\sigma) - \mathcal{G}^P_{\Delta}(\sigma)\partial_{z'}\Phi(x',t',z'))|_{z' \rightarrow 0}\label{green00}
\end{align}
we get
\begin{align}
    \Phi(x,t,z)
    &= \int \mathrm{d}t' \mathrm{d}^{d-1}x' \sqrt{g'}\Big((z'^{\Delta}\phi_{0}(x') + z'^{d-\Delta}j_{0}(x'))\Delta k_{\Delta} \frac{\sigma^{-\Delta}}{z'} \nonumber \\ &- k_{\Delta}\frac{\sigma^{-\Delta}}{z'}(\Delta z'^{\Delta}\phi_{0}(x') + (d-\Delta)z'^{d - \Delta}j_{0}(x'))\Big) |_{z' \rightarrow 0} \nonumber \\
    &= \int \mathrm{d}t' \mathrm{d}^{d-1}x' z'^{-d + 1}\Big((2\Delta - d)k_{\Delta}\frac{\sigma^{-\Delta}}{z'}z'^{d-\Delta}j_{0}(x')\Big)
\end{align}
This gives us the following reconstruction kernel (noting that $(2\Delta - d)k_{\Delta} = 2 a'_{d\Delta}$) for the non-normalizable mode
\begin{equation}
    K_{2}(x,t,z,x',t') = 2 a'_{d \Delta}\lim_{z' \rightarrow 0}(\sigma z')^{-\Delta} 
    \label{ker2green}
\end{equation}
Happily, this result matches with \eqref{spclkK2}. The Poincar\'e kernels computed via either method (mode sum or chordal Green's function) match, and they match (up to the factor of 2 and the integration domain) with the global kernels computed via either method.



For the normalizable mode, the chordal distance approach in even-dimensional AdS was undertaken in \cite{Hamilton:2006az}. We will not repeat it here, except to note that there also, in Poincar\'e AdS there was an extra factor of two. We encountered this factor in our discussion in section \ref{spacelikekernel}. In global AdS \cite{Hamilton:2006az, Balasubramanian:1998sn} the normalizable kernel is given by \cite{Hamilton:2006az}
\begin{align}
     K_{1} = a_{d \Delta}\lim_{\rho' \rightarrow \frac{\pi}{2}}(\sigma \cos\rho')^{\Delta - d}\theta(\text{spacelike}). 
\end{align}
Note that spacelike-ness ensures that $\sigma \cos\rho'$ is always a positive quantity. So it can be replaced by $|\sigma \cos\rho'|$: 
\begin{align}
    K_{1} = a_{d \Delta} \lim_{\rho' \rightarrow \frac{\pi}{2}}|\sigma \cos\rho'|^{\Delta - d}\theta(\text{spacelike}).
\end{align}
At this stage, we can convert to Poincar\'e coordinates \cite{Bayona:2005nq}. If we consider the entire kernel integral including the boundary field, the boundary field will acquire the phases discussed in section \ref{spacelikekernel} after the antipodal mapping to Poincar\'e. 
Again, these phases can be absorbed into the definition of the kernel and  then by the addition of a ``trivial'' function \eqref{f1} we can remove the non-spacelike pieces while producing an extra factor of two in the spacelike region: 
\begin{align}
    K_{1} &= 2 a_{d \Delta}\lim_{z' \rightarrow 0}(\sigma z')^{\Delta - d}\theta(\text{spacelike}).
\end{align}
This is the same as \eqref{spclkK1}, as expected. Note that the step function in the last expression restricts the kernel to the spacelike region of the Poincar\'e patch, as opposed to the spacelike region of the global patch. The philosophy here is that we wish to view the global and Poincar\'e coordinates to mesh together in a nice geometric way, and antipodal matching is the natural way to do it. As a result, the phase structure of the trivial function \eqref{f1} precisely cancels the extra phases arising from the antipodal matching.

We conclude this section by summarizing the definitions of the coefficients that show up in our kernels.
\begin{itemize}
\item $a'_{d \Delta}$ =  The coefficient for the non-normalizable kernel in even AdS (defined in global coordinates, so the Poincar\'e coefficient is twice this).  
\item $c'_{d \Delta}$ = The coefficient for the non-normalizable kernel in odd AdS (which is the same in Poincar\'e and global coordinates).
\item $a_{d \Delta}$ = The coefficient for the normalizable kernel in even AdS (defined in global coordinates, so the Poincar\'e coefficient is twice this).
\item $c_{d \Delta}$ = The coefficient for the normalizable kernel in odd AdS (which is the same in Poincar\'e and global coordinates).
\end{itemize}

\section{Inside the Breitenlohner-Freedman Window} 

Our results so far are somewhat formal. This is because the kernel integrals are not convergent for all relevant values of $\Delta$ (even though they are convergent for infinite ranges of values of $\Delta$). Therefore they need analytic continuation to be fully defined. This was emphasized in \cite{Nadal, Aoki}, where analytic continuation prescriptions were presented that were used above the unitarity bound for the normalizable kernel. In this section, we will use the same analytic continuation argument to argue that both kernels can be made simultaneously well-defined within the Brienlohner-Freedman (BF) window. This is useful because, in this regime of scaling dimensions,  we expect the source mode to also be fully understood as a CFT operator (in the Legendre transformed CFT \cite{Kleb}). 

The key observation is simple. Even though the authors of \cite{Nadal, Aoki} do not state it in this way, it is easy to see that their analytic continuation argument is used when the power of $\sigma $ in the kernel is 
\bea 
{\rm power} \ge -\frac{d}{2} -1  
\eea
When we have both the normalizable and non-normalizable kernels, this leads to two simultaneous conditions which together yield
\bea
\frac{d}{2}-1 \le \Delta \le \frac{d}{2}+1
\eea
which comfortably contains the BF window. 

The fact that the lower end of the range extends below the lower end of the BF window makes one suspect that the analytic continuation argument of \cite{Nadal} is not maximal and can perhaps be extended further. In fact, it turns out that even though \cite{Nadal} uses their result above the unitarity bound, the argument can, in fact, be extended to generic values of $\Delta$, making the kernel well-defined on generic points on the complex $\Delta$-plane.  

Let us conclude this section by noting one nice feature of the BF window when we are considering both kernels together. Consider our mode solution in the Poincar\'e patch \eqref{a1}, which is valid when $\omega^2 > |k|^2$. Even though we did not emphasize it, our argument for omitting the $\omega^2 < |k|^2$ modes was that this leads to a Bessel $I$ function. The Bessel $I$, even though it has the correct (normalizable and non-normalizable) behavior at the boundary, blows up at the Poincar\'e horizon. From the CFT point of view, this elimination of certain modes may seem ad-hoc. For the normalizable mode, however, since it is mapped to a CFT operator, it is plausible that it only contains modes that satisfy $\omega^2 > |k|^2$.\footnote{In the vacuum, this constraint arises simply by Fourier transforming the CFT two-point Wightman function to momentum space, while demanding Lorentz invariance. We leave it as an exercise for the reader to check this. This argument holds for the vacuum 2-point function. It is believed that around black hole backgrounds, this constraint no longer applies, and that all modes appear  \cite{Liu}. But direct evidence for this, is only available from bulk calculations. We thank Dan Kabat for discussions on these questions.} For the non-normalizable mode, on the other hand, since it is associated with a source, generically, there is no obvious CFT reason that can justify the omission of the $\omega^2 < |k|^2$ modes. This situation changes in the BF window because we know that the source is also be a CFT operator, albeit in a different (holographic) CFT \cite{Kleb} -- therefore, it is natural from the perspective of the BF window that the non-normalizable modes are also subject to the same restriction as the normalizable modes. 

\section{Discussion and Open Questions}

We have already summarized our main results in the introduction, so we will conclude by reviewing some (but not all) of them and also mentioning some open questions. Note that while the focus of our discussion in this paper has been on the non-normalizable mode, some of our observations on the normalizable mode fill some gaps in the literature. 

We used the mode sum approach on the Poincar\'e patch to obtain results for the two kernels via mode sum integrals in arbitrary even and odd dimensions. 
The even and odd-dimensional cases have technical differences (presumably related to the nature of Huygens' principle in odd vs. even dimensions). We expect that we can re-write these expressions in an AdS covariant form but present explicit demonstrations of this only in certain (half-) integer $\Delta$ cases for the normalizable mode. For this we developed an $i \epsilon$ prescription that leads to a generalization of an argument used for integer $\Delta \ge 2$ cases in AdS$_3$ in \cite{Hamilton:2006az}. Our $i \epsilon$ prescription involves a slight complexification of a boundary coordinate. In that sense, it has moral similarities to the discussion in \cite{Hamilton:2006fh} (see also our appendix \ref{complexboundary}), where an integral over a complex boundary spatial coordinate was introduced in the context the Poincar\'e patch kernel. This Poincar\'e integral had connections to the complexification required in the Rindler kernel \cite{Hamilton:2006az, Hamilton:2006fh,Hamilton:2007wj}. Rindler reconstruction is presently understood by viewing the kernel as a distribution \cite{Morrison:2014jha, Almheiri:2014lwa} that is useful for extracting correlators (and not directly the bulk operators). It would be interesting to investigate these connections further and understand where our prescription fits into this landscape. 


In the even AdS case, the AdS covariant form of the kernel can be restricted to a spacelike region. This makes a natural connection to the antipodal map noted in \cite{Hamilton:2006az} from the global coordinates. We identified an antipodal map for the non-normalizable mode as well in global coordinates to connect with our discussion. The spacelike non-normalizable kernel can, in fact, also be obtained from a spacelike chordal Green's function approach -- the two methods are very different, but the two results match precisely. We have also presented various auxiliary results in the text and in the appendix that may be of some interest, we will emphasize one here -- we used a simple Lorentzian version of Witten's original Euclidean argument to fix the normalization of the spacelike kernel for the non-normalizable mode in the chordal distance language.


An outstanding technical question that seems to have not gotten adequate attention is the comparison between Poincar\'e and global kernels. It is tantalizing that the Poincar\'e kernels we obtained via mode sums have a natural re-writing in an AdS covariant form (with immediate connections to global), but only up to some extra terms. It will be interesting to find a general argument (valid for all relevant\footnote{Note that unitarity constraints etc. suggest that the argument need not be valid for arbitrary $\Delta$.} $\Delta$) for dropping these extra terms, or alternatively, come up with an understanding of why they need not be dropped. Our $i \epsilon$ prescription as well as the complexification of \cite{Hamilton:2006az} seem to suggest that to make a general statement regarding this, we may require some form of analytic continuation.

\section{Acknowledgments}

We thank Dan Kabat for discussions, helpful explanations and comments on the draft. BB thanks Tanay Pathak for helpful discussions. BB is supported by the Ministry of Human Resource Development, Govt. of India through the Prime Ministers' Research Fellowship. The work of DS is supported by the DST-FIST grant number SR/FST/PSI-225/2016 and SERB MATRICS grant MTR/2021/000168. DS would like to thank the CHEP group at Indian Institute of Science (IISc.) for their kind hospitality and support during the initial part of this project.

\appendix

\section{Normalizable Chordal Green's Function in General Dimensions}

In this appendix, we will try to generalize the chordal function approach of \cite{Hamilton:2006az} for the normalizable mode, to general dimensions. We will fail, and see that the kernel vanishes in any (real) dimension other than when it is even.

To begin with, we will review the even-dimensional calculation \cite{Hamilton:2006az}. In terms of the chordal distance in Euclidean AdS,
\begin{equation}
	\sigma(z,x;z',x') = \frac{z^{2} + z'^{2} + |x-x'|^{2}}{2 z z'}
\end{equation}
 the wave equation reduces to the following form:
\begin{equation}
(\sigma^{2}-1)\phi''(\sigma) + (d+1)\sigma \phi'(\sigma) - \Delta(\Delta - d)\phi(\sigma) = 0
\end{equation}
The general solution is 
\begin{equation}
\phi(\sigma) = (\sigma^{2}-1)^{-\mu/2}(c_{1}P^{\mu}_{\nu}(\sigma) + c_{2}Q^{\mu}_{\nu}(\sigma))
\end{equation}
where $\mu = \frac{d-1}{2}$ and $\nu = \Delta - \frac{d+1}{2}$ and $P$ and $Q$ are associated Legendre functions. In the even-dimensional case, where $\mu$ is an integer, the Legendre polynomials behave in the following way in the limit $\sigma \rightarrow 1$ (which corresponds to coincident points)
\begin{equation}\label{lim}
P^{\mu}_{\nu}(\sigma) \sim \frac{2^{-\mu/2}\Gamma(\nu+\mu+1)}{\mu! \Gamma(\nu-\mu + 1)}(\sigma - 1)^{\mu/2}, \;\; \;\; Q^{\mu}_{\nu}(\sigma) \sim 2^{\mu/2-1}e^{i \pi \mu}\Gamma(\mu)(\sigma - 1)^{-\mu/2}
\end{equation}
The interesting fact to note here is that the asymptotic behavior of $P^{\mu}_{\nu}$ given above holds only for $\mu = m \in \text{Integer}$, while the behavior of $Q^{\mu}_{\nu}$ holds for any $\mu$ as long as $\mu + \nu = \Delta - 1$ is not a negative integer.

There are two points we need to address when $\mu$ is not integer. The first problem is that we need to determine a solution that replaces $P^{\mu}_\nu$ with the correct short distance behavior \eqref{lim}. The second is that $Q^{\mu}_{\nu}$ is no longer real. 

The latter problem is easily fixed, it is known that 
\bea
\mathbf{Q}^{\mu}_{\nu}(\sigma) = \frac{e^{-i\mu\pi}}{\Gamma(\mu + \nu + 1)}Q^{\mu}_{\nu}(\sigma)
\eea
is real even when $\mu$ is not an integer. So we will work with $\mathbf{Q}^{\mu}_{\nu}$ instead of $Q^\mu_\nu$ as the second independent solution.  

To identify the other independent solution, we first
note the hypergeometric function representation of $P^{m}_{\nu}$. From here on, $m$ will denote an integer value, while $\mu$ can mean either integer or non-integer. 
\begin{equation}\label{integ}
P^{m}_{\nu}(x) = \frac{\Gamma(\nu+m+1)}{2^{m}\Gamma(\nu - m +1)\Gamma(m+1)}(x^{2}-1)^{m/2}\;_{2}F_{1}(\nu + m + 1, m-\nu; m + 1, \frac{1-x}{2})
\end{equation}
The advantage of this form is that it is immediate to read off the asymptotic behavior of the function when the argument goes near unity, from the fact that hypergeometric function goes to 1 in that limit. Let us also note the useful relations,\footnote{All the identities used in this subsection are taken from \cite{Arthur, Stegun} and the Digital Library of Mathematical Functions. }
\begin{equation}\label{ax1}
P^{\mu}_{\nu}(x) = \frac{\Gamma(\nu + \mu + 1)}{\Gamma(\nu - \mu + 1)}P^{-\mu}_{\nu}(x) + \frac{2 \sin(\mu \pi)e^{-i \mu \pi}}{\pi}Q^{\mu}_{\nu}(x),
\end{equation}
as well as
\begin{equation}\label{ax3}
\;_{2}F_{1}(a,b;c; z) = (1-z)^{c - a - b}\;_{2}F_{1}(c-a,c-b; c; z).
\end{equation}
and 
\begin{equation}\label{ax2}
P^{-\mu}_{\nu}(x) = \frac{1}{\Gamma(\mu+1)}\Big(\frac{x-1}{x+1}\Big)^{\mu/2}\;_{2}F_{1}(-\nu, \nu+1 ; \mu + 1; \frac{1-x}{2}),
\end{equation}
Note again that the last formula is useful when we want to fix the short distance behavior. Using \eqref{ax3} in \eqref{ax2}, we get the following expression (note that $c = \mu + 1$, $a = -\nu$ and $b = \nu + 1$)
\begin{align}
P^{-\mu}_{\nu}(x) &= \frac{1}{\Gamma(\mu + 1)}\Big(\frac{x-1}{x+1}\Big)^{\mu/2}\Big(\frac{1+x}{2}\Big)^{\mu}\;_{2}F_{1}(\mu + \nu + 1, \mu - \nu; \mu + 1; \frac{1-x}{2}) \nonumber \\
&= \frac{1}{2^{\mu}\Gamma(\mu + 1)}(x^{2}-1)^{\mu/2}\;_{2}F_{1}(\mu + \nu + 1, \mu - \nu; \mu + 1; \frac{1-x}{2})\label{ax4}
\end{align}
Using \eqref{ax4} and \eqref{ax1} , we see the interesting combination
\begin{align}
P^{\mu}_{\nu}(x) - \frac{2 \sin(\mu \pi)e^{-i \mu \pi}}{\pi}Q^{\mu}_{\nu}(x) &= \frac{\Gamma(\nu + \mu + 1)}{\Gamma(\nu - \mu + 1)}P^{-\mu}_{\nu}(x) \nonumber \\
&= \frac{\Gamma(\nu + \mu + 1)(x^{2}-1)^{\mu/2}}{2^{\mu}\Gamma(\mu + 1)\Gamma(\nu - \mu + 1)}\;_{2}F_{1}(\mu + \nu + 1, \mu - \nu; \mu + 1; \frac{1-x}{2}) \label{pmu}
\end{align}
One can compare it to \eqref{integ} and see that these are exactly the same, with $m$ replaced by $\mu$. The short distance behavior is manifest. This suggests that we utilize the following two {\em real} functions as the two independent solutions, when we are away from integer $\mu$, if we want the short-distance behavior on the right hand sides of \eqref{lim}:
\begin{align}
\phi_{1}(\sigma) &= (\sigma^{2}-1)^{-\mu/2}\Big(P^{\mu}_{\nu}(\sigma) + \lambda \mathbf{Q}^{\mu}_{\nu}(\sigma) \Big)\\
\phi_{2}(\sigma) &= (\sigma^{2}-1)^{-\mu/2}\mathbf{Q}^{\mu}_{\nu}(\sigma)
\end{align}
where $\lambda = - \frac{2}{\pi}\Gamma(\mu + \nu + 1)\sin\pi\mu$. 
The general solution is
\begin{align}
\phi({\sigma}) = (\sigma^{2} - 1)^{-\mu/2}(c_{1}\phi_1(\sigma) + c_{2}\phi_2(\sigma)).\label{soln1}
\end{align}

\subsection{Fixing Constants by Analytic Continuation to the Cut}

We will attempt fix the constants by demanding specific short distance behavior and by demanding that the Green's function vanish in the timelike region. This will be our spacelike Green's function.
 

With \eqref{soln1} we can demand the same short distance behavior for the Euclidean Green's function for generic $\mu$, as was demanded for integer $\mu$ \cite{Hamilton:2006az}:\footnote{When we write $d$, this means that we do not necessarily have to even work with integer dimensions. Note in particular that the volume of a sphere in $d$-dimensions can be defined using Gamma functions etc. for non-integer $d$.}
\begin{equation}
G_{E}(r) \sim - \frac{1}{(d-1)\text{Vol}(S^{d})r^{d-1}}, \;\;\;\; \text{as} \;\; r \rightarrow 0
\end{equation}
Noting that $\sigma \sim 1 + \frac{r^{2}}{2 R^{2}}$ (in global coordinates), we have the following information about the constraints
\begin{equation}
c_{1} = \text{arbitrary} \;\;\;\; c_{2} = -\frac{1}{2^{\mu - 1}(d-1)\text{Vol}(S^{d})\Gamma(\mu)R^{d-1}}
\end{equation}
Now, we wick rotate to Lorentizan signature $G_{M}(\sigma) = i \phi(\sigma + i \epsilon)$, and we restrict to $-1 < \sigma < 1$. Above the cut, the behavior of $P^{\mu}_{\nu}$ and $Q^{\mu}_{\nu}$ are known, and are as follows
\begin{align}
P^{\mu}_{\nu}(x + i \epsilon) &= i^{-\mu}\hat{P}^{\mu}_{\nu}(x), \\
\mathbf{Q}^{\mu}_{\nu}(x + i \epsilon) &= \frac{i^{\mu}}{\Gamma(\mu + \nu + 1)}\Big(\hat{Q}^{\mu}_{\nu}(x) - \frac{i \pi}{2}\hat{P}^{\mu}_{\nu}(x)\Big),
\end{align}
where $\hat{P}$ and $\hat{Q}$ denote Ferrers functions of the first and second kind. All we need to keep in mind about these functions is that they are real in the cut $-1<x<1$. Inserting this in \eqref{soln1} gives us 
\begin{align}
G_{M}(\sigma) = i(-1)^{-\mu/2}(1-\sigma^{2})^{-\mu/2}\Big(c_{1}\Big(i^{-\mu}\hat{P}^{\mu}_{\nu}(\sigma) &+ \frac{i^{\mu}\lambda }{\Gamma(\mu + \nu + 1)}\{\hat{Q}^{\mu}_{\nu}(\sigma) - \frac{i \pi}{2} \hat{P}^{\mu}_{\nu}(\sigma)\}\Big)\nonumber \\ &+ \frac{c_{2}}{\Gamma(\mu +\nu + 1)}i^{\mu}\{\hat{Q}^{\mu}_{\nu}(\sigma) - \frac{i \pi}{2}\hat{P}^{\mu}_{\nu}(\sigma)\}\Big).
\end{align}
This can be re-written as
\begin{equation}
G_{M}(\sigma) = i (1-\sigma^{2})^{-\mu/2}\Big( ((-1)^{-\mu}c_{1} - \frac{i \pi (\lambda c_{1}+ c_{2})}{2\Gamma(\mu+\nu+1)})\hat{P}^{\mu}_{\nu}(\sigma) + \frac{c_{1}\lambda + c_{2}}{\Gamma(\mu + \nu + 1)}\hat{Q}^{\mu}_{\nu}(\sigma)\Big)
\end{equation}

At this stage, we consider the real part of $G_{M}$ as our sought-after Green's function, since we are dealing with real boundary data and real fields.\footnote{We could have also considered imaginary part, which also leads to a real Green's function, but this does not substantively change the conclusion as can be checked.} Picking only the real part of $G_{M}$ and demanding that it vanishes in the cut region (which corresponds to the timelike region) gives us the following result
\begin{align}
\text{Re}( i(-1)^{-\mu}c_{1} + \frac{ \pi (\lambda c_{1}+ c_{2})}{2\Gamma(\mu+\nu+1)}) &= 0 \label{const1}\\
\text{Re}(i c_{1}\lambda + i c_{2}) &= 0 \label{const2}
\end{align}
Note that in the case where $\mu$ is an integer (even-dimensional AdS), we have $\lambda = 0$. Therefore, we can write this as
\begin{align}
\text{Re}(i c_{1} + \frac{(-1)^{\mu}\pi c_{2}}{2}) &= 0 \\
\text{Re}(i c_{2}) &= 0
\end{align}
The second equation is trivially satisfied, since $c_{2}$ is real, and the first equation gives the result noted in \cite{Hamilton:2006az}
\begin{align}
\text{Re}(i c_{1}) + \frac{(-1)^{\mu}\pi c_{2}}{2} = 0
\end{align}

But we are interested in arbitrary $\mu$ for the remainder of the calculation. Writing $c_{1} = \alpha + i \beta$, we get the following constraint from \eqref{const2}
\begin{align}
\frac{2}{\pi}\Gamma(\mu + \nu + 1)\beta\sin\pi\mu &= 0 \label{beta}
\end{align}
Note that \eqref{beta} does not tell us anything at all when $\mu$ is an integer. However, when $\mu$ is non-integer, $\beta = 0$. Therefore, $c_{1}$ is purely real. Now 
let us recall the solution $\phi(\sigma)$
\begin{equation}
\phi(\sigma) = (\sigma^{2}-1)^{-\mu/2}(c_{1}(P^{\mu}_{\nu}(\sigma) + \lambda \mathbf{Q}^{\mu}_{\nu}(\sigma)) + c_{2} \mathbf{Q}^{\mu}_{\nu}(\sigma))
\end{equation}
The analytic continuation $G_{M}(\sigma) = i \phi(\sigma + i \epsilon)$ into the spacelike region does not cause any functional difference in $P, \mathbf{Q}$ since now $\sigma >  1$, and thus outside the cut $-1< \sigma < 1$. Hence the epsilon is unimportant and we simply have the following analytic continuation
\begin{equation}
G_{M}(\sigma) = i (\sigma^{2}-1)^{-\mu/2}(c_{1}(P^{\mu}_{\nu}(\sigma) + \lambda \mathbf{Q}^{\mu}_{\nu}(\sigma)) + c_{2} \mathbf{Q}^{\mu}_{\nu}(\sigma))
\end{equation}
Again, one can see that for even dimensions $\mu \in \mathbf{Z}$, $\lambda$ is 0. 
By noting that $P^{\mu}_{\nu}$, $\lambda \mathbf{Q}^{\mu}_{\nu}$ and $c_{2} \mathbf{Q}^{\mu}_{\nu}$ are all real, we can see that the real part of the Green's function will be
\begin{equation}
\text{Re}(G_{M}(\sigma)) = -\beta (\sigma^{2}-1)^{-\mu/2}(P^{\mu}_{\nu}(\sigma) + \lambda \mathbf{Q}^{\mu}_{\nu}(\sigma))
\end{equation}
with $\beta$ given by \eqref{beta}. Therefore, the real spacelike Green's function vanishes except for integer $\mu$, i.e. except for even AdS.

\section{Two General Integrals}\label{BApp}

The kind of integral that we come across quite often in mode sum integrals in empty AdS is the following
\begin{align}
    K = \int_{|q|^{2} \geq 0} \frac{d^{d}q}{(2\pi)^{d}}e^{i q. (x-x')}z^{\frac{d}{2}}|q|^{\mu}\zeta_{\nu}(|q|z) \label{org}
\end{align}
where $\zeta_{\nu}$ is some Bessel function. We write some fairly general result for this integral. We consider both Lorentzian ($K_{L}$) and Euclidean ($K_{E}$) AdS .

We begin with the Lorentzian case.
\begin{align}
    K_{L} &= \int_{|q|^{2} \geq 0} \frac{d^{d}q}{(2\pi)^{d}}e^{i q. (x-x')}z^{\frac{d}{2}}|q|^{\mu}\zeta_{\nu}(|q|z) \nonumber \\
    &= \int_{\omega > |k|}\frac{\mathrm{d}\omega \mathrm{d}^{d-1}\vec{k}}{(2\pi)^{d}}e^{-i\omega(t-t')}e^{i \vec{k}.(\vec{x} - \vec{x}')} z^{\frac{d}{2}}(\sqrt{\omega^{2} - |k|^{2}})^{\mu}\zeta_{\nu}(\sqrt{\omega^{2} - |k|^{2}}z)
\end{align}
Writing $\vec{k}.(\vec{x} - \vec{x}') = k \Delta x \cos\theta$, $\theta \in (0, \pi)$
\begin{align}
     K_{L} &= \int_{\omega > |k|}\frac{\mathrm{d}\omega \mathrm{d}^{d-1}\vec{k}}{(2\pi)^{d}}e^{-i\omega(t-t')}e^{i \vec{k}.(\vec{x} - \vec{x}')} z^{\frac{d}{2}}(\sqrt{\omega^{2} - |k|^{2}})^{\mu}\zeta_{\nu}(\sqrt{\omega^{2} - |k|^{2}}z) \nonumber \\
     &= \frac{\Omega_{d-3}}{(2\pi)^{d}}\int_{\omega > k}\mathrm{d}\omega \mathrm{d}k e^{-i\omega \Delta t}k^{d-2}z^{\frac{d}{2}}(\sqrt{\omega^{2} - k^{2}})^{\mu}\zeta_{\nu}(\sqrt{\omega^{2} - k^{2}}z)\int_{0}^{\pi}(\sin\theta)^{d-3}e^{i k \Delta x \cos\theta}\mathrm{d}\theta  \label{loreq1}
\end{align}
where $\Omega_{d-3} = \frac{2 \pi^{\frac{d}{2}-1}}{\Gamma(\frac{d}{2}-1)}$. The integral over $\theta$ uses the result
\begin{align}
     I_{\theta} &= \int_{0}^{\pi}\sin^{d-3}\theta_{1} e^{i k \Delta x \cos\theta_{1}}\mathrm{d}{\theta_{1}} \nonumber \\
    &= \int_{-1}^{1} (1-t^{2})^{\frac{d}{2}-2}e^{i k \Delta x t}\mathrm{d}t \nonumber \\
    &= \sqrt{\pi}\frac{\Gamma(\frac{d}{2}-1)}{\Gamma(\frac{d-1}{2})}\;_{0}F_{1}(\frac{d-1}{2}, -\frac{ (k\Delta x)^{2}}{4}) \nonumber \\
    &= \sqrt{\pi}\Gamma(\frac{d}{2}-1)2^{\frac{d-3}{2}}\frac{J_{\frac{d-3}{2}}(k\Delta x)}{(k\Delta x)^{\frac{d-3}{2}}} \label{itheta}
\end{align}

Therefore, \eqref{loreq1} becomes
\begin{align}
    K_{L} = \frac{\Omega_{d-3}\sqrt{\pi}2^{\frac{d-3}{2}}\Gamma(\frac{d}{2}-1)}{(2\pi)^{d}(\Delta x)^{\frac{d-3}{2}}}\int_{\omega > k}\mathrm{d}\omega \mathrm{d}k e^{-i\omega \Delta t}k^{\frac{d-1}{2}}z^{\frac{d}{2}}(\sqrt{\omega^{2} - k^{2}})^{\mu}\zeta_{\nu}(\sqrt{\omega^{2} - k^{2}}z)J_{\frac{d-3}{2}}(k\Delta x)
\end{align}
Note that since  $\omega > k$, we choose the parametrization
\begin{align}
    \omega = a \cosh y \;\;\;\; k = a \sinh y \;\;\;\; 0 < a, y < \infty
\end{align}
The Jacobian of this transformation is $a$.

Therefore, we get the following result
\begin{align}
    K_{L} = \frac{\Omega_{d-3}\sqrt{\pi}2^{\frac{d-3}{2}}\Gamma(\frac{d}{2}-1)}{(2\pi)^{d}(\Delta x)^{\frac{d-3}{2}}} \int_{a = 0}^{\infty} \mathrm{d}a  z^{\frac{d}{2}} a^{\mu + \frac{d+1}{2}}\zeta_{\nu}(a z)\int_{y = 0}^{\infty} \mathrm{d}y e^{-i a \Delta t \cosh y}(\sinh y)^{\frac{d-1}{2}}J_{\frac{d-3}{2}}(a\Delta x \sinh y)
\end{align}
To evaluate the $y$ integral $I_{y}$, we proceed as follows
\begin{align}
    I_{y} &= \int_{0}^{\infty}\mathrm{d}y (\sinh y)^{\frac{d-1}{2}}e^{-i a \Delta t \cosh y}J_{\frac{d-3}{2}}(a\Delta x \sinh y) \nonumber \\
    &= \int_{1}^{\infty}\mathrm{d}x (\sqrt{x^{2} - 1} )^{\frac{d-3}{2}}e^{-i x a\Delta t }J_{\frac{d-3}{2}}(a\Delta x \sqrt{x^{2} - 1})
\end{align}
Define $\alpha = a \Delta t$ and $\beta  = a \Delta x$
\begin{equation}
    I_{y} = \int_{1}^{\infty}\mathrm{d}x (\sqrt{x^{2} - 1} )^{\frac{d-3}{2}}e^{-i x \alpha }J_{\frac{d-3}{2}}(\beta \sqrt{x^{2} - 1})
\end{equation}
At this stage, it is necessary to introduce  a small complex piece to $\alpha$ in order to make the integral convergent. Specifically, we elevate $\alpha$ to $\tilde{\alpha} = \alpha - i\epsilon$, with $\epsilon > 0$. This gives us the following integral
\begin{equation}
    I_{y} = \int_{1}^{\infty}\mathrm{d}x (\sqrt{x^{2} - 1} )^{\frac{d-3}{2}}e^{- x (\epsilon +  i\alpha) }J_{\frac{d-3}{2}}(\beta \sqrt{x^{2} - 1})
\end{equation}
We use the following identity \cite{grad}
\begin{equation}
    \int_{1}^{\infty}(x^{2}-1)^{\nu/2}e^{-\alpha x}J_{\nu}(\beta \sqrt{x^{2}-1})\mathrm{d}x = \sqrt{\frac{2}{\pi}}\beta^{\nu}(\alpha^{2} + \beta^{2})^{-\frac{\nu}{2} - \frac{1}{4}}K_{\nu + \frac{1}{2}}(\sqrt{\alpha^{2} + \beta^{2}})
\end{equation}
Using this identity and identifying $\tilde{\alpha} =  \alpha - i\epsilon$ gives us
\begin{equation}
    I_{y} =  \sqrt{\frac{2}{\pi}}\beta^{\frac{d-3}{2}}(\beta^{2} + (i)^{2} \tilde{\alpha}^{2})^{\frac{2-d}{4}}K_{\frac{d}{2}-1}(\sqrt{\beta^{2} - \alpha^{2}})
\end{equation}
Inserting the expressions for $\alpha$ and $\beta$, we get (we replace $\Delta t = t - t'$ with $\Delta \tilde{t} = t' - t$ by extracting a $-$ve sign, to keep a consistent convention throughout the paper). 
\begin{equation}
    I_{y} =  \sqrt{\frac{2}{\pi}}a^{-\frac{1}{2}}(\Delta x)^{\frac{d-3}{2}}\left(\Delta x^{2} - (\Delta \tilde{t} + i\epsilon)^{2}\right)^{\frac{2-d}{4}}K_{\frac{d}{2}-1}(a \sqrt{\Delta x^{2} - (\Delta \tilde{t} + i\epsilon)^2 })
\end{equation}
Hence, we use this to write the expression for $K_{L}$. We use the notation $X \equiv \sqrt{\Delta x^{2} - (\Delta \tilde{t} + i\epsilon)^{2}} = \sqrt{(x - x')^{2} - (t' - t + i\epsilon)^{2}}$. 
\begin{align}
K_{L} = \frac{\sqrt{2}\Omega_{d-3}\sqrt{\pi}2^{\frac{d-3}{2}}\Gamma(\frac{d}{2}-1)}{(2\pi)^{d}X^{\frac{d}{2}-1}\sqrt{\pi}} z^{\frac{d}{2}}\int_{a = 0}^{\infty}  a^{\mu + \frac{d}{2}}\zeta_{\nu}(a z)K_{\frac{d}{2}-1}(a X) \mathrm{d}a
\end{align}
From here on we will refrain from explicitly mentioning the $i\epsilon$ term, but it is to be understood that it is necessary to include this term to handle singularities

The coefficients in the front simplify to give us the following result
\begin{align}
    K_{L} = \int_{|q|^{2} \geq 0} \frac{d^{d}q}{(2\pi)^{d}}e^{i q. (x-x')}z^{\frac{d}{2}}|q|^{\mu}\zeta_{\nu}(|q|z) = \frac{1}{\pi (2\pi)^{\frac{d}{2}}}\frac{z^{\frac{d}{2}}}{X^{\frac{d}{2}-1}}\int_{a = 0}^{\infty}  a^{\mu + \frac{d}{2}}\zeta_{\nu}(a z)K_{\frac{d}{2}-1}(a X) \mathrm{d}a \label{loreqfinal}    
\end{align}

We turn to the Euclidean case. This is considerably simpler because all the momentum vector components are on the same footing. The integral is then written as 
\begin{align}
     K_{E} &= \int_{q \geq 0} \frac{d^{d}q}{(2\pi)^{d}}e^{i q. (x-x')}z^{\frac{d}{2}}|q|^{\mu}\zeta_{\nu}(|q|z) \nonumber \\
    &= \frac{\Omega_{d-2}}{(2\pi)^{d}}\int_{q \geq 0} q^{d-1} \mathrm{d}q \mathrm{d}\theta (\sin\theta)^{d-2}e^{i q \Delta x \cos \theta} z^{\frac{d}{2}}q^{\mu}\zeta_{\nu}(q z)
\end{align}
where $\Omega_{d-2} = \frac{2 \pi^{\frac{d-1}{2}}}{\Gamma(\frac{d-1}{2})}$. The $\theta$ integral then gives us (following \eqref{itheta})
\begin{align}
    I_{\theta} = \sqrt{\pi}\Gamma(\frac{d-1}{2})2^{\frac{d}{2}-1}\frac{J_{\frac{d}{2}-1}(q \Delta x)}{(q \Delta x)^{\frac{d}{2}-1}}
\end{align}
Therefore, we get
\begin{align}
    K_{E} = \frac{\Omega_{d-2}\sqrt{\pi}2^{\frac{d}{2}-1}\Gamma(\frac{d-1}{2})}{(2\pi)^{d}(\Delta x)^{\frac{d}{2}-1}}z^{\frac{d}{2}}\int_{q>0} q^{\frac{d}{2} + \mu} \zeta_{\nu}(q z)J_{\frac{d}{2}-1}(q \Delta x)\mathrm{d}q
\end{align}
The coefficient out front simplifies to give us (with $X = \Delta x$)
\begin{align}
    K_{E} = \int \frac{d^{d}q}{(2\pi)^{d}}e^{i q. (x-x')}z^{\frac{d}{2}}|q|^{\mu}\zeta_{\nu}(|q|z) = \frac{1}{(2\pi)^{\frac{d}{2}}}\frac{z^{\frac{d}{2}}}{X^{\frac{d}{2}-1}}\int_{q=0}^{\infty}q^{\mu + \frac{d}{2}} \zeta_{\nu}(q z)J_{\frac{d}{2}-1}(q X)\mathrm{d}q \label{euceqnfinal}
\end{align}
Using these standard results \eqref{loreqfinal},\eqref{euceqnfinal} we can compute the integrals of the type \eqref{org} for any Bessel function $\zeta_{\nu}$ in the Lorentzian and Euclidean cases respectively.

\section{Identities Involving Hypergeometric/Gamma Functions}

In this section we will employ variable transformation relations for the results in \eqref{ker1.7}-\eqref{ker1.8}. Certain transformations are distinct for $d$ even or odd. Therefore, we treat these cases separately. The general strategy would be to employ the transformation (following  \cite{Hamilton:2006az}) to the hypergeometric function
\begin{align}
	\;_{2}F_{1}(a,b;c;z) &= \frac{\Gamma(c)\Gamma(b-a)}{\Gamma(b)\Gamma(c-a)}(-z)^{-a}\;_{2}F_{1}(a, 1-c+a; 1-b+a; \frac{1}{z})\nonumber \\ &+  \frac{\Gamma(c)\Gamma(a-b)}{\Gamma(a)\Gamma(c-b)}(-z)^{-b}\;_{2}F_{1}(b, 1-c+b;1-a+b; \frac{1}{z}) \label{id2}
\end{align}
However, this relation holds only when $a-b$ is not an integer. In the case where $a-b$ is an integer, we resort to the following series sum 
\begin{align}
	\frac{\;_{2}F_{1}(a,a+m;c;z)}{\Gamma(c)} &=\frac{(-1)^{m}(1-z)^{-a-m}}{\Gamma(a)\Gamma(c-a-m)}\sum_{k=0}^{\infty}\frac{(a+m)_{k}(c-a)_{k}}{\Gamma(k+1)\Gamma(k+m+1)}(1-z)^{-k}(\ln(1-z) + h_{n})  \nonumber \\
	&+\frac{(1-z)^{-a}}{\Gamma(a+m)\Gamma(c-a)}\sum_{k=0}^{m-1}\frac{(a)_{k}(c-a-m)_{k}\Gamma(m-k)}{\Gamma(k+1)}(z-1)^{-k} \label{id4}
\end{align}
where $h_{n} = \psi(k+1) + \psi(1+k+m) - \psi(a+m+k) - \psi(c-a+k)$.The expression is valid for $|z-1| > 1$ and $|\text{ph}(1-z)|<\pi$. In the following subsection, we will find the expressions to have the argument $z$ such that it precisely satisfies these conditions.

We will also need to use the following expansion for a binomial 
\begin{align}
	(1+x)^{\alpha} = - \frac{\sin\pi \alpha}{\pi}\Gamma(1+\alpha)\sum_{n=0}^{\infty}\frac{\Gamma(n-\alpha)}{\Gamma(n+1)}(-x)^{n} \label{form0}
\end{align}
It is fairly straightforward to derive this relation. All one needs for this is the usual series representation for $(1+x)^{\alpha}$, and the relation between $\Gamma(1-z)$ and $\Gamma(z)$ for $z$ not an integer. These are as follows
\begin{align}
	(1+x)^{\alpha} &= \sum_{n=0}^{\infty}\frac{\Gamma(\alpha + 1)}{\Gamma(n+1)\Gamma(\alpha - n + 1)}x^{n} \label{form1} \\
	\Gamma(1-z)\Gamma(z) &= \frac{\pi}{\sin\pi z} \label{form2}
\end{align}
Using \eqref{form2} and replacing $z$ by $n-z$, we get the relation
\begin{equation}
	\Gamma(n-z)\Gamma(1-n+z) = (-1)^{n-1}\Gamma(z)\Gamma(1-z) \label{form3}
\end{equation}
Using \eqref{form3} in \eqref{form1}, we get \eqref{form0}. Now, we can turn to the odd and even $d+1$ cases separately.

\section{Spatial $i\epsilon$-Prescription}\label{iepsilon}

In expressions like \eqref{K1_3.9}, we encounter certain extra terms beyond those in the final form of the kernel. We expect they should vanish, for various reasons discussed in the main text. 

From a simple series expansion in the first term of \eqref{K1_3.9}, we see that it consists of powers of $(T^{2} - R^{2})$, while the second term of \eqref{K1_3.9} has powers of $(T^{2} - R^{2} - Z^{2}) $. Here, $T = t' - t$ and $R = |x' - x|$. The exponent in each case need not be integer. The argument we present in this appendix will enable us to drop terms that are polynomials in (negative, fractional) powers of $(T^{2} - R^{2})$, while retaining those in $(T^{2} - R^{2} - Z^{2})$.  Naively this is enough to drop the first line while retaining the second. But the trouble is that the first line of \eqref{evenhyp} contains a hypergeometric function and therefore generically it is not a polynomial of this type. It can have zeroes or poles not just in $(T^{2} - R^{2})$ but also $(T^{2} - R^{2} - Z^{2})$, that are invisible from the naive power series expansion around the origin. So our argument in this section applies only in those special cases where the troublesome contributions from $(T^{2} - R^{2} - Z^{2})$ are absent. We have checked that examples of this type  arise when half-integer $\Delta \ge \frac{d}{2}$ in even-dimensional AdS as well as integer $\Delta \ge d$ in odd AdS, both for the normalizable mode.\footnote{For the non-normalizable mode the analogous values lead to wave equation solutions which are Bessel functions of the second kind, whose reconstruction kernels are presented in appendix \ref{integernu}.} For the more general case an argument that goes beyond what we discuss here will be necessary. 




Consider the kernel integrated against the positive frequency boundary mode
\begin{align}
I = \int K_1(z,x;x')\phi_{0 +}(x',t')\mathrm{d}^{d-1}\vec{x}' d t'. \label{modeint}
\end{align}
Using the Fourier modes of the boundary field,
\begin{align}
\phi_{0 +} (x,t) = \int \tilde{\phi}_{0 +} (\omega, k)e^{-i \omega t}e^{i\vec{k}.\vec{x}}\mathrm{d}^{d-1}\vec{k} \mathrm{d}\omega \label{modedecomp}
\end{align}
the following two types of integrals of interest emerge (we restrict to $AdS_4$ concreteness, but a similar argument can be made in other dimensions as well):
\begin{align}
I_{A} &= \int_{T, X, Y = -\infty}^{\infty} (T^{2} - X^{2} - Y^{2})^{\lambda}e^{- i \omega T}e^{i k_x X}e^{i k_y Y}\mathrm{d}X \mathrm{d}Y \mathrm{d}T \label{IA} \\
I_{B} &= \int_{T, X, Y = -\infty}^{\infty} (T^{2} - X^{2} - Y^{2} - Z^{2})^{\lambda}e^{- i \omega T}e^{i k_x X}e^{i k_y Y}\mathrm{d}X \mathrm{d}Y \mathrm{d}T \label{IC}
\end{align}
These expressions are eventually integrated over $\omega, \vec{k}$ (from \eqref{modedecomp}), and since $\tilde{\phi}_{0 +}$ is $0$ for $\omega < |k|$, this condition can be applied when performing the integrals \eqref{IA}-\eqref{IC}.

It is useful to perform the integral \eqref{IA} (and similarly \eqref{IC}) in two parts. In the region $Y > 0$, the integral $I_A$ can be written with the usual parametrization $X = R \cos\phi,\; Y = R \sin\phi$, where $\phi$ is defined with respect to the $+$ ve $X$-axis and $R$ lies in the range $(0, \infty)$.
\begin{align}
I_A|_{Y > 0} = \int_{T = -\infty}^{\infty}\int_{R = 0}^{\infty}\int_{\phi = 0}^{\pi} R\,(T^{2} - R^{2})^{\lambda}e^{- i \omega T}e^{i k_x R\cos\phi}e^{i k_y R\sin\phi}d R\, \mathrm{d}\phi\,\mathrm{d}T \label{IA2}
\end{align}
In the region $Y < 0$, a different parametrization is chosen. With the angle $\phi'$ defined with respect to the $-$ve $X$-axis, the parametrization $X = R' \cos\phi',\; Y = R'\sin\phi'$ is used, but (crucially!) with $R' \in (-\infty,0)$. The integral \eqref{IA} becomes
\begin{align}
I_A|_{Y < 0} = \int_{T = -\infty}^{\infty} \int_{R' = -\infty}^{0}\int_{\phi' = 0}^{\pi} R'\,(T^{2} - R'^{2})^{\lambda}e^{- i \omega T}e^{i k_x R'\cos\phi'}e^{i k_y R'\sin\phi'}dR'\,\mathrm{d}\phi'\,\mathrm{d}T \label{IA3}
\end{align}
Note that the range of $R'$ is correspondingly adjusted to $(-\infty, 0)$ in order to compensate for this choice of angle $\phi'$.

Suppressing the primes in the dummy variables in \eqref{IA3}, we can now add \eqref{IA2} and \eqref{IA3} trivially to provide an expression for the full integral \eqref{IA} where the range of $R$ is the entire real line: 
\begin{align}
I_A =\int_{T = -\infty}^{\infty} \int_{R = -\infty}^{\infty}\int_{\phi = 0}^{\pi} R\,(T^{2} - R^{2})^{\lambda}e^{- i \omega T}e^{i k_x R\cos\phi}e^{i k_y R\sin\phi}\mathrm{d}R\,\mathrm{d}\phi\,\mathrm{d} \mathrm{d}T \label{IApen}
\end{align}
This now explains our motivation behind choosing the peculiar parametrization in the previous paragraph -- had we chosen an ordinary polar coordinate system, the radial variable would be strictly positive. Instead, we wish to treat it as a real variable spanning $(-\infty, \infty)$ so that the integral can be computed by closing the contour in the upper or lower half $R$-plane. Note that in the above integral, the variables $R$ and $\phi$ should {\em not} be confused with the usual polar coordinates. In particular $R$ spans the entire real line, and $\phi$ ranges only over $(0,\pi)$ and not $(0,2 \pi)$. It should also be clear that a similar construction can be done in higher dimensions, by spanning the sphere via two separate coordinate systems -- one based on the North pole and the other, the South pole. 

Denote $k_x = k\cos\alpha$ and $k_y = k\sin\alpha$, with $k > 0$ and $\alpha \equiv \arctan(k_y/k_x)$. This puts the integral \eqref{IApen} in the following form
\begin{align}
I_A = \int_{T = -\infty}^{\infty}\int_{R = -\infty}^{\infty}\int_{\phi = 0}^{\pi} R(T^{2} - R^{2})^{\lambda}e^{- i \omega T}e^{i k R\cos(\phi - \alpha)}\mathrm{d}\phi \mathrm{d}R \mathrm{d}T \label{IAfin}
\end{align}
Similarly, the integral \eqref{IC} can be written as
\begin{align}
I_B = \int_{T = -\infty}^{\infty}\int_{R = -\infty}^{\infty}\int_{\phi = 0}^{\pi}R (T^{2} - R^{2} - Z^{2})^{\lambda}e^{- i \omega T}e^{i k R\cos(\phi - \alpha)}\mathrm{d}\phi\mathrm{d}R  \mathrm{d}T \label{ICfin}
\end{align}


In terms of the $U = T - R$, $V = T + R$ coordinates the integral \eqref{IAfin} becomes
\begin{align}
I_{A} &= \int_{U, V = -\infty}^{\infty} \int_{\phi = 0}^{\pi} \left( \frac{V - U}{2}\right)(U V)^{\lambda}e^{- i \omega_{-} U}e^{-i \omega_{+} V}\mathrm{d}\phi \mathrm{d}U \mathrm{d}V \label{IB21}
\end{align}
where $\omega_{\pm} = \omega \pm k \cos(\phi - \alpha) > 0$. The range of integration is $U, V \in (-\infty, \infty)$.

Positivity of $\omega_{\pm}$ implies that the integration contour has to be closed in the lower half plane (LHP) for both $U$ and $V$. The integral \eqref{IB21} would acquire a non-trivial value if there is any pole/branch point of the integrand inside the integration contour. Since the integrand has singularities on the real line at (noting that $d>2$) 
\begin{align}
U V = 0, \label{pbp1}
\end{align}
we will introduce an $i\epsilon$-prescription as $R \rightarrow R \pm i\epsilon$ to handle them. This shifts the coordinates $U$ and $V$ by $\mp i\epsilon$ and $\pm i\epsilon$ respectively. This is reflected in the following modification to the pole/branch point condition \eqref{pbp1}
\begin{align}
(U \mp i \epsilon)(V \pm i\epsilon) = 0
\end{align}
With this condition, the integral \eqref{IB21} can be done first on either $U$ or $V$, depending on which term comes with a $ - i\epsilon$ shift. 

Let us consider the case where the pole/branch point is given by $(U - i\epsilon)(V + i \epsilon) = 0$. This arises from the choice $R \rightarrow R + i\epsilon$. In terms of the $U$ coordinate, the location of the pole/branch point is $U = i \epsilon$. It is located in the positive imaginary axis, and hence outside the contour of integration for $U$ (which is the LHP). So the function is analytic in $U$, and the integral over $U$ will give $0$.

Let us also consider \eqref{IC} to ensure that it does {\em not} vanish. In the $U, V$ language, it becomes
\begin{align}
I_{B} &=  \int_{U, V = -\infty}^{\infty} \int_{\phi = 0}^{\pi}  \left( \frac{V - U}{2}\right)(U V - Z^{2})^{\lambda}e^{- i \omega_{-} U}e^{-i \omega_{+} V}\mathrm{d}\phi \mathrm{d}U \mathrm{d}V \label{IB31}
\end{align} 
The pole/branch point condition is (considering the $i\epsilon$ prescription)
\begin{align}
(U \mp i \epsilon)(V \pm i\epsilon) - Z^{2} = 0 \label{van1}
\end{align}
Considering the $R \rightarrow R + i\epsilon$ case, the following condition for the pole/branch point emerges
\begin{align}
&(U - i\epsilon)(V + i\epsilon) = Z^{2} \nonumber \\
&\implies U = \frac{Z^{2}}{V} - i \epsilon\frac{Z^{2}}{V^{2}} + i\epsilon 
\end{align}
The singularity is outside the contour of integration of $U$ for $\frac{Z^{2}}{V^{2}} < 1$, and would be inside the contour for $\frac{Z^{2}}{V^{2}} > 1$. This condition will be satisfied for $V^{2} < Z^{2}$. Since the integration \eqref{IB31} is also over the full range of $V$, there will be some range of $V$ for which this inequality is satisfied. Thus, there is some pole/branch point inside the contour of integration, which ensures that the integral does not vanish.

The key point is that these results are achieved as long as $U$ and $V$ are shifted by $i\epsilon$ with opposite signs. Accordingly one can choose to integrate along either $U$ or $V$, whichever comes with the $-i\epsilon$ in \eqref{van1}. Since $\omega_{\pm}$ in \eqref{IB31} are always positive (as $\omega > |k|$), the contour is closed in the LHP of the chosen integration variable. While this particular demonstration is done for $AdS_4$, similar arguments will hold for general $AdS_{d+1}$ as long as the extra terms are polynomials, as we discussed.

We suspect that the spatial $i \epsilon$-prescription that we have presented in this section is related to the Wick rotation of spatial coordinates over which the integral is done, as was discussed in \cite{Hamilton:2006fh}. Note in particular that the specific choice of the sign of the epsilon prescription was not important for the success of our calculation. It will be nice to understand the connection with the Wick rotation prescription, better. For completeness, we present explicit formulas in general dimensions for the kernel with the Wick rotated spatial coordinates, in the next section. It turns out that the mode sum integrals simplify when we do this, effectively allowing us to bypass some of the subtleties we discussed in detail. But of course the price is that now the integrals are over imaginary spatial coordinates.

\section{Kernels with Complex Boundary Coordinates}\label{complexboundary}

In this appendix, we  complexify the boundary coordinates in general dimensions and obtain an expression for the kernel in terms of the complex boundary coordinates. It is a generalization of  the result in \cite{Hamilton:2006az}, where it was done for $AdS_{3}$, to general dimensions. The generalization is not entirely trivial, so we present the calculation explicitly for the normalizable mode. The non-normalizable case follows similarly.

One can write the solution of the wave equation $\phi$ as follows 
\begin{align}
    \phi(t,x,z) &=\frac{2^{\nu}\Gamma(1+\nu)}{(2\pi)^{d}} \int_{\omega > |k|}\mathrm{d}\omega \mathrm{d}^{d-1}k z^{d/2}\frac{J_{\nu}(z\sqrt{\omega^{2} - k^{2}})}{(\sqrt{\omega^{2} - k^{2}})^{\nu}} \int \mathrm{d}t' \mathrm{d}^{d-1}x' e^{-i \omega (t- t')}e^{i \vec{k}.(\vec{x} - \vec{x'})}\phi_{0}(t',x')  \nonumber \\
    &= \frac{2^{\nu}\Gamma(1+\nu)}{(2\pi)^{d}} \int_{\omega > |k|}\mathrm{d}\omega \mathrm{d}^{d-1}k e^{-i \omega t}\;e^{i \vec{k}.\vec{x}}z^{d/2}\frac{J_{\nu}(z\sqrt{\omega^{2} - k^{2}})}{(\sqrt{\omega^{2} - k^{2}})^{\nu}} \tilde{\phi}_{0}(\omega, k)\label{eqn0}
\end{align}
where we have used the Fourier decomposition of $\phi_{0}(x',t')$
\begin{align}
    \tilde{\phi}_{0}(\omega,k) = \int \mathrm{d}t' \mathrm{d}^{d-1}x' e^{i \omega t'}e^{- i \vec{k}.\vec{x}'}\phi_{0}(x',t')
\end{align}

To write this in terms of complex boundary coordinates, we need show that
\begin{align}
    \frac{J_{\nu}(z \sqrt{\omega^{2} - |k|^{2}})}{(\sqrt{\omega^{2} - k^{2}})^{\nu}} \propto I =  \int_{t'^{2} + y'^{2} < z^{2}} \mathrm{d}t' \mathrm{d}^{d-1}y' (z^{2} - t'^{2} - y'^{2})^{\nu - \frac{d}{2}}e^{-i \omega t'}e^{- \vec{k}.\vec{y'}}\label{eqnm1}
\end{align}
We start by writing the integral $I$ as 
\begin{align}
    I &= \int_{t'^{2} + y'^{2} < z^{2}} \mathrm{d}t' (z^{2} - t'^{2} - y'^{2})^{\nu - \frac{d}{2}}e^{-i \omega t'} \int e^{-\vec{k}.\vec{y'}}\mathrm{d}^{d-1}y' \nonumber \\
    &= \Omega_{0}\int_{t'^{2} + y'^{2} < z^{2}} \mathrm{d}t' (z^{2} - t'^{2} - y'^{2})^{\nu - \frac{d}{2}}e^{-i \omega t'} \int e^{-k y' \cos\theta}(\sin\theta)^{d-3}y'^{d-2}\mathrm{d}y'\mathrm{d}\theta
\end{align}
where $\Omega_{0} = \frac{2 \pi^{\frac{d}{2} - 1}}{\Gamma(\frac{d}{2} - 1)}$ and $y' = |\vec{y'}|$. That is, we cast the $y'$ coordinates in their polar representation.

Using the following integral representation of the Bessel $I$
\begin{align}
    I_{\nu}(z) = \frac{(\frac{1}{2}z)^{\nu}}{\sqrt{\pi}\Gamma(\nu+\frac{1}{2})}\int_{0}^{\pi} e^{\pm z \cos\theta}(\sin\theta)^{2\nu}\mathrm{d}\theta
\end{align}
we have the result
\begin{align}
    I = \Omega_{0}\frac{2^{\frac{d-3}{2}}\pi\Gamma(\frac{d}{2}-1)}{\sqrt{\pi}}\int_{t'^{2} + y'^{2} < z^{2}} \mathrm{d}t' (z^{2} - t'^{2} - y'^{2})^{\nu - \frac{d}{2}}e^{-i \omega t'} \int y'^{d-2}\frac{I_{\frac{d-3}{2}}(k y')}{(k y')^{\frac{d-3}{2}}}\mathrm{d}y'
\end{align}
Therefore, we need to tackle the following integral ( we use $c = \Omega_{0}\frac{2^{\frac{d-3}{2}}\sqrt{\pi}\Gamma(\frac{d}{2}-1)}{ k^{\frac{d-3}{2}}} = \frac{(2\pi)^{\frac{d-1}{2}}}{k^{\frac{d-3}{2}}}$ for simplicity)
\begin{align}
     I &=\Omega_{0}\frac{2^{\frac{d-3}{2}}\pi\Gamma(\frac{d}{2}-1)}{\sqrt{\pi}}\int_{t'^{2} + y'^{2} < z^{2}} \mathrm{d}t' (z^{2} - t'^{2} - y'^{2})^{\nu - \frac{d}{2}} y'^{d-1}\frac{I_{\frac{d-3}{2}}(k y')}{(k y')^{\frac{d-3}{2}}}e^{-i \omega t'}\mathrm{d}y' \nonumber \\
     &= c\int_{t'^{2} + y'^{2} < z^{2}} \mathrm{d}t'\mathrm{d}y' (z^{2} - t'^{2} - y'^{2})^{\nu - \frac{d}{2}} y'^{\frac{d-1}{2}}I_{\frac{d-3}{2}}(k y')e^{-i \omega t'} 
\end{align}

At this stage it is convenient to redefine the variables $t' = u z$ and $y' = v z $.
\begin{align}
    I = c z^{2\nu - \frac{d-3}{2}}\int_{u^{2} + v^{2} < 1} \mathrm{d}u\mathrm{d}v(1 - u^{2} - v^{2})^{\nu - \frac{d}{2}} v^{\frac{d-1}{2}}I_{\frac{d-3}{2}}(k v z)e^{-i \omega u z}
\end{align}
Due to the constraint $u^{2} + v^{2} < 1$, we can use the parametrization $u = a \cos\theta$, $v = a \sin\theta$.
\begin{align}
    I = c z^{2\nu - \frac{d-3}{2}}\int_{a=0}^{1}\int_{\theta=0}^{\pi} a^{\frac{d+1}{2}}(1 - a^{2})^{\nu - \frac{d}{2}} (\sin\theta)^{\frac{d-1}{2}}I_{\frac{d-3}{2}}(k a z \sin\theta)e^{-i \omega a z \cos\theta}\mathrm{d}a\mathrm{d}\theta \label{eqn1}
\end{align}

We perform the $\theta$ integral first
\begin{align}
    A = \int_{\theta = 0}^{\pi}(\sin\theta)^{\mu+1}I_{\mu}(\alpha \sin\theta)e^{-i\beta \cos\theta}\mathrm{d}\theta
\end{align}
where we have used the notation $\mu =\frac{d-3}{2}$, $\alpha = k a z$ and $\beta = \omega a z$. Using the relation $I_{\mu}(z) = i^{-\mu}J_{\mu}(i z)$, the integral $A$ becomes (with $\tilde{\alpha} \equiv i \alpha$)
\begin{align}
    A = i^{-\mu}\int_{0}^{\pi}(\sin\theta)^{\mu+1}J_{\mu}(\tilde{\alpha}\sin\theta)e^{-i \beta \cos\theta}\mathrm{d}\theta 
\end{align}

Splitting the integral into $0-\pi/2$ and $\pi/2-\pi$, and using $\theta \rightarrow \pi - \theta$ for the second integral, we get the following result
\begin{align}
    A = 2 i^{-\mu}\int_{0}^{\pi/2}(\sin\theta)^{\mu+1}J_{\mu}(\tilde{\alpha}\sin\theta)\cos(\beta\cos\theta)\mathrm{d}\theta 
\end{align}
The following result is used to replace the $\cos(\beta \cos\theta)$
\begin{align}
    J_{-\frac{1}{2}}(z) = \sqrt{\frac{2}{\pi z}}\cos z
\end{align}
With this substitution, the integral $A$ equals
\begin{align}
    A = 2 i^{-\mu}\sqrt{\frac{\pi \beta}{2}} \int_{0}^{\pi /2}(\sin\theta)^{\mu+1}J_{\mu}(\tilde{\alpha}\sin\theta)\cos^{1/2}(\beta\cos\theta)J_{-\frac{1}{2}}(\beta\cos\theta) \mathrm{d}\theta
\end{align}

Here we shall employ the following identity
\begin{align}
    \int_{0}^{\pi/2}J_{\nu}(z_{1}\sin\theta)J_{\mu}(z_{2}\cos\theta)\sin^{\nu+1}\theta \cos^{\mu+1}\theta \mathrm{d}\theta = \frac{z_{1}^{\nu}z_{2}^{\mu}J_{\nu + \mu + 1}(\sqrt{z_{1}^{2} + z_{2}^{2}})}{(\sqrt{z_{1}^{2} + z_{2}^{2}})^{\mu + \nu + 1}} \;\; \forall \;\; \text{Re}(\mu) > -1, \; \text{Re}(\nu) > -1
\end{align}
Using this, we can write the integral $A$ as
\begin{align}
    A = 2 i^{-\mu}\sqrt{\frac{\pi \beta}{2}}\frac{\beta^{-\frac{1}{2}}\tilde{\alpha}^{\mu}J_{\mu + \frac{1}{2}}(\sqrt{\beta^{2} + \tilde{\alpha}^{2}})}{(\sqrt{\beta^{2} + \tilde{\alpha}^{2}})^{\mu + \frac{1}{2}}}
\end{align}
Inserting $\tilde{\alpha} = i \alpha$ and $\mu = \frac{d-3}{2}$, this expression simplifies to 
\begin{align}
    A = \sqrt{2\pi}\alpha^{\mu}\frac{J_{\frac{d}{2}-1}(\sqrt{\beta^{2} - \alpha^{2}})}{(\sqrt{\beta^{2} - \alpha^{2}})^{\frac{d}{2}-1}}
\end{align}
Using the explicit expressions of $\beta$ and $\alpha$
\begin{align}
    A = \sqrt{\frac{2\pi}{a z}}k^{\frac{d-3}{2}}\frac{J_{\frac{d}{2}-1}(a z\sqrt{\omega^{2} - k^{2}})}{(\sqrt{\omega^{2} - k^{2}})^{\frac{d}{2}-1}}
\end{align}

We insert $A$ in $I$ \eqref{eqn1} to get
\begin{align}
    I = \sqrt{2\pi}c k^{\frac{d-3}{2}} z^{2\nu - \frac{d}{2} + 1}\int_{0}^{1}a^{\frac{d}{2}}(1-a^{2})^{\nu - \frac{d}{2}}\frac{J_{\frac{d}{2}-1}(a z \sqrt{\omega^{2} - k^{2}})}{(\sqrt{\omega^{2} - k^{2}})^{\frac{d}{2}-1}}\mathrm{d}a \label{eqn2}
\end{align}
To evaluate this, we use the following result
\begin{align}
    \int_{0}^{1}x^{\nu+1}(1-x^{2})^{\mu}J_{\nu}(b x)\mathrm{d}x = 2^{\mu}\Gamma(\mu+1)b^{-\mu - 1}J_{\mu + \nu + 1}(b)
\end{align}
Employing this to \eqref{eqn2}, we get the result
\begin{align}
    I &= \sqrt{2\pi}c k^{\frac{d-3}{2}} z^{2\nu - \frac{d}{2} + 1}2^{\nu - \frac{d}{2}}\Gamma(\nu - \frac{d}{2}+1)\frac{J_{\nu}(z\sqrt{\omega^{2} - k^{2}})}{z^{\nu - \frac{d}{2} + 1}(\sqrt{\omega^{2} - k^{2}})^{\nu}} \nonumber \\
    &= 2^{\nu}\pi^{\frac{d}{2}}\Gamma(\nu -\frac{d}{2} + 1)z^{\nu}\frac{J_{\nu}(z\sqrt{\omega^{2} - k^{2}})}{(\sqrt{\omega^{2} - k^{2}})^{\nu}}\label{eqn3}
\end{align}
From \eqref{eqn3} and \eqref{eqnm1}, one can write
\begin{align}
    \frac{J_{\nu}(z\sqrt{\omega^{2} - k^{2}})}{(\sqrt{\omega^{2} - k^{2}})^{\nu}} = \frac{1}{2^{\nu}\pi^{\frac{d}{2}}\Gamma(\nu -\frac{d}{2} + 1)z^{\nu}} \int_{t'^{2} + y'^{2} < z^{2}} \mathrm{d}t' \mathrm{d}^{d-1}y' (z^{2} - t'^{2} - y'^{2})^{\nu - \frac{d}{2}}e^{-i \omega t'}e^{- \vec{k}.\vec{y'}}
\end{align}

Plugging it back into \eqref{eqn0}, we get ( denoting $C = \frac{}{}\frac{2^{\nu}\Gamma(1+\nu)}{(2\pi)^{d} 2^{\nu}\pi^{\frac{d}{2}}\Gamma(\nu -\frac{d}{2} + 1)} = \frac{\Gamma(1+\nu)}{(2\pi)^{d}\pi^{\frac{d}{2}}\Gamma(\nu - \frac{d}{2}+1)}$ for simplicity)
\begin{align}
    \phi(x,t,z) &=  C \int_{\omega > |k|}\mathrm{d}\omega \mathrm{d}^{d-1}k\; e^{-i \omega t}e^{i \vec{k}.\vec{x}}z^{d/2}\frac{J_{\nu}(z\sqrt{\omega^{2} - k^{2}})}{(\sqrt{\omega^{2} - k^{2}})^{\nu}} \tilde{\phi}_{0}(\omega, k) \nonumber \\
    &= C \int_{t'^{2} + y'^{2} < z^{2}} d t' \mathrm{d}^{d-1}y' \Big(\frac{z^{2} - t'^{2} - y'^{2}}{z}\Big)^{\nu - \frac{d}{2}}\int \mathrm{d}\omega\mathrm{d}^{d-1}k\; e^{-i\omega(t+t')}e^{i\vec{k}.(\vec{x} + i\vec{y}')}\phi_{0}(\omega,k) \nonumber \\
    &= (2\pi)^{d}C \int_{t'^{2} + y'^{2} < z^{2}} d t' \mathrm{d}^{d-1}y' \lim_{z' \rightarrow 0}(2 z' \sigma(t,x,z; t +t', x+ i y', z'))^{\nu - \frac{d}{2}}\phi_{0}(t + t', x + i y')
\end{align}

From this expression, we can read off the kernel $K_1$ corresponding to the normalizable mode. Recall that for normalizable mode, we have $\nu = \Delta - \frac{d}{2}$ and for non-normalizable mode we have $\nu = \frac{d}{2} - \Delta$ (also $\phi_{0}$ is replaced by $j_{0}$). Therefore, the kernel for the normalizable mode $K_{1}$ and the non-normalizable mode $K_{2}$ are as follows (using $ (2\pi)^{d}C = \frac{\Gamma(1+\nu)}{\pi^{\frac{d}{2}}\Gamma(\nu - \frac{d}{2}+1)}$)
\begin{align}
    K_{1}(z,x,t; z', t+t', x+ i y') &=  \frac{\Gamma(1+\Delta - \frac{d}{2})}{\pi^{\frac{d}{2}}\Gamma(\Delta - d +1)}\lim_{z' \rightarrow 0}(2 z' \sigma(t,x,z; t +t', x+ i y', z'))^{\Delta - d} \\
    K_{2}(z,x,t; z', t+t', x+ i y') &=  \frac{\Gamma(1-\Delta + \frac{d}{2})}{\pi^{\frac{d}{2}}\Gamma(1-\Delta)}\lim_{z' \rightarrow 0}(2 z' \sigma(t,x,z; t +t', x+ i y', z'))^{-\Delta}
\end{align}

\section{Non-normalizable Mode for Integer $\nu$}\label{integernu}

	The kernel integral for integer $\nu \equiv p$ is given by 
	\begin{align}
		K_{2}(z,x;x') &= -\int \frac{|q|^{p}\pi}{2^{p}\Gamma(p)}e^{i q.(x-x')}z^{\frac{d}{2}}\{\frac{-\gamma + \psi(p+1)- 2 \ln(|q|/2)}{\pi}J_{p}(|q|z) + Y_{p}(|q|z)\}\frac{\mathrm{d}^{d}q}{(2\pi)^{d}}
	\end{align}
	This can be broken into three integrals, each of which we will evaluate separately:
	\begin{align}
		\tilde I_{1} &= -\frac{(-\gamma + \psi(p+1) + 2\ln(2))}{2^{p} (2\pi)^{d} \Gamma(p) }\int |q|^{p}e^{i q. (x-x')}z^{d/2}J_{p}(|q|z)\mathrm{d}^{d}q \\
		\tilde I_{2} &= \frac{2}{2^{p}\Gamma(p) (2\pi)^{d} }\int |q|^{p}\ln(|q|)e^{i q. (x-x')}z^{d/2}J_{p}(|q|z)\mathrm{d}^{d}q \\
		\tilde I_{3} &= -\frac{\pi}{2^{p} (2\pi)^{d}\Gamma(p) }\int |q|^{p}e^{i q. (x-x')}z^{d/2}Y_{p}(|q|z)\mathrm{d}^{d}q
	\end{align}
	We begin with the integral $\tilde I_{3}$. 
	\subsection{Integral $\tilde I_3$}
	To perform this integral, we write out $q$ in terms of it's components $(\omega,\vec{k})$. For simplicity, we ignore the prefactor $-\frac{\pi}{2^{p} (2\pi)^{d}\Gamma(p) }$ for the moment. This factor will be included at the end.
	\begin{align}
		I_{3} = \int_{\omega >|k|}(\sqrt{\omega^{2} - |k|^{2}})^{p}e^{-i\omega \Delta t}e^{i \vec{k}.\Delta\vec{x}}z^{d/2}Y_{p}(z\sqrt{\omega^{2} - |k|^{2}})\mathrm{d}\omega \mathrm{d}^{d-1}\vec{k}
	\end{align}
	where we have used the notation $\Delta t = t - t',\, \Delta\vec{x} = \vec{x} - \vec{x'}$. As before, this integral simplifies to the following (where we denote $|k|$ by $k$)
	\begin{align}
		I_{3} = \int_{\omega >k}\int_{\theta = 0}^{\pi}(\sqrt{\omega^{2} - k^{2}})^{p}e^{-i\omega \Delta t}e^{i k\Delta x \cos\theta}z^{d/2}Y_{p}(z\sqrt{\omega^{2} - k^{2}})(\sin\theta)^{d-3}k^{d-2}\mathrm{d}\omega \mathrm{d}k \mathrm{d}\theta
	\end{align}
	We first evaluate the $\theta$ integral. As we have observed before
	\begin{align}
		\int_{0}^{\pi}\sin^{d-3}\theta e^{i k \Delta x \cos\theta}\mathrm{d}{\theta} = \sqrt{\pi}\Gamma(\frac{d}{2}-1)2^{\frac{d-3}{2}}\frac{J_{\frac{d-3}{2}}(k\Delta x)}{(k\Delta x)^{\frac{d-3}{2}}}
	\end{align}
	Using this expression in $I_3$ gives us
	\begin{align}
		I_{3} = \sqrt{\pi}\Gamma(\frac{d}{2}-1)2^{\frac{d-3}{2}}\int_{\omega >k}(\sqrt{\omega^{2} - k^{2}})^{p}e^{-i\omega\Delta t}z^{d/2}Y_{p}(z\sqrt{\omega^{2} - k^{2}})\frac{J_{\frac{d-3}{2}}(k \Delta x)}{(k \Delta)^{\frac{d-3}{2}}}k^{d-2}\mathrm{d}\omega \mathrm{d}k
	\end{align}
	The condition $\omega > k$ is easily parametrized by the variable choice $\omega= s \cosh y. \, k = s \sinh y$ with $0 \leq s,\,y \leq \infty$. This gives the following form of the integral
	\begin{align}
		I_{3} = \frac{\sqrt{\pi}\Gamma(\frac{d}{2}-1)2^{\frac{d-3}{2}}}{(\Delta x)^{\frac{d-3}{2}}} \int_{0}^{\infty}s^{p}e^{-i s\Delta t \cosh y}z^{d/2}Y_{p}(s z)J_{\frac{d-3}{2}}(s \Delta x \sinh y)(s \sinh y)^{\frac{d-1}{2}}s \mathrm{d}s\mathrm{d}y
	\end{align}
	The $y$ integral can be evaluated to give the following result
	\begin{align}
		\int_{y = 0}^{\infty}(\sinh y)^{\frac{d-1}{2}}e^{-i s \Delta t \cosh y}J_{\frac{d-3}{2}}(s \Delta x \sinh y)\mathrm{d}y = \sqrt{\frac{2}{\pi}}s^{-\frac{1}{2}}(\Delta x)^{\frac{d-3}{2}}X^{\frac{2-d}{2}}K_{\frac{d}{2}-1}(s X)
	\end{align}
	where $X = \sqrt{\Delta x^{2} - \Delta t^{2}}$ and the usual $i\epsilon$ light-cone regularization in $t$ is implicit. The integral $I_3$ now becomes
	\begin{align}
		I_{3} = \sqrt{2}\Gamma(\frac{d}{2}-1)2^{\frac{d-3}{2}}X^{1 - \frac{d}{2}}z^{d/2}\int_{0}^{\infty}s^{p + \frac{d}{2}}Y_{p}(s z)K_{\frac{d}{2}-1}(a X)\mathrm{d} s \label{I3}
	\end{align}
	The non-trivial bit is doing this integral. To evaluate this integral, we resort to the connection between Bessel $Y$ and $K$.
	\begin{mdframed}
		\subsubsection*{Connecting Bessel Y and K}
		To arrive at a relation between the Bessel $Y$ and $K$ functions, we turn to the following relations
		\begin{align}
			Y_{\alpha}(x) &= \frac{1}{2 i}\left(H^{(1)}_{\alpha}(x) -  H^{(2)}_{\alpha}(x)\right) \notag \\
			K_{\alpha}(x) &= \begin{cases}
				&\frac{\pi}{2}i^{\alpha + 1}H^{(1)}_{\alpha}(i x)\;\;\;\;\;\;\;\;\; -\pi < \Arg{(x)} \leq \frac{\pi}{2} \notag \\
				&\frac{\pi}{2}(-i)^{\alpha + 1}H^{(2)}_{\alpha}(i x)\;\;\;\; -\frac{\pi}{2} < \Arg{(x)} \leq \pi
			\end{cases}
		\end{align}
		The objective is to write $H^{(1,2)}_\alpha$ in terms of $K_\alpha$, while keeping in mind that the argument of $Y_\alpha$ has phase $0$. Since we already know that the argument of $Y_\alpha$ is $s z > 0$, we use the same for the rest of the calculation. So, we can invert the relation between $H^{(1,2)}$ and $K$ to write the following
		\begin{align}
			H^{(1)}_\alpha ( s z) &= \frac{2}{\pi}(-i)^{\alpha + 1}K_{\alpha}(-i s z)\;\;\;\; -\pi < \Arg{(-i s z)} \leq \frac{\pi}{2} \\
			H^{(2)}_\alpha ( s z) &= \frac{2}{\pi}(i)^{\alpha + 1}K_{\alpha}(i s z) \;\;\;\;\;\;\;\;\;\;\;\;\;\; -\frac{\pi}{2} < \Arg{(i s z)} \leq \pi
		\end{align}
		Note that both the conditions in the equations above hold, since $\Arg{(-i s z)} = \Arg{(-i)} + \Arg{(s z)} = -\frac{\pi}{2}$ and $\Arg{(i s z)} = \Arg{(i)} + \Arg{(s z)} = \frac{\pi}{2}$, which satisfy the corresponding inequalities. Hence we can now use these to write
		\begin{align}
			Y_{\alpha}(s z) = \frac{1}{2 i}\left(\frac{2}{\pi}(-i)^{\alpha + 1}K_{\alpha}(-i s z) -  \frac{2}{\pi}(i)^{\alpha + 1}K_{\alpha}(i s z)\right)
		\end{align}
		And this reduces to (setting $\alpha = p$ to keep up with the notation in the rest of the text)
		\begin{equation}
			Y_{p}( s z) = -\frac{1}{\pi} e^{p \pi i /2}(K_{p}(i s z) + (-1)^{p}K_{p}(-i s z))
		\end{equation}
	\end{mdframed}
The relation in the box connecting $Y$ and $K$ Bessel functions is not directly available in the tables we are aware of. But we have numerically double-checked our final result above for various values of the parameters. 
	This leads us to the following two integrals from \eqref{I3}
	\begin{align}
		I_{3,\,1} &= \int_{0}^{\infty}s^{p + \frac{d}{2}}K_{p}(-i s z)K_{\frac{d}{2}-1}(s X) \mathrm{d}s \notag \\
		I_{3,\,2} &= \int_{0}^{\infty}s^{p + \frac{d}{2}}K_{p}(i s z)K_{\frac{d}{2}-1}(s X) \mathrm{d}s \notag 
	\end{align}
	To evaluate these, we turn to the following integral identity
	\begin{align}\label{iden2}
		\int_{0}^{\infty}x^{-\lambda}K_{\mu}(a x)K_{\nu}(b x)\mathrm{d}x &= \frac{2^{-2 -\lambda}a^{-\nu + \lambda - 1}b^{\nu}}{\Gamma(1-\lambda)}\Gamma(\frac{1- \lambda + \mu + \nu}{2})\Gamma(\frac{1-\lambda - \mu + \nu}{2}) \nonumber \\
		&\times \Gamma(\frac{1-\lambda + \mu - \nu}{2})\Gamma(\frac{1-\lambda - \mu -\nu}{2}) \nonumber \\
		&\times \;_{2}F_{1}(\frac{1- \lambda + \mu + \nu}{2}, \frac{1-\lambda - \mu + \nu}{2}; 1 - \lambda; 1 - \frac{b^{2}}{a^{2}}) \nonumber \\
		&\;\;\;\; \forall \;\; \text{Re}(a + b) > 0, \;\; \text{Re}(\lambda) < 1 - |\text{Re}(\mu)| - |\text{Re}(\nu)|
	\end{align}
	The conditions for the identity are satisfied. The integrals become
	\begin{align}
		I_{3,\, 1} &= \frac{2^{p + \frac{d}{2} - 2}(-i z)^{-p - d}X^{\frac{d}{2}-1}}{\Gamma(1 + p + \frac{d}{2})}\Gamma(p + \frac{d}{2})\Gamma(\frac{d}{2})\Gamma(p+1)\;_{2}F_{1}\left(p + \frac{d}{2}, \frac{d}{2};1 + p + \frac{d}{2}; 1 + \frac{X^{2}}{z^{2}}\right) \\
		I_{3,\, 2} &= \frac{2^{p + \frac{d}{2} - 2}(i z)^{-p - d}X^{\frac{d}{2}-1}}{\Gamma(1 + p + \frac{d}{2})}\Gamma(p + \frac{d}{2})\Gamma(\frac{d}{2})\Gamma(p+1)\;_{2}F_{1}\left(p + \frac{d}{2}, \frac{d}{2};1 + p + \frac{d}{2}; 1 + \frac{X^{2}}{z^{2}}\right)
	\end{align}
	Therefore, we have the following integral result (by noting that $I_{3,\,1} = (-1)^{-p-d}I_{3,\,2}$
	\begin{align}
		&\int_{0}^{\infty}s^{p + \frac{d}{2}}Y_{p}(s z)K_{\frac{d}{2}-1}(s X)\mathrm{d} s = -\frac{e^{\frac{p \pi i}{2}}}{\pi}(1+(-1)^{d})I_{3,\,2} \notag \\
		&= -\frac{(1+(-1)^{d}) e^{\frac{p \pi i}{2}}}{\pi}\frac{2^{p + \frac{d}{2} - 2}(i z)^{-p - d}X^{\frac{d}{2}-1}}{\Gamma(1 + p + \frac{d}{2})}\Gamma(p + \frac{d}{2})\Gamma(\frac{d}{2})\Gamma(p+1)\;_{2}F_{1}\left(p + \frac{d}{2}, \frac{d}{2};1 + p + \frac{d}{2}; 1 + \frac{X^{2}}{z^{2}}\right) \notag \\
		&= -\frac{(1+(-1)^{d})2^{p + \frac{d}{2}-2}i^{-d}z^{-p - d}X^{\frac{d}{2}-1}}{(p  +\frac{d}{2})\pi}\Gamma(p+1)\Gamma(\frac{d}{2})\;_{2}F_{1}\left(p + \frac{d}{2}, \frac{d}{2};1 + p + \frac{d}{2}; 1 + \frac{X^{2}}{z^{2}}\right) \notag \\
		&= - \cos\left(\frac{\pi d}{2}\right)\frac{2^{p + \frac{d}{2}-1}z^{-p - d}X^{\frac{d}{2}-1}}{(p  +\frac{d}{2})\pi}\Gamma(p+1)\Gamma(\frac{d}{2})\;_{2}F_{1}\left(p + \frac{d}{2}, \frac{d}{2};1 + p + \frac{d}{2}; 1 + \frac{X^{2}}{z^{2}}\right)
	\end{align}
	Re-instating the omitted pre-factors, this expression therefore allows us to write $\tilde I_{3}$ as
	\begin{align}
		\tilde I_{3} = &\frac{\pi}{2^{p} (2\pi)^{d}\Gamma(p) }\sqrt{2}\Gamma(\frac{d}{2}-1)2^{\frac{d-3}{2}}X^{1 - \frac{d}{2}}\cos\left(\frac{\pi d}{2}\right)\frac{2^{p + \frac{d}{2}-1}z^{-p - \frac{d}{2}}X^{\frac{d}{2}-1}}{(p  +\frac{d}{2})\pi}\Gamma(p+1)\Gamma(\frac{d}{2}) \notag \\ &\times\;_{2}F_{1}\left(p + \frac{d}{2}, \frac{d}{2};1 + p + \frac{d}{2}; 1 + \frac{X^{2}}{z^{2}}\right)
	\end{align}
	This simplifies to 
	\begin{align}
		\tilde I_{3} = \frac{\pi ^{-d} p \cos \left(\frac{\pi  d}{2}\right) \Gamma \left(\frac{d}{2}-1\right) \Gamma \left(\frac{d}{2}\right)}{4 (p + \frac{d}{2})}z^{-p - \frac{d}{2}}\;_{2}F_{1}\left(p + \frac{d}{2}, \frac{d}{2};1 + p + \frac{d}{2}; 1 + \frac{X^{2}}{z^{2}}\right)
	\end{align}
	
	\subsection{Integral $\tilde I_2$}
	In this sub-section, we will evaluate the integral $\tilde I_{2}$, which we reiterate below 
	\begin{align}
		\tilde I_{2} = \frac{2}{2^{p}\Gamma(p) (2\pi)^{d} }\int |q|^{p}\ln(|q|)e^{i q. (x-x')}z^{d/2}J_{p}(|q|z)\mathrm{d}^{d}q
	\end{align}
	Following the same process as for $I_{3}$, this integral reduces to (again, leaving the original prefactors for now)
	\begin{align}
		I_{2} = \sqrt{2}\Gamma(\frac{d}{2}-1)2^{\frac{d-3}{2}}X^{1 - \frac{d}{2}}z^{d/2}\int_{0}^{\infty}s^{p + \frac{d}{2}}\ln(s)J_{p}(s z)K_{\frac{d}{2}-1}(s X)\mathrm{d} s \label{I2}
	\end{align}
	The integral that we have to evaluate is 
	\begin{align}
		\int_{0}^{\infty}s^{p + \frac{d}{2}}\ln(s)J_{p}(s z)K_{\frac{d}{2}-1}(s X)\mathrm{d} s
	\end{align}
	This is an integral of the following form
	\begin{align}
		\int_{0}^{\infty} x^{-\lambda}\ln(x)K_{\mu}(a x)J_{\nu}(b x)\mathrm{d}x = -\frac{\partial}{\partial \lambda}\int_{0}^{\infty} x^{-\lambda}K_{\mu}(a x)J_{\nu}(b x)\mathrm{d}x
	\end{align}
	We use the integral
	\begin{align}
		\int_{0}^{\infty} x^{-\lambda}K_{\mu}(a x)J_{\nu}(b x)\mathrm{d}x &= \frac{b^{\nu}}{2^{\lambda + 1}a^{\nu - \lambda + 1}\Gamma(1+\nu)}\Gamma(\frac{\nu - \lambda + \mu + 1}{2})\Gamma(\frac{\nu - \lambda - \mu + 1}{2})\nonumber \\ &\times\;_{2}F_{1}(\frac{\nu - \lambda + \mu + 1}{2}, \frac{\nu - \lambda - \mu + 1}{2}; \nu + 1; -\frac{b^{2}}{a^{2}}) \nonumber \\
		&\;\;\;\;\;\; \forall \;\; \text{Re}(a \pm i b) > 0 \;\;\&\;\; \text{Re}(\nu - \lambda + 1) > |\text{Re}(\mu)| \label{KJ}
	\end{align}
	Taking the derivative of this gives us the following result
	\begin{align}
		&\int_{0}^{\infty} x^{-\lambda}\ln(x)K_{\mu}(a x)J_{\nu}(b x)\mathrm{d}x = -\frac{2^{-\lambda -2}}{\Gamma (\nu +1)} b^{\nu } a^{\lambda -\nu -1} \Gamma(\kappa_{-})\Gamma(\kappa_{+})\Bigg(G^{(1)}_{a}(\kappa_{-},\kappa_{+},\nu+1,-\frac{b^{2}}{a^{2}})\notag \\ + &G^{(1)}_{b}(\kappa_{-},\kappa_{+},\nu+1,-\frac{b^{2}}{a^{2}}) +\;_{2}F_{1}(\kappa_{-},\kappa_{+},\nu+1,-\frac{b^{2}}{a^{2}})(\ln(\frac{4}{a^{2}}) + \psi(\kappa_{-}) + \psi(\kappa_{+}))\Bigg)
	\end{align}
	where we have used the shorthand $\kappa_{\pm} = \frac{-\lambda \pm \mu + \nu + 1}{2}$. The functions $G^{(1)}_{a, b}$ indicate the derivative of $\;_{2}F_{1}(a,b;c;z)$ with respect to the parameters $a$ or $b$ \cite{Ancarani_2009}. These functions are represented by the following series
	\begin{align}
		G^{(1)}_{a}(a,b;c;z) &= \sum_{n = 0}^{\infty}\frac{(a)_{n}(b)_{n}}{(c)_{n}}\left( \psi(a + n) - \psi(a) \right)\frac{z^{n}}{\Gamma(n+1)} \label{Ga}\\
		G^{(1)}_{b}(a,b;c;z) &= \sum_{n = 0}^{\infty}\frac{(a)_{n}(b)_{n}}{(c)_{n}}\left( \psi(b + n) - \psi(b) \right)\frac{z^{n}}{\Gamma(n+1)} \label{Gb}
	\end{align}
	Now we can plug in the values of $\lambda, \mu,\nu, a \,\& b$. From these values we note that $\kappa_{+} = p  + \frac{d}{2}$ and $\kappa_{-} = p + 1$. Thus the integral becomes
	\begin{align}
		&\int_{0}^{\infty}s^{p + \frac{d}{2}}\ln(s)J_{p}(s z)K_{\frac{d}{2}-1}(s X)\mathrm{d} s= -2^{p + \frac{d}{2} -2} z^{p} X^{-2 p - \frac{d}{2}-1} \Gamma(p  + \frac{d}{2})\Bigg(G^{(1)}_{a}(p + 1,p  + \frac{d}{2},p+1,-\frac{z^{2}}{X^{2}})\notag \\ + &G^{(1)}_{b}(p + 1,p  + \frac{d}{2},p+1,-\frac{z^{2}}{X^{2}}) +\;_{2}F_{1}(p + 1,p  + \frac{d}{2},p+1,-\frac{z^{2}}{X^{2}})(\ln(\frac{4}{X^{2}}) + \psi(p + 1) + \psi(p  + \frac{d}{2}))\Bigg)
	\end{align}
	A bit of simplification happens, since two of the entries of the hypergeomtric functions and their derivatives are the same. The sums in \eqref{Ga}-\eqref{Gb} can be explicitly evaluated to give
	\begin{align}
		G^{(1)}_a (a, b; a; z) &=  \sum_{n = 0}^{\infty}(b)_{n}(\psi(a + n) - \psi(a))\frac{z^{n}}{\Gamma(n+1)}\\
		G^{(1)}_b (a,b; a; z) &= -(1- z)^{-b}\ln(1 - z) \\
		\,_{2}F_{1}(a, b; a, z) &= (1 - z)^{-b}
	\end{align}
	We are not aware of a closed form expression for $G^{(1)}_a (a, b; a; z)$. Putting back the pre-factors, the final form of $\tilde I_{2}$ is
	\begin{align}
	\tilde I_{2} &= -\frac{z^{p + \frac{d}{2}}\pi ^{-d} \Gamma \left(\frac{d}{2}-1\right) \Gamma \left(\frac{d}{2}+p\right)}{4 \Gamma (p)}\Bigg(G^{(1)}_{a}(p + 1,p  + \frac{d}{2},p+1,-\frac{z^{2}}{X^{2}})- \left(1 + \frac{z^{2}}{X^{2}}\right)^{-p - \frac{d}{2}}\ln\left(1 + \frac{z^{2}}{X^{2}}\right) + \notag \\ &\left(1  + \frac{z^{2}}{X^{2}} \right)^{-p - \frac{d}{2}}\left(\ln(\frac{4}{X^{2}}) + \psi(p + 1) + \psi(p  + \frac{d}{2})\right)\Bigg)X^{-2 p - d}
	\end{align}
	
	\subsection{Integral $\tilde I_1$}
	Finally we turn to the integral $\tilde I_{1}$, which we repeat below
	\begin{align}
		I_{1} = -\frac{(-\gamma + \psi(p+1) + 2\ln(2))}{2^{p} (2\pi)^{d} \Gamma(p) }\int |q|^{p}e^{i q. (x-x')}z^{d/2}J_{p}(|q|z)\mathrm{d}^{d}q
	\end{align}
	Leaving the prefactors aside for the moment
	\begin{align}
		I_{1} &= \int |q|^{p}e^{i q. (x-x')}z^{d/2}J_{p}(|q|z)\mathrm{d}^{d}q \notag \\ &= \sqrt{2}\Gamma(\frac{d}{2}-1)2^{\frac{d-3}{2}}X^{1 - \frac{d}{2}}z^{d/2}\int_{0}^{\infty}s^{p + \frac{d}{2}}J_{p}(s z)K_{\frac{d}{2}-1}(s X)\mathrm{d} s
	\end{align}
	The $s$-integral is evaluated by considering \eqref{KJ}. From there, we get
	\begin{align}
		\int_{0}^{\infty}s^{p + \frac{d}{2}}J_{p}(s z)K_{\frac{d}{2}-1}(s X)\mathrm{d} s &= \frac{z^{p}}{2^{-p - \frac{d}{2}+1}X^{2 p + \frac{d}{2}+1}}\Gamma(p + \frac{d}{2})\Gamma(p + 1)\;_{2}F_{1}(p + 1, p + \frac{d}{2}; p + 1; - \frac{z^{2}}{X^{2}}) \notag \\
		&= \frac{z^{p}}{2^{-p - \frac{d}{2}+1}X^{2 p + \frac{d}{2}+1}}\Gamma(p + \frac{d}{2})\Gamma(p + 1)\left(1 + \frac{z^{2}}{X^{2}}\right)^{-p - \frac{d}{2}}
	\end{align}
	Thus the full integral $\tilde I_{1}$ becomes
	\begin{align}
	\tilde I_{1} = \frac{1}{4} \pi ^{-d} p \Gamma \left(\frac{d}{2}-1\right) \Gamma \left(\frac{d}{2}+p\right) \left(2 \gamma -\psi(p+1)-\log (4)\right)\left( \frac{X^{2} + z^{2}}{z^{2}} \right)^{-p - \frac{d}{2}}.
	\end{align}
	\subsection{The Final Kernel}
	Finally, we put together the integrals $\tilde I_{1}$, $\tilde I_{2}$ and $\tilde I_{3}$ to write the full kernel $K_{2}$ for the non-normalizable mode for integer $\nu = p$. We also use the more familiar notation of $\Delta, d$ by writing $p = \Delta - \frac{d}{2}$ wherever it appears. We also introduce the $\sigma z'$ notation wherever possible.\footnote{The limit $\lim_{z' \rightarrow 0}$ is understood in the following expressions.} The three integrals were evaluated to the following forms
	\begin{align}
	\tilde I_{1} &= \frac{1}{8} \pi ^{-d} (d-2 \Delta ) \Gamma \left(\frac{d}{2}-1\right) \Gamma (\Delta ) \left(2 \gamma - \psi({\Delta - \frac{d}{2}+1})-\log (4)\right)\left( 2\sigma z'\right)^{-\Delta} \\
	\tilde I_{2} &= -\frac{\pi ^{-d} \Gamma \left(\frac{d}{2}-1\right) \Gamma (\Delta )}{4 \Gamma \left(\Delta -\frac{d}{2}\right)}\Bigg( \left(2\sigma z' \right)^{-\Delta}\left(\ln(\frac{2}{z^{2}}) + \psi(\Delta - \frac{d}{2} + 1) + \psi(\Delta) - \ln(\sigma z')\right) \notag \\ &+ \frac{z^{\Delta}}{X^{2\Delta}}G^{(1)}_{a}(\Delta - \frac{d}{2} + 1,\Delta,\Delta - \frac{d}{2}+1,-\frac{z^{2}}{X^{2}})\Bigg) \\
	\tilde I_{3} &= -\frac{\pi ^{-d} (d-2 \Delta ) \cos \left(\frac{\pi  d}{2}\right) \Gamma \left(\frac{d}{2}-1\right) \Gamma \left(\frac{d}{2}\right)}{8 \Delta }z^{-\Delta}\;_{2}F_{1}\left(\Delta, \frac{d}{2};1 + \Delta; 1 + \frac{X^{2}}{z^{2}}\right)
	\end{align}
	The final kernel $K_2$ is written simply as $K_{2} =\tilde  I_{1} + \tilde I_{2} + \tilde I_{3}$.

\section{Euclidean AdS Wave Equation in Terms of Chordal Distance}
\label{chordalwaveeqn}

We recall the PoincarPoincar\'ee chordal distance as follows
\begin{equation}
    \sigma(z,x;z',x') = \frac{z^{2} + z'^{2} + |x-x'|^{2}}{2 z z'} \label{chordPformG}
\end{equation}
We need to evaluate the partial derivatives, which are as follows
\begin{align}
    \frac{\partial \Phi}{\partial z} &= \Big(\frac{1}{z'} - \frac{\sigma}{z}\Big)\frac{\partial\Phi}{\partial \sigma} \\
    \frac{\partial^{2}\Phi}{\partial z^{2}} &= \Big( \frac{2 \sigma}{z^{2}} - \frac{1}{z z'} \Big)\frac{\partial \Phi}{\partial \sigma} + \Big( \frac{1}{z'} - \frac{\sigma}{z} \Big)^{2}\frac{\partial^{2}\Phi}{\partial \sigma^{2}}\\
    \frac{\partial^{2}\Phi}{\partial \vec{x}^{2}} &= \Big( \frac{2 \sigma}{z z'} - \frac{1}{z^{2}} - \frac{1}{z'^{2} }\Big) \frac{\partial^{2}\Phi}{\partial \sigma^{2}} + \frac{d}{z z'}\frac{\partial \Phi}{\partial \sigma}
\end{align}
Plugging this into the Euclidean wave equation
\begin{equation}
\Big(z^{2}\partial^{2}_{z} - z(d-1)\partial_{z} + z^{2}\partial^{2}_{\vec{x}} - m^{2}\Big)\Phi(x,t,z) = 0
\end{equation}
we get
\begin{align}
    (2 \sigma - \frac{z}{z'})\Phi' + (\frac{z}{z'} - \sigma)\Phi'' - (d-1)(\frac{z}{z'} - \sigma)\Phi' + (\frac{2\sigma z}{z'} - 1 - \frac{z^{2}}{z'^{2}})\Phi'' + d\frac{z}{z'}\Phi' - \Delta(\Delta - d)\Phi = 0
\end{align}
where we have used the notation 
\begin{align*}
    \Phi' &= \frac{d \Phi}{d \sigma} \\
    \Phi'' &= \frac{d^{2} \Phi}{d \sigma^{2}}\\
    m^{2} &= \Delta(\Delta - d)
\end{align*}
And from this we get
\begin{equation}
(\sigma^{2} - 1)\frac{d^{2}\Phi(\sigma)}{d \sigma^{2}} + (d+1)\sigma \frac{d\Phi(\sigma)}{d\sigma} - \Delta(\Delta - d)\Phi(\sigma) = 0
\end{equation}


\end{document}